%% file: od_estimation.tex
\documentclass{article}

\usepackage[a4paper,margin=1in]{geometry}
\usepackage{microtype}
\usepackage{caption}
\usepackage{fancyhdr}
\usepackage[utf8]{inputenc}
\usepackage[T1]{fontenc}
\usepackage{amsmath,amssymb,bm}
\usepackage{booktabs,array,longtable}
\usepackage{graphicx}
\usepackage{algorithm}
\usepackage{algpseudocode}
\usepackage{pdflscape}
\usepackage[colorlinks=true,allcolors=blue]{hyperref}
\hypersetup{pdftitle={Calibrating demand models versus adjusting OD matrices with traffic counts: a controlled comparison},pdfauthor={Yue Li, Shujuan Chen, Ying Jin}}

\usepackage[numbers,sort&compress]{natbib}

\newcommand{\tbl}[2]{\caption{#1}\noindent\makebox[\linewidth]{#2}}
\newcommand{\keywords}[1]{\par\medskip\noindent\textbf{Keywords:} #1}

\newcommand{\CPC}{\mathrm{CPC}}
\newcommand{\NRMSE}{\mathrm{NRMSE}}
\newcommand{\Sset}{\mathcal{S}}
\newcommand{\Hset}{\mathcal{H}}
\newcommand{\Eset}{\mathcal{E}}
\newcommand{\Zset}{\mathcal{Z}}

\title{Calibrating demand models versus adjusting OD matrices with traffic counts: a controlled comparison}

\author{Yue Li, Shujuan Chen\thanks{Corresponding author: \texttt{sc2343@cam.ac.uk}}, and Ying Jin\\
Martin Centre for Architectural and Urban Studies\\
University of Cambridge, Cambridge, United Kingdom}

\date{}

\begin{document}

\maketitle

\begin{abstract}
Traffic counts are routinely used to calibrate travel demand models, particularly when the origin–destination (OD) matrix is not observed directly. An important modelling decision is whether to adjust individual OD flows or the parameters of the demand model that generates them. The latter approach uses fewer parameters and preserves a relationship between zonal characteristics and demand, but its accuracy depends on the assumed model structure. We compare three established approaches using the same prior information, traffic counts and equilibrium assignment. They estimate five production–gravity coefficients, origin and destination adjustment factors, or individual OD flows. Synthetic experiments on the Eastern Massachusetts and Chicago Sketch networks allow recovery to be assessed against known demand under different forms of model misspecification. Calibrating the five-coefficient model gives better OD recovery and held-out link-flow prediction when demand follows its assumptions or a two-purpose gravity process. Its advantage declines or reverses when destination choice or production relationships vary across zones. Cellwise adjustment achieves the closest fit to calibration counts in 65 of 66 evaluated conditions, yet it recovers OD demand least accurately of the three approaches in 45 of them. These findings show that model parameterisation matters beyond calibration fit. For transport planning, traffic counts should be used alongside independent evidence on trip production and distribution when selecting and validating a demand model.
\end{abstract}


\section{Introduction}\label{sec:intro}

Travel demand models (TDMs) are widely used to assess road networks, land-use development and transport policies. A conventional TDM relates socioeconomic and land-use characteristics to trip production, trip distribution, mode choice and traffic assignment \citep{OrtuzarWillumsen2011}. The resulting origin–destination (OD) matrix is an important input to network analysis. Because complete OD demand is rarely observed, modellers commonly adjust an initial matrix using traffic counts. This improves agreement between modelled and observed traffic, but it does not necessarily improve the estimated journeys between origins and destinations.

Traffic counts measure the total flow on a link without identifying the OD pairs using it. A limited number of counts therefore cannot determine a large OD matrix uniquely, and many matrices can reproduce the same counts \citep{Bell1983,YangEtAl1991,Hazelton2003}. OD estimation methods address this problem using prior matrices, regularisation and assumptions about demand \citep{VanZuylenWillumsen1980,CascettaNguyen1988,Hazelton2003}. The choice of parameterisation determines which parts of the demand can change during calibration. Cellwise estimation adjusts individual OD flows. Factor models adjust shared origin and destination effects. Demand-model calibration instead updates a small set of trip-production or distribution coefficients, and the model then generates the whole matrix \citep{Hogberg1976,Willumsen1981,TaminWillumsen1989,CascettaRusso1997}.

These choices have different consequences. A low-dimensional demand model can use counts on a few roads to update common relationships across the network. It also retains a connection between demand and socioeconomic inputs, which is useful when the model is applied to changed planning conditions. However, the same structure limits what can be recovered. A single distance-decay coefficient cannot describe all origin-specific travel preferences, and common production coefficients cannot represent every local difference. A flexible OD estimator can accommodate more variation, but its additional parameters may be weakly constrained by sparse counts. It can then reproduce the counts closely while recovering the OD matrix poorly \citep{Hazelton2003,Bierlaire2002}. This trade-off is relevant to both base-year calibration and the reliability of subsequent scenario analysis.

Methods for OD estimation and count-based model calibration are well established. Entropy and information-based methods select the matrix closest to a prior matrix among those consistent with the counts \citep{VanZuylenWillumsen1980,LopezOspinaEtAl2022}. Generalised least squares and Bayesian estimators weigh count residuals against deviations from the prior \citep{Cascetta1984,Maher1983,CascettaNguyen1988}. Bilevel and gradient-based formulations include equilibrium assignment in the estimation \citep{Fisk1988,YangEtAl1992,Spiess1990,Yang1995,LundgrenPeterson2008}. Most of these methods adjust OD flows directly. Structured estimators instead exploit known relationships among OD flows \citep{BierlaireToint1995}. More recent studies estimate demand parameters through computational graphs or use lower-dimensional representations of large OD matrices \citep{WuEtAl2018,MaEtAl2020,GuardaEtAl2024,PengEtAl2025}.

Comparative studies usually evaluate a new estimator against established alternatives on a benchmark network, examine identification and solution quality for a given formulation, or predict traffic at unobserved locations \citep{FrederixEtAl2013,AntoniouEtAl2016,RosRocaEtAl2021,GuardaQian2025}. The parameterisation itself has received less controlled attention. A remaining practical concern is how the assumptions used to parameterise demand affect recovery when the true demand differs from those assumptions. In particular, it is not well established how the advantage of a low-dimensional demand model depends on the form and severity of misspecification and on count coverage, when every method uses the same counts, prior information and equilibrium assignment. A closer count fit may conceal an incorrect OD matrix, while a restricted model may become biased as its coefficients compensate for omitted demand patterns. Such a comparison requires known OD demand, which real networks do not provide.

This study compares three established estimation approaches under common prior information and equilibrium assignment. The first calibrates five coefficients of a production–gravity model. The second adjusts origin and destination factors, and the third adjusts individual OD flows. We use synthetic demand on the Eastern Massachusetts and Chicago Sketch networks so that recovery can be evaluated against known OD matrices. The data-generating processes include matched demand, multiple trip purposes, heterogeneous destination choice and spatially varying production effects. We vary the available count coverage and evaluate both OD recovery and prediction on unobserved links.

The study examines when structural assumptions improve recovery, when misspecification reverses the comparison, and whether calibration fit indicates demand accuracy. The objective is to assess how the choice of TDM parameterisation affects OD recovery when traffic counts are used for calibration. The contribution is a paired, controlled comparison that separates the choice of demand parameterisation from the quality of fit to observed counts, with negative results reported alongside positive ones. We discuss the implications for validating OD matrices and applying calibrated models under changed conditions, while distinguishing these implications from a direct test of real-world forecasting performance.

\section{Estimation methods}\label{sec:methods}

\subsection{Common setting}\label{sec:common}

Let $q_{ij}$ denote the number of trips from zone $i$ to zone $j$, and let $\mathbf q$ contain demand for the supported OD pairs $\Omega$. Traffic assignment maps OD demand to link flows $\mathbf x=\Phi(\mathbf q)$. We use deterministic user equilibrium with the fixed link cost functions supplied by the benchmark networks \citep{Wardrop1952,Sheffi1985}. The full assignment formulation and solver tolerances are given in Appendix Sections~\ref{sec:assignment} and \ref{app-sec:numerics}.

Counts $y_a$ are available on a set of calibration links $\Sset$, while links in $\Hset$ are held out for evaluation. All three approaches use the same prior matrix $\mathbf q^{(0)}$, network and normalised count loss,
\begin{equation}
L(\mathbf q)=\frac{1}{|\Sset|}\sum_{a\in\Sset}\left(\frac{\Phi_a(\mathbf q)-y_a}{s_y}\right)^2,\qquad
s_y=\max\left\{10^{-9},\left(\frac{1}{|\Sset|}\sum_{a\in\Sset}y_a^2\right)^{1/2}\right\}.
\label{eq:loss}
\end{equation}
Each method minimises this loss with a penalty that controls deviation from its prior. The difference is which demand quantities can be estimated from the counts.

\subsection{Five-coefficient production–gravity model}\label{sec:generator}

The production–gravity model represents trip production in zone $i$ as
\begin{equation}
O_i=\kappa N_i\exp\Big(\gamma_0+\sum_{k=1}^{3}\gamma_k z_{ik}\Big),
\label{eq:production}
\end{equation}
where $N_i$ is population, $z_{ik}$ are standardised zonal covariates (car availability, working-age share and employed-adult share) and $\kappa$ is a fixed exposure scale that is not estimated. Production is distributed by an origin-constrained gravity model with a geometric accessibility correction,
\begin{equation}
q_{ij}=O_i\,\frac{A_jK_{ij}/H_j}{\sum_{k:(i,k)\in\Omega}A_kK_{ik}/H_k},\qquad K_{ij}=e^{-b\,c_{ij}},\qquad H_j=\sum_{\ell:(\ell,j)\in\Omega}K_{\ell j},
\label{eq:gravity}
\end{equation}
where $A_j$ is destination attractiveness, $c_{ij}$ is the normalised free-flow travel time and $b>0$ is the decay coefficient. The accessibility term $H_j$ reduces the relative attractiveness of destinations reached at low cost from many origins. It does not impose destination totals. The five estimated coefficients are $\bm\theta=(\gamma_0,\gamma_1,\gamma_2,\gamma_3,\eta)$, where $b=e^{\eta}$. A quadratic penalty controls departure from the prior coefficients,
\begin{equation}
R_G(\bm\theta)=\frac{1}{5}\sum_{k=1}^{5}\left(\frac{\theta_k-\theta_k^{(0)}}{\sigma_k}\right)^{2},\qquad \bm\sigma=(0.20,0.20,0.20,0.20,0.25),
\label{eq:genpen}
\end{equation}
and the coefficients are bounded by $|\gamma_0|\le2$, $|\gamma_k|\le1.5$ and $0.2\le b\le8$. The OD matrix is generated from the fitted coefficients rather than estimated cell by cell.

\subsection{Origin–destination factor adjustment}\label{sec:factor}

The intermediate estimator rescales the prior matrix by an overall level, origin and destination factors and a decay adjustment,
\begin{equation}
q_{ij}=q^{(0)}_{ij}\exp\left(\tau+u_i+v_j-\Delta b\,c_{ij}\right).
\label{eq:factor}
\end{equation}
Two weighted normalisation constraints, $\sum_iO^{(0)}_iu_i=0$ and $\sum_jD^{(0)}_jv_j=0$ with the prior origin and destination totals as weights, remove the redundant parameterisation. This leaves $2|\Zset|$ free coordinates on the fully supported experimental networks. The penalty is
\begin{equation}
R_F=\frac{1}{5}\Big[(\tau/\sigma_\tau)^2+\sum_i(u_i/\sigma_u)^2+\sum_j(v_j/\sigma_v)^2+(\Delta b/\sigma_{\Delta b})^2\Big],
\label{eq:facpen}
\end{equation}
with scales chosen so that the prior coefficient uncertainty of the production–gravity model maps onto matched marginal uncertainty in these coordinates (Appendix Section~\ref{sec:configs}). The factors are bounded by $|\tau|,|u_i|,|v_j|\le5$ and the total decay by $0.2\le b^{(0)}+\Delta b\le8$. The factor family contains the matched truth, but its factors are free zone effects that do not follow the demographic covariates.

\subsection{Cellwise OD adjustment}\label{sec:cellwise}

The most flexible estimator adjusts every supported OD cell and penalises the generalised relative entropy to the prior matrix,
\begin{equation}
R_C(\mathbf q)=\frac{\sum_{(i,j)\in\Omega}\left[q_{ij}\log(q_{ij}/q^{(0)}_{ij})-q_{ij}+q^{(0)}_{ij}\right]}{\sum_{(i,j)\in\Omega}q^{(0)}_{ij}},\qquad q_{ij}\ge10^{-12}q^{(0)}_{ij}.
\label{eq:cell}
\end{equation}
It has $|\Omega|$ free coordinates. A squared-deviation penalty is evaluated as a secondary variant (Appendix Section~\ref{sec:additional}).

\subsection{Calibration and numerical solution}\label{sec:calibration}

Each method minimises $L(\mathbf q)+\lambda R$ over its own parameters. The regularisation weight $\lambda$ is selected by cross-validation on the calibration counts. Traffic assignment is re-solved for each candidate demand estimate. To reduce computational cost, the optimisers use frozen-route derivatives as approximate sensitivities of link flows to OD demand \citep{Spiess1990}. The production–gravity estimator uses damped Gauss–Newton (Levenberg–Marquardt) steps with analytic demand derivatives \citep{Levenberg1944,Marquardt1963}. The factor and cellwise estimators use L-BFGS-B to minimise successive frozen-route approximations \citep{ByrdEtAl1995}. An update is accepted only when the full objective decreases after a new equilibrium solve.

Frozen-route derivatives omit route changes and do not guarantee stationarity of the equilibrium objective. We evaluate this numerical limitation separately. Appendix Sections~\ref{sec:jacobian}–\ref{app-sec:numerics} document the derivatives, optimisation algorithm, convergence criteria and additional checks.

The prior generator, without calibration, provides a reference. Two production–gravity ablations separately test calibration of demographic slopes and impedance decay. Their definitions and results are given in the appendices.

\section{Experimental design}\label{sec:design}

\subsection{Networks and observations}\label{sec:networks}

We use the Eastern Massachusetts (EMA) and Chicago Sketch networks from the Transportation Networks for Research collection \citep{TNTP}. The EMA network is a highway subnetwork whose 74 nodes all act as zones \citep{ZhangEtAl2016}. Chicago Sketch, developed by the Chicago Area Transportation Study, has 387 zones represented by pure centroids, 933 nodes and 2,950 directed links. Routes may not pass through Chicago centroids, and the 774 centroid connectors are excluded from observation. Only link supply is taken from the benchmarks. All OD demand is synthetic. Table~\ref{tab:networks} gives the observation design and the estimator dimensions.

\begin{table}[htbp]\small
\tbl{Networks, observation design and estimator dimensions. Coverage is a share of all eligible directed links. The held-out set is fixed within each world and is disjoint from every calibration set.}
{\begin{tabular}{lrr}\toprule
 & EMA & Chicago Sketch\\\midrule
Zones / nodes / directed links & 74 / 74 / 258 & 387 / 933 / 2,950\\
Links eligible for observation & 258 & 2,176\\
Held-out links (20\% of eligible) & 52 & 436\\
Calibration links at 5, 10, 20, 40\% coverage & 13, 26, 52, 104 & 109, 218, 436, –\\
Supported OD pairs (cellwise coordinates) & 5,402 & 149,382\\
OD-factor coordinates & 148 & 774\\
Production–gravity coefficients & 5 & 5\\\bottomrule
\end{tabular}}
\label{tab:networks}
\end{table}

Each network has 20 independently generated synthetic worlds and two prior realisations per world. Within a world, a random permutation of the eligible links defines the held-out set (its first 20\%) and nested calibration sets (prefixes of the remainder). Count noise is multiplicative, $y_a=\max\{0,x^*_a(1+0.05\,\epsilon_a)\}$ with $\epsilon_a\sim N(0,1)$, where $x^*_a$ is the equilibrium flow of the true demand. One noise draw per link and world is shared by all estimators and calibration sets.

\subsection{Synthetic inputs and prior information}\label{sec:inputs}

Population and attractiveness are lognormal with medians 1,000 and 500 and log-scale standard deviations 0.45 and 0.65. The three demographic shares are bounded logistic transformations of Gaussian fields that combine an independent component with a spatially smoothed one. Smoothing uses an exponential kernel in symmetric free-flow travel time. OD impedance $c_{ij}$ is the free-flow shortest-path time divided by its median over supported pairs. The exposure scale $\kappa$ is set once per world so that the matched demand produces a network travel-time inflation of 1.25 (total travel time over total free-flow travel time), and it is then held fixed for every truth and estimator in that world. The realised inflation ranges from 1.245 to 1.255 across the 40 worlds.

Prior coefficients are drawn around the matched coefficients, $\bm\theta^{(0)}=\mathrm{proj}(\bm\theta^*+\bm\varepsilon)$ with independent normal errors of standard deviations $\bm\sigma$ in Equation~\eqref{eq:genpen}, projected onto the coefficient bounds. The same prior is used under every truth, including truths that the restricted model cannot represent. The prior therefore carries comparable information across conditions, and it favours the restricted model when truth is matched. All equations, parameter values and random streams are given in Appendix Section~\ref{sec:synthetic}, where each stored world is reproduced from these equations to machine precision.

\subsection{Demand-generating processes}\label{sec:truths}

Synthetic truth either follows the fitted model or departs from it in one component, at graded severity (Table~\ref{tab:truths}). The estimators and their settings are the same under every truth. Severity is reported both by the nominal parameter and by the realised deviation, the median OD common part of trips (Equation~\eqref{eq:cpc}) between each truth and the matched truth of the same world.

\begin{table}[htbp]
\small
\tbl{Synthetic demand conditions and observation coverage. The fitted production–gravity model is unchanged across conditions. Full data-generating equations and parameter values are in Appendix Section~\ref{sec:truthsS}.}
{\begin{tabular}{>{\raggedright\arraybackslash}p{0.20\linewidth}>{\raggedright\arraybackslash}p{0.48\linewidth}>{\raggedright\arraybackslash}p{0.24\linewidth}}
\toprule
Condition & Difference from the fitted demand model & Networks and coverage\\
\midrule
Matched & Same production and gravity distribution & EMA 5–40\%, Chicago~5–20\%\\
Two-purpose & Two purposes with different destination attractiveness and decay & EMA 5–40\%, Chicago~5–20\%\\
Decay contrast & Stronger difference between the two purpose-specific decay parameters & EMA 10–40\%\\
Mixture & Combination of matched and two-purpose demand & EMA 10–40\%\\
Heterogeneous destination choice & Origin-specific distance decay and OD interactions, without accessibility correction & EMA 10–40\%, Chicago 10 and 20\% for strongest condition\\
Production misspecification & Omitted demographic interaction or spatially varying production effect, applied to two-purpose demand & EMA 10–40\%\\
\bottomrule
\end{tabular}}
\label{tab:truths}
\end{table}

The two-purpose truth keeps total origin production but distributes it by two gravity components with different attractiveness and decay. The decay-contrast variants increase the difference between the two purposes, with decays of 0.5 and 5 or 0.25 and 8 in place of 1 and 3. The destination-choice family replaces the accessibility-corrected gravity model by a multinomial logit destination choice. Its decay varies by origin with car availability at strength $\zeta\in\{0,0.5,1\}$, and a smoothed random OD interaction with weight $\rho\in\{0,0.3,0.7\}$ adds pair-specific utility. Even its anchor ($\zeta=\rho=0$) differs from the fitted distribution, because it omits the accessibility correction. The two production families multiply the two-purpose production by an omitted bounded interaction of strength $\delta\in\{0.25,0.5,1\}$ or by a car-availability effect of strength $\phi\in\{0.5,1\}$ that varies smoothly across space. These are controlled departures, not calibrated descriptions of real behaviour. Two further sensitivities, at 10\% coverage, bias two prior coefficients ($\gamma_1$ and $\eta$ shifted by $+0.30$) or give the estimators link capacities perturbed by mean-one lognormal factors with log-scale standard deviation 0.15. Truth and evaluation then use the nominal capacities. These sensitivities are reported in Appendix Section~\ref{sec:additional}.

\subsection{Tuning and evaluation}\label{sec:evaluation}

Penalty weights are selected by three-fold cross-validation within the calibration set, using validation count error only. The candidate weights are $\{0,0.01,0.1,1,10\}$ for the production–gravity model and its ablations, and $\{0.001,0.01,0.1,1,10,\infty\}$ for the matrix estimators, where $\infty$ returns the prior matrix. Ties favour the larger weight, and selection of the largest finite weight triggers one tenfold extension. The selected weight is refitted on the whole calibration set. Evaluation budgets, tolerances and bounds are listed in Appendix Section~\ref{sec:configs}.

OD recovery is measured by the common part of trips,
\begin{equation}
\CPC(\hat{\mathbf q},\mathbf q^*)=\frac{2\sum_{(i,j)\in\Omega}\min(\hat q_{ij},q^*_{ij})}{\sum_{(i,j)\in\Omega}\hat q_{ij}+\sum_{(i,j)\in\Omega}q^*_{ij}},
\label{eq:cpc}
\end{equation}
which equals one for exact recovery. Prediction on unobserved links is measured by the normalised root mean squared error on the held-out links,
\begin{equation}
\NRMSE_{\Hset}=\frac{\big(|\Hset|^{-1}\sum_{a\in\Hset}(\hat x_a-x^*_a)^2\big)^{1/2}}{|\Hset|^{-1}\sum_{a\in\Hset}x^*_a},
\label{eq:nrmse}
\end{equation}
where $\hat x_a$ is the equilibrium flow of the estimated demand on the nominal network and $x^*_a$ is the noise-free true flow. Calibration fit is the same NRMSE computed on the calibration links against the noisy counts. All flows used for evaluation are solved to a relative gap of $10^{-6}$. None of the 22,640 evaluation assignments failed to converge.

\subsection{Paired design and inference}\label{sec:inference}

All estimators in a world share inputs, prior, counts, held-out links and assignment. The statistical unit is the world. The two prior realisations are averaged within each world before paired differences are formed, and a world enters a comparison only if both priors are present for both estimators. Every reported contrast is oriented so that a positive value favours the production–gravity model: generator minus comparator for CPC, and comparator minus generator for error measures. We report the mean paired difference across the 20 worlds with a 95\% percentile interval from 10,000 bootstrap resamples of worlds, and $W$, the number of worlds favouring the generator. Method performance is summarised separately by the median across worlds. An interval excluding zero is described as favouring one estimator. Otherwise the comparison is described as inconclusive.

The evidence comes from two programmes. The confirmation programme was specified and frozen before its worlds were generated. It covers the matched and two-purpose truths at 5, 10 and 20\% coverage on both networks, with prespecified primary comparisons at 10\%, and uses bootstrap seed 20261028. The post-confirmation programme was specified after the EMA confirmation results had been seen. It adds 40\% coverage, every misspecification family, the prior and capacity sensitivities and the numerical diagnostics, and it uses bootstrap seed 20261031. Its comparisons are exploratory. Where the programmes overlap, their 1,780 shared records are identical. Conditions run on Chicago were limited by the cost of cellwise estimation on 149,382 OD pairs. Table~\ref{tab:truths} lists them. All 2,600 datasets completed without failures.

\section{Results}\label{sec:results}

\subsection{Recovery under matched and two-purpose demand}\label{sec:res_anchor}

Table~\ref{tab:primary} reports the prespecified comparisons of the confirmation programme at 10\% coverage. Under matched truth, the production–gravity model recovers the OD matrix more accurately than both matrix estimators on both networks. Its median OD CPC is 0.963 on EMA and 0.981 on Chicago, against 0.864 and 0.888 for OD-factor adjustment and 0.837 and 0.831 for cellwise adjustment. The mean paired advantage over cellwise adjustment is 0.123 on EMA and 0.141 on Chicago. Under the two-purpose truth the advantages are smaller but remain in favour of the restricted model: 0.064 and 0.048 over factor adjustment, and 0.075 and 0.090 over cellwise adjustment. Held-out link-flow error shows the same ordering. All 24 prespecified confirmation endpoints, which also include a scenario-projection endpoint reported in Appendix Section~\ref{sec:projection}, favour the generator.

\begin{table}[htbp]\footnotesize
\tbl{Primary comparisons at 10\% coverage (confirmation programme, 20 worlds). Panel A: median over worlds of the world-mean metric. Panel B: mean paired difference, positive favouring the five-coefficient generator, with the 95\% world-bootstrap interval beneath and $W$, the number of worlds favouring the generator, in parentheses. Bold: interval excludes zero. Calibration NRMSE is computed on the calibration links against the noisy counts.}
{\input{tables/tab_primary.tex}}
\label{tab:primary}
\end{table}

Better OD recovery does not require a closer fit to the observed counts. Cellwise adjustment has the lowest calibration error in each of these strata (Table~\ref{tab:primary}, Panel A). On EMA its median calibration NRMSE is 0.010 under matched truth, against 0.057 for the restricted model, but its median OD CPC is 0.126 lower. OD-factor adjustment fits the Chicago counts less closely than the restricted model and also recovers OD demand less accurately.

\begin{figure}[htbp]
\includegraphics[width=\linewidth]{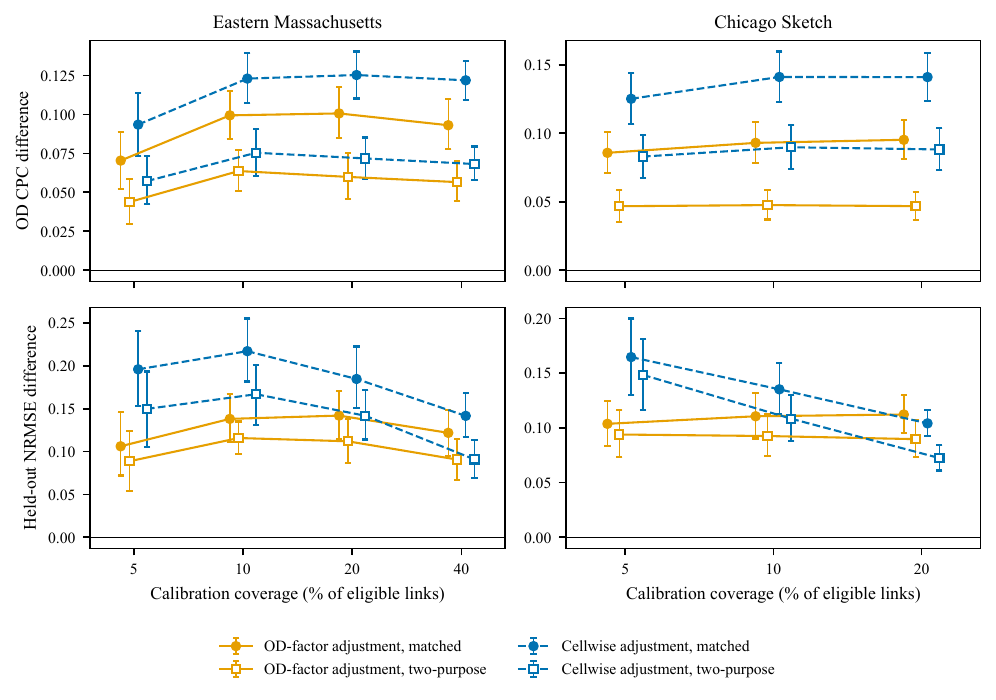}
\caption{Mean paired difference between the five-coefficient generator and each matrix estimator under matched and two-purpose truth, by calibration coverage. Upper panels: OD CPC (generator minus comparator). Lower panels: held-out link-flow NRMSE (comparator minus generator). Positive values favour the generator. Bars are 95\% world-bootstrap intervals over 20 worlds. The 40\% coverage was evaluated on EMA only.}
\label{fig:anchor}
\end{figure}

Figure~\ref{fig:anchor} shows the comparisons across coverage. The OD CPC advantage over both matrix estimators is stable between 10\% and 40\% coverage on EMA and between 5\% and 20\% on Chicago. It is smaller at 5\% on EMA, where only 13 links are counted. The held-out NRMSE advantage over cellwise adjustment narrows as coverage increases: on EMA under matched truth it falls from 0.217 at 10\% to 0.142 at 40\%, and on Chicago from 0.165 at 5\% to 0.104 at 20\%. Every interval in Figure~\ref{fig:anchor} lies above zero. A biased prior widens the advantage of the restricted model, and a capacity error in the estimators' assignment narrows it on EMA without reversing it (Appendix Section~\ref{sec:additional}).

\subsection{Destination-choice misspecification and estimator reversals}\label{sec:res_dest}

\begin{figure}[htbp]
\includegraphics[width=\linewidth]{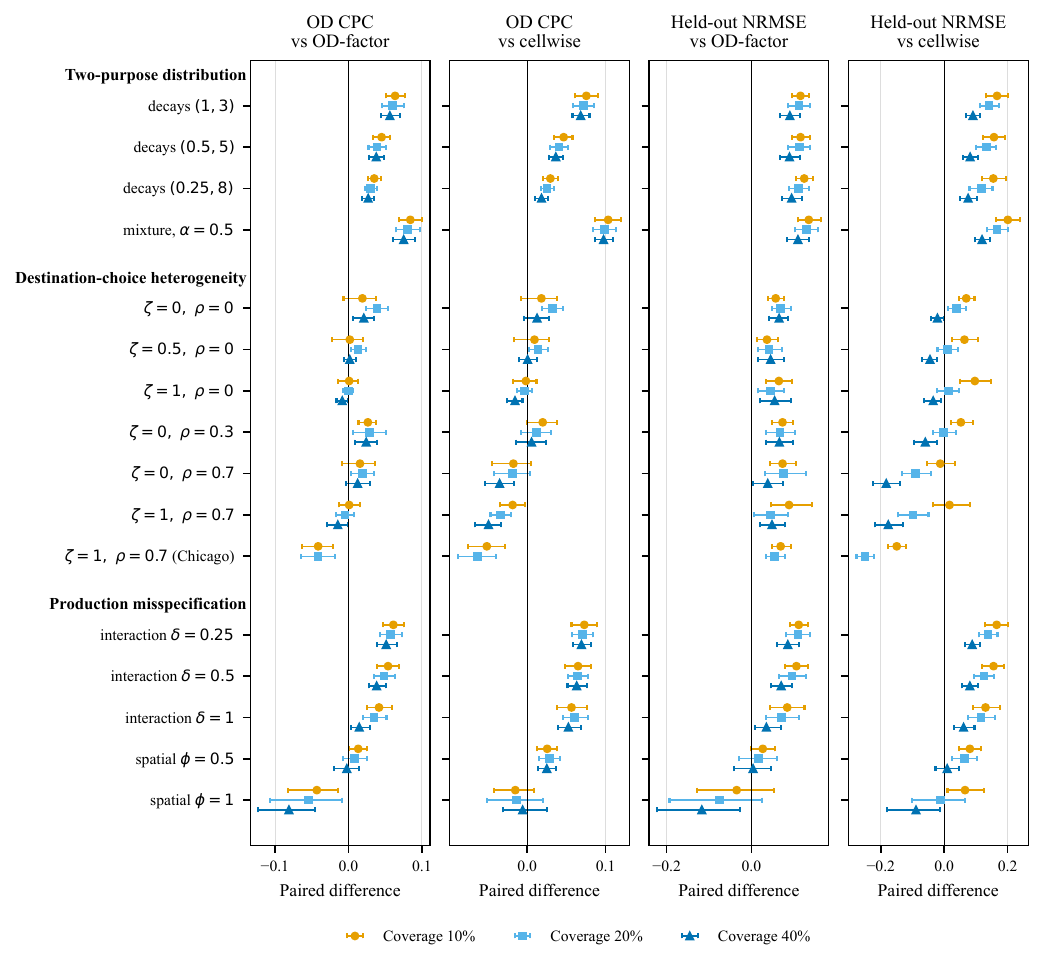}
\caption{Mean paired differences under misspecified truths, with 95\% world-bootstrap intervals over 20 worlds, by calibration coverage. Positive values favour the five-coefficient generator. Destination-choice heterogeneity: $\zeta$ scales origin-specific decay and $\rho$ the OD interaction. Production misspecification is applied to the two-purpose truth. All rows are EMA except the row marked Chicago.}
\label{fig:misspec}
\end{figure}

Figure~\ref{fig:misspec} and Table~\ref{tab:misspec} show the misspecification results. Increasing the contrast between the two purpose-specific decays reduces the advantage of the restricted model but does not reverse it. With decays of 0.25 and 8, the OD CPC advantage over cellwise adjustment is 0.030 at 10\% and 0.018 at 40\% coverage, and every interval in this family lies above zero.

Destination-choice heterogeneity has a stronger effect. Removing the accessibility correction from the true distribution, with no further heterogeneity ($\zeta=\rho=0$), already reduces the OD CPC advantage over cellwise adjustment at 10\% coverage from 0.123 under matched truth to 0.018, with an interval spanning zero. Across the 18 EMA strata of this family, OD CPC favours the restricted model over cellwise adjustment in 3, is inconclusive in 10 and favours cellwise adjustment in 5. The reversals occur with strong heterogeneity and higher coverage. In the strongest condition ($\zeta=1$, $\rho=0.7$) the cellwise estimator recovers OD demand more accurately at every coverage, with mean differences of $-0.019$, $-0.034$ and $-0.049$ at 10, 20 and 40\%. Only 2 of 20 worlds favour the restricted model at 40\%. Held-out NRMSE favours cellwise adjustment in 8 strata, mostly at 40\% coverage. Against OD-factor adjustment the picture differs: OD CPC favours the factor model only in 2 strata, and held-out NRMSE favours the restricted model in all 18.

\begin{table}[htbp]\footnotesize
\tbl{Selected misspecified strata. Cells as in Table~\ref{tab:primary}, Panel B. Dev.: realised OD CPC of the truth against the matched truth (median over worlds). Truth labels as in Table~\ref{tab:truths}. Every stratum is listed in Appendix Table~\ref{stab:allcontrasts}.}
{\input{tables/tab_misspec.tex}}
\label{tab:misspec}
\end{table}

On Chicago the strongest condition reverses the comparison against both matrix estimators. At 10\% and 20\% coverage, cellwise adjustment has higher OD CPC by 0.052 and 0.064, and factor adjustment by 0.041 at both coverages (Table~\ref{tab:misspec}). Cellwise adjustment also predicts held-out flows more accurately, by 0.150 and 0.250 in NRMSE, with one and zero worlds favouring the restricted model. The restricted model still predicts held-out flows more accurately than factor adjustment. The realised deviation from the matched truth is similar on the two networks (0.626 on Chicago and 0.619 on EMA), but the reversal is larger on Chicago.

None of the estimators recovers this demand well. At 10\% coverage on Chicago the median OD CPC is 0.615 for cellwise adjustment, 0.608 for factor adjustment and 0.571 for the restricted model, against 0.591 for the prior. Calibration to counts moves the restricted model away from the true matrix: its OD CPC is lower than that of its own prior by 0.027 (95\% interval $-0.051$ to $-0.003$) at 10\% and by 0.027 at 20\%. The calibration counts are reproduced closely, with a median calibration NRMSE of 0.514 for the restricted model against 0.769 for its prior, while the coefficients absorb the misspecification.

\subsection{Production misspecification}\label{sec:res_prod}

An omitted bounded interaction in production reduces the advantage of the restricted model, which nevertheless recovers OD demand more accurately than both matrix estimators in all nine strata of this family. The weakest case is $\delta=1$ at 40\% coverage, with an advantage of 0.015 over factor adjustment (14 of 20 worlds). A car-availability effect that varies across space has a different consequence. At $\phi=1$ the OD-factor estimator recovers OD demand more accurately than the restricted model at all three coverages, with mean differences of $-0.043$, $-0.054$ and $-0.081$ at 10, 20 and 40\% (Table~\ref{tab:misspec}). At 40\% coverage both matrix estimators also predict held-out flows more accurately. The OD CPC comparison against cellwise adjustment is inconclusive at each coverage. Free origin factors can absorb zone-specific production deviations that the common slopes cannot represent, without fitting every OD cell. The ablations (Appendix Section~\ref{sec:additional}) show which coefficients gain information from the counts. At 10\% coverage, fitting the demographic slopes improves OD recovery over the slope-fixed ablation in 21 of 25 strata, the exceptions being the conditions with $\rho=0.7$ and the strongest spatial production effect. Fitting the decay improves it in 13 of 25 strata and is inconclusive under every destination-choice condition and the stronger production misspecifications.

\subsection{Observation coverage, calibration fit and recovery}\label{sec:res_fit}

Coverage affects the estimators differently. Under matched and two-purpose truth the OD CPC advantage of the restricted model changes little with coverage, while its held-out advantage over cellwise adjustment narrows (Figure~\ref{fig:anchor}). Under misspecification, coverage favours the flexible estimators: in the destination-choice family, the paired OD CPC difference against cellwise adjustment is lower at 40\% than at 10\% coverage in each of the six conditions, and in the strongest spatial production condition the deficit against factor adjustment grows from 0.043 to 0.081 (Figure~\ref{fig:misspec}).

\begin{figure}[htbp]
\includegraphics[width=\linewidth]{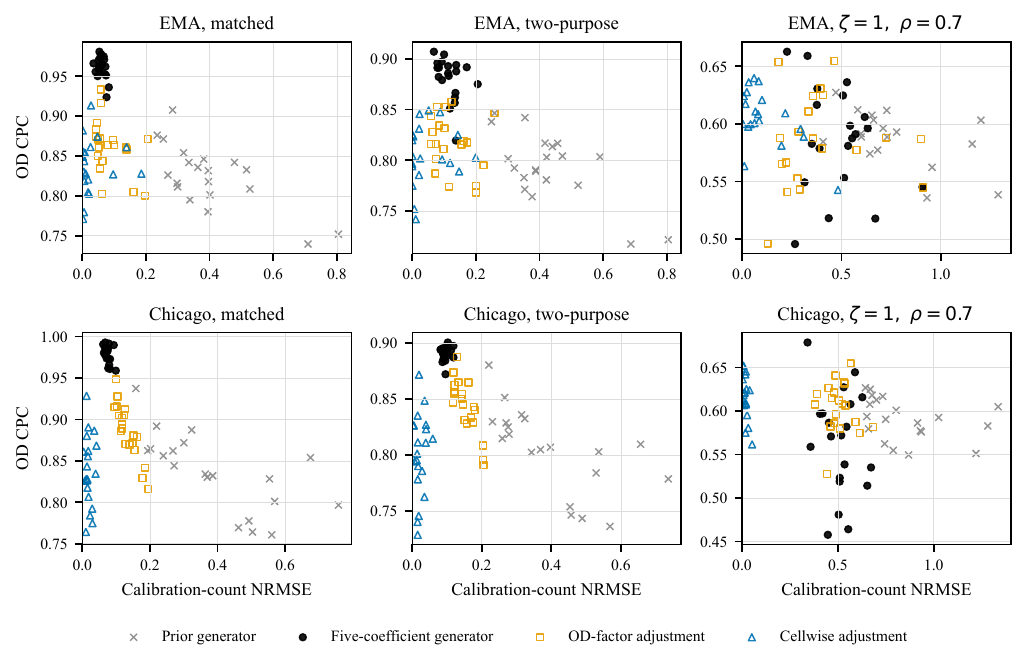}
\caption{Calibration-count NRMSE against OD CPC at 10\% coverage. Each point is one world (mean over its two priors). $\zeta=1$, $\rho=0.7$ denotes the strongest destination-choice heterogeneity. Cellwise adjustment reproduces the counts most closely in every panel, while its OD recovery is lower than that of the generator under matched and two-purpose truth.}
\label{fig:fit}
\end{figure}

Calibration fit is a poor guide to recovery across estimators. Figure~\ref{fig:fit} plots each world's calibration NRMSE against its OD CPC at 10\% coverage. Cellwise adjustment has the lowest median calibration NRMSE in 65 of the 66 evaluated strata. The exception is matched truth on Chicago with capacity error, where the restricted model has the lowest. Yet the restricted model has the highest median OD CPC in 54 of the 66 strata, and cellwise adjustment has the lowest in 45. Within an estimator, the regularisation path tells a similar story. On EMA at 10\% coverage, the median Spearman correlation between calibration NRMSE and OD CPC across the cellwise penalty weights is 0.00 under both matched and two-purpose truth. The calibration error spans almost two orders of magnitude along that path, but the median OD CPC range is only 0.008. For the restricted model the correlation is $-1.00$ and $-0.95$: closer count fit along its path goes with better recovery (Appendix Section~\ref{sec:additional}).

The held-out flow advantage shows little relationship with distance from the nearest calibration link. The median per-dataset Spearman correlation between this distance and the restricted model's absolute-error advantage lies between $-0.065$ and $0.187$ across the EMA strata and between $-0.002$ and $0.054$ on Chicago (Appendix Table~\ref{stab:d7}).

\subsection{Numerical robustness}\label{sec:numerics}

Table~\ref{tab:numerics} summarises the numerical diagnostics of the restricted model, evaluated on its stored fits. The five-column count Jacobian, scaled by the prior standard deviations, has full effective rank at the prior, at the truth and at the fit in 11,911 of 11,920 finite-difference evaluations. The nine exceptions are all on EMA, six of them at fits under the pair-interaction truth ($\zeta=0$, $\rho=0.7$), and six of the nine at the looser equilibrium tolerance. The frozen-route Jacobian is a poor local approximation of the equilibrium Jacobian: the median relative difference is 0.55 on EMA and 0.89 on Chicago. The stored fits are consequently stationary for the surrogate but not for the equilibrium objective. A Gauss–Newton step with the finite-difference Jacobian predicts a median relative objective decrease of 2.0\% on EMA and 2.5\% on Chicago.

\begin{table}[htbp]\footnotesize
\tbl{Numerical diagnostics of the five-coefficient generator. Polishing refers to projected Gauss–Newton iterations with the central-difference equilibrium Jacobian from each stored fit. \textsuperscript{a}Equilibrium relative gap $10^{-8}$ (EMA) or $10^{-7}$ (Chicago). \textsuperscript{b}Mean paired contrasts of the generator against both matrix estimators whose classification (favours generator, inconclusive, favours comparator) changes when the generator is polished, with the OD-factor fits also polished on EMA.}
{\input{tables/tab_numerics.tex}}
\label{tab:numerics}
\end{table}

To measure what this gap means for recovery, every stored fit of the restricted model was polished on the equilibrium objective with the finite-difference Jacobian, and the OD-factor fits were polished in the same way on EMA. The realised objective decrease has stratum medians of 0.6–7.3\% on EMA and 1.1–5.9\% on Chicago. The median change in OD CPC per stratum lies between $-0.003$ and $0.009$, about a tenth of the primary contrasts. Of 240 paired contrasts, 7 change classification and none changes from favouring the restricted model to favouring a comparator. Two EMA comparisons with cellwise adjustment under the pair-interaction truth ($\rho=0.7$, 10\% and 20\%) move from inconclusive to favouring cellwise adjustment, so polishing strengthens the reversals rather than removing them. Five-start re-optimisation leaves the OD CPC within 0.031 of the stored fit.

We report the stored fits for each estimator as the primary results. The algorithms use frozen-route sensitivities and accept updates after evaluating the full equilibrium objective. Polishing only the restricted model would compare estimators solved with different numerical effort. On EMA, where both the restricted model and the factor model were polished, polishing only the former inflates its advantage over factor adjustment by 0.0017 in OD CPC and 0.0043 in held-out NRMSE on average. The Chicago factor fits were not polished, because each Gauss–Newton step would require 774 finite-difference columns. On Chicago, polishing the restricted model leaves the reversal against factor adjustment unchanged ($-0.041$ and $-0.042$ OD CPC at 10\% and 20\% coverage). The cellwise fits were not polished. A finite-difference stationarity gate during protocol development found that further optimisation of the cellwise and factor fits along restricted directions changed OD CPC and held-out NRMSE by less than 0.01, but this was assessed on four development datasets only (Appendix Section~\ref{app-sec:numerics}).

\section{Discussion}\label{sec:discussion}

The results demonstrate that the parameterisation of a travel demand model affects what can be recovered from traffic counts. When demand follows the production–gravity structure, estimating five shared coefficients gives more accurate OD matrices and held-out link flows than adjusting between 148 and 149,382 demand coordinates. The advantage also persists when the truth combines two gravity-based trip purposes. In these conditions, the restricted structure makes effective use of limited observations, even when its count fit is less accurate.

The same restriction can be harmful when important demand relationships are omitted. Origin-specific distance sensitivity and OD interactions favour cellwise adjustment, while spatial variation in production effects favours origin–destination factors. These differences are not explained by parameter counts alone. The more flexible methods perform better when their additional coordinates represent variation that the restricted model cannot capture. Under strong destination-choice misspecification on Chicago, calibrating the five-coefficient model improves the count fit but reduces OD accuracy relative to its uncalibrated prior. Additional counts cannot correct a structural error that the fitted model is unable to represent.

This finding matters for conventional TDM calibration. The quality of an assigned flow fit does not determine whether trip productions, destinations or OD flows have been recovered correctly. Cellwise adjustment fits the calibration counts best in nearly all tested conditions, but frequently produces less accurate OD matrices. Conversely, a small number of demand parameters should not be taken as evidence of reliability. Their value depends on whether the assumed production and distribution relationships are appropriate for the study area. Calibration results should therefore be assessed alongside independent information, such as travel surveys, trip-end observations and evidence of spatially varying travel behaviour. Held-out counts provide a useful test of link-flow prediction, but cannot alone establish OD accuracy.

The implications extend to planning scenarios. A calibrated demand model may be applied to changed population or land-use inputs, whereas an adjusted OD matrix generally needs an additional growth or pivot rule. A shared parameterisation provides an explicit response mechanism, but this is useful only if the relationships remain applicable when inputs change. The projection experiments in Appendix Section~\ref{sec:projection} illustrate this distinction under prescribed demographic changes. They do not establish temporal transferability in real cities, where behaviour, network supply and the quality of future inputs may also change.

The study uses synthetic demand on two benchmark networks. The priors are informative, the observation locations are sampled at random, and the true and estimated assignment models are largely consistent. Count locations chosen to target OD information could change the value of additional counts for each method \citep{ChootinanEtAl2005,ZhouList2010}. The misspecification experiments cover selected changes in production and destination choice, not the full complexity of observed travel behaviour. Numerical optimisation also relies on approximate assignment sensitivities, although the additional convergence checks leave the principal comparisons unchanged. The results should therefore be interpreted as controlled evidence about the effects of demand-model assumptions, rather than as a validated method-selection rule for real networks.

\section{Conclusions}\label{sec:conclusion}

This study compared three established approaches to calibrating OD demand from sparse traffic counts. A five-coefficient production–gravity model provided better demand recovery under matched and two-purpose demand. Its advantage declined and sometimes reversed when destination choice or production varied across zones. Factor and cellwise adjustment then recovered patterns excluded by the restricted demand model, particularly as count coverage increased. The estimator with the closest count fit was often not the estimator with the most accurate OD matrix.

These results show why model parameterisation is an important calibration choice. Traffic counts provide evidence about assigned link flows, but cannot independently validate the demand relationships that generate them. Both flexible matrix adjustment and restrictive demand-model calibration therefore require support from other travel-demand information, particularly when the calibrated model is intended for subsequent planning scenarios.

\section*{Acknowledgements}
Computational resources were provided by the Cambridge Service for Data Driven Discovery (CSD3), operated by the University of Cambridge Research Computing Service, with support from Dell EMC and Intel under EPSRC Tier-2 funding (EP/T022159/1) and DiRAC funding from the Science and Technology Facilities Council.

\section*{Funding}
This research was supported by the Cambridge Commonwealth, European and International Trust. Additional support was provided by the Martin Centre for Architectural and Urban Studies and Cambridge Public Health.

\section*{Data and code availability}
The benchmark networks are available from the Transportation Networks for Research repository \citep{TNTP}. The synthetic data and analysis code are available from the corresponding author upon reasonable request.

\clearpage
\appendix
\renewcommand{\thesection}{A\arabic{section}}
\renewcommand{\thesubsection}{A\arabic{section}.\arabic{subsection}}
\renewcommand{\theequation}{A\arabic{equation}}
\renewcommand{\thetable}{A\arabic{table}}
\renewcommand{\thefigure}{A\arabic{figure}}
\renewcommand{\thealgorithm}{A\arabic{algorithm}}
\setcounter{equation}{0}
\setcounter{table}{0}
\setcounter{figure}{0}
\setcounter{algorithm}{0}
\section*{Technical appendices}
\noindent The appendices provide full model definitions, gradient calibration, numerical diagnostics, synthetic-data generation and additional results.\par

\section{Model definitions and notation}\label{sec:models}

\begin{table}[htbp]\small
\tbl{Notation.}
{\begin{tabular}{p{0.23\linewidth}>{\raggedright\arraybackslash}p{0.70\linewidth}}\toprule
Symbol & Meaning\\\midrule
$\Zset$, $\Eset$, $\Omega$ & Zones, directed links, supported OD pairs (ordered pairs of distinct zones connected in both directions)\\
$\mathbf q=(q_{ij})$, $\mathbf q^{(0)}$, $\mathbf q^*$, $\hat{\mathbf q}$ & OD demand on $\Omega$, prior matrix, synthetic truth, estimate\\
$\Phi$, $\mathbf x$, $t_a$, $t^0_a$, $C_a$ & Equilibrium assignment map, link flows, link cost, free-flow time, capacity\\
$\Sset$, $\Hset$, $y_a$, $P$ & Calibration links, held-out links, noisy counts, selection matrix of $\Sset$\\
$N_i$, $A_j$, $z_{ik}$, $\kappa$ & Population, attractiveness, standardised covariate $k$ of zone $i$, fixed exposure scale\\
$c_{ij}$ & Free-flow shortest-path time divided by its median over $\Omega$\\
$\bm\theta=(\gamma_0,\ldots,\gamma_3,\eta)$, $b=e^\eta$ & Generator coefficients and decay\\
$\bm\theta^{(0)}$, $\bm\sigma$, $\lambda$ & Prior coefficients, prior scales, penalty weight\\
$\tau$, $u_i$, $v_j$, $\Delta b$ & OD-factor level, origin factors, destination factors and decay adjustment\\
$R_{a,ij}$ & Frozen-route share of the demand of $(i,j)$ on link $a$\\
$\bm\theta^*$, $b^*$ & Matched truth coefficients $(0,0.35,0.20,0.25,\log2)$ and decay 2\\
$w_i$, $b^{(1)}$, $b^{(2)}$ & Two-purpose share and purpose decays\\
$\zeta$, $\rho$, $\omega_{ij}$ & Origin-decay heterogeneity, OD-interaction amplitude, OD-interaction field\\
$\delta$, $\phi$, $\chi_i$ & Production interaction amplitude, spatial production amplitude, spatial field\\
$\mathbf W$, $d_{ij}$, $h_d$ & Row-normalised smoothing kernel, symmetric free-flow time, kernel bandwidth\\\bottomrule
\end{tabular}}
\label{stab:notation}
\end{table}

\subsection{Network and equilibrium assignment}\label{sec:assignment}

Link costs are $t_a(x_a)=t^0_a[1+0.15(x_a/C_a)^4]$ on every link of both networks, as supplied in the benchmark files. For demand $\mathbf q$, the equilibrium link flows solve the Beckmann programme
\begin{equation}
\Phi(\mathbf q)=\arg\min_{\mathbf x\in\mathcal F(\mathbf q)}\sum_{a\in\Eset}\int_0^{x_a}t_a(s)\,ds,
\end{equation}
where $\mathcal F(\mathbf q)$ is the set of link flows generated by nonnegative path flows that satisfy the OD demand. On Chicago Sketch each zone node is split into a source copy carrying its outgoing connectors and a sink copy carrying its incoming connectors, so that no route passes through a centroid. The EMA zone nodes are ordinary road nodes. Convergence is measured by the relative gap
\begin{equation}
g(\mathbf x)=\frac{\sum_at_a(x_a)x_a-\sum_{(i,j)\in\Omega}q_{ij}\,\pi_{ij}(\mathbf x)}{\sum_at_a(x_a)x_a},
\end{equation}
where $\pi_{ij}(\mathbf x)$ is the shortest-path time between $i$ and $j$ at link costs $t(\mathbf x)$. The solver is described in Section~\ref{app-sec:numerics}.

\subsection{Five-coefficient production–gravity generator}\label{sec:gen}

Production and distribution are
\begin{equation}
O_i(\bm\theta)=\kappa N_i\exp\Big(\gamma_0+\sum_{k=1}^3\gamma_kz_{ik}\Big),\quad
q_{ij}(\bm\theta)=O_i\,\pi_{ij},\quad
\pi_{ij}=\frac{A_jK_{ij}/H_j}{\sum_{k:(i,k)\in\Omega}A_kK_{ik}/H_k},
\label{seq:gen}
\end{equation}
with $K_{ij}=e^{-bc_{ij}}$ on $\Omega$, $H_j=\sum_{\ell:(\ell,j)\in\Omega}K_{\ell j}$ and $b=e^\eta$. Rows sum to production, $\sum_jq_{ij}=O_i$. Destination totals are outputs. Unsupported and intrazonal pairs carry no demand. The covariates are car availability ($k=1$), working-age share ($k=2$) and employed-adult share ($k=3$), each standardised by its base-year mean and standard deviation. The exposure $\kappa$ is fixed per world (Section~\ref{sec:synthetic}). The coefficient bounds are $\gamma_0\in[-2,2]$, $\gamma_k\in[-1.5,1.5]$ for $k=1,2,3$ and $b\in[0.2,8]$. The penalty is $R_G(\bm\theta)=\frac15\sum_{k=1}^5((\theta_k-\theta^{(0)}_k)/\sigma_k)^2$ with $\bm\sigma=(0.20,0.20,0.20,0.20,0.25)$.

\subsection{Prior generator}\label{sec:prior}

Each world has two prior coefficient vectors $\bm\theta^{(0)}$ (Section~\ref{sec:synthetic}). The prior matrix of all estimators is $\mathbf q^{(0)}=\mathbf q(\bm\theta^{(0)})$ from Equation~\eqref{seq:gen}. The prior generator is evaluated without count fitting as a reference.

\subsection{OD-factor adjustment}\label{sec:factorS}

The factor model is
\begin{equation}
q_{ij}=q^{(0)}_{ij}\exp(\tau+u_i+v_j-\Delta b\,c_{ij}),\qquad \sum_iw^O_iu_i=0,\qquad \sum_jw^D_jv_j=0,
\label{seq:factor}
\end{equation}
with $w^O_i=O^{(0)}_i/\sum_kO^{(0)}_k$ and $w^D_j=D^{(0)}_j/\sum_kD^{(0)}_k$ the shares of the prior origin and destination totals. The constraints are imposed by writing $\mathbf u=\mathbf B_u\bm\psi_u$ and $\mathbf v=\mathbf B_v\bm\psi_v$, where the columns of $\mathbf B_u$ and $\mathbf B_v$ are orthonormal bases of the null spaces of $\mathbf w^{O\top}$ and $\mathbf w^{D\top}$. The free coordinates are $\bm\psi=(\tau,\bm\psi_u,\bm\psi_v,\Delta b)$, of dimension $1+(|\Zset|-1)+(|\Zset|-1)+1=2|\Zset|$ on a connected support: 148 on EMA and 774 on Chicago. Before each dataset is fitted, the code verifies that the map from $\bm\psi$ to $\log\mathbf q$ is injective, that is, that the smallest eigenvalue of $\mathbf D^\top\mathbf D$ is positive for $\mathbf D=\partial\log\mathbf q/\partial\bm\psi$. The penalty is
\begin{equation}
R_F(\bm\psi)=\frac15\Big[(\tau/\sigma_\tau)^2+\sum_iu_i^2/\sigma_u^2+\sum_jv_j^2/\sigma_v^2+(\Delta b/\sigma_{\Delta b})^2\Big].
\end{equation}
The divisor 5 is the dimension of the generator, retained so that the factor penalty uses the same convention as $R_G$ (a sum over coordinates divided by 5). It is not the factor dimension. The scales are $(\sigma_\tau,\sigma_u,\sigma_v)=(0.414,0.340,0.090)$ on EMA and $(0.449,0.330,0.063)$ on Chicago, and $\sigma_{\Delta b}=0.25\,b^{(0)}$. They were obtained before the confirmation programme by projecting the prior coefficient uncertainty of the generator onto these coordinates, as medians over five development worlds and both priors. They match marginal scales, not the full prior covariance. The bounds are $|\tau|,|u_i|,|v_j|\le5$ and $0.2\le b^{(0)}+\Delta b\le8$, where $b^{(0)}=e^{\eta^{(0)}}$. The demand derivatives are analytic: $\partial\log q_{ij}/\partial\tau=1$, $\partial\log q_{ij}/\partial u_i=1$, $\partial\log q_{ij}/\partial v_j=1$ and $\partial\log q_{ij}/\partial\Delta b=-c_{ij}$.

\subsection{Cellwise adjustment}\label{sec:cellS}

The primary cellwise estimator optimises $\mathbf q$ directly with the generalised relative entropy
\begin{equation}
R_C(\mathbf q)=\frac{\sum_{\Omega}[q_{ij}\log(q_{ij}/q^{(0)}_{ij})-q_{ij}+q^{(0)}_{ij}]}{\sum_\Omega q^{(0)}_{ij}},
\end{equation}
in the relative coordinates $v_{ij}=q_{ij}/q^{(0)}_{ij}\ge10^{-12}$. All supported prior cells are strictly positive. A secondary variant uses the squared deviation $R_{C2}(\mathbf q)=\sum_\Omega(q_{ij}-q^{(0)}_{ij})^2/\sum_\Omega(q^{(0)}_{ij})^2$ with $q_{ij}\ge0$. It was run on worlds 2000–2004 at 10\% coverage under the matched and two-purpose truths.

\subsection{Generator ablations}\label{sec:ablation}

The slope-fixed ablation fits $(\gamma_0,\eta)$ and holds $(\gamma_1,\gamma_2,\gamma_3)$ at the prior. The decay-fixed ablation fits $(\gamma_0,\ldots,\gamma_3)$ and holds $\eta$ at the prior. Both use the generator's grid, cross-validation, loss, penalty (over all five coefficients, the fixed ones contributing zero) and budgets. The fixed coefficients are removed from the free set of the Levenberg–Marquardt system (Algorithm~\ref{alg:lm}).

\section{Analytic demand Jacobian}\label{sec:jacobian}

The derivatives of Equation~\eqref{seq:gen} with respect to the production coefficients follow from $q_{ij}=O_i\pi_{ij}$ and $\partial\log O_i/\partial\gamma_0=1$, $\partial\log O_i/\partial\gamma_k=z_{ik}$, because $\pi_{ij}$ does not depend on $\bm\gamma$:
\begin{equation}
\frac{\partial q_{ij}}{\partial\gamma_0}=q_{ij},\qquad \frac{\partial q_{ij}}{\partial\gamma_k}=z_{ik}\,q_{ij},\quad k=1,2,3.
\label{seq:dgamma}
\end{equation}
The decay enters both the kernel and the accessibility denominator. Write $W_{ij}=A_jK_{ij}/H_j$, so that $\pi_{ij}=W_{ij}/\sum_kW_{ik}$. Since $\partial K_{ij}/\partial b=-c_{ij}K_{ij}$,
\begin{equation}
\frac{\partial\log H_j}{\partial b}=-\frac{\sum_{\ell:(\ell,j)\in\Omega}c_{\ell j}K_{\ell j}}{H_j}=-\bar c_j(b),\qquad
\frac{\partial\log W_{ij}}{\partial b}=\bar c_j(b)-c_{ij}\equiv e_{ij},
\end{equation}
where $\bar c_j(b)$ is the kernel-weighted mean cost into destination $j$. Differentiating the row normalisation gives $\partial\log\pi_{ij}/\partial b=e_{ij}-\sum_k\pi_{ik}e_{ik}$, so
\begin{equation}
\frac{\partial q_{ij}}{\partial b}=q_{ij}\Big[e_{ij}-\sum_{k:(i,k)\in\Omega}\pi_{ik}e_{ik}\Big],\qquad
\frac{\partial q_{ij}}{\partial\eta}=b\,\frac{\partial q_{ij}}{\partial b}.
\label{seq:deta}
\end{equation}
The production derivatives do not affect destination shares, and the decay derivative does not change origin totals, because $\sum_j\partial q_{ij}/\partial b=0$. Stacking Equations~\eqref{seq:dgamma} and \eqref{seq:deta} on $\Omega$ gives the $|\Omega|\times5$ demand Jacobian $\mathbf B=\partial\mathbf q/\partial\bm\theta$. The implementation (\texttt{generator.demand\_and\_jac}) computes exactly these expressions.

\subsection{Verification}
We implemented Equations~\eqref{seq:dgamma}–\eqref{seq:deta} independently of the experiment code and compared them with central finite differences of Equation~\eqref{seq:gen} (step $10^{-6}\max\{1,|\theta_k|\}$) at the matched coefficients, both priors and three random points, in EMA worlds 2000 and 2001 and Chicago world 2000. The largest relative error (maximum absolute error over the largest absolute derivative) was $5.4\times10^{-10}$, and the largest relative difference from the experiment code was $3.0\times10^{-16}$. Omitting the dependence of $H_j$ on $b$, that is, using $\partial\log W_{ij}/\partial b=-c_{ij}$, gives a decay derivative in error by 15–79\% in the same worlds. The accessibility term is therefore essential.

\section{Gradient-based calibration algorithm}\label{sec:algorithm}

\subsection{Residuals, scaling and the frozen-route Jacobian}

For calibration counts $\mathbf y_{\Sset}$ with $n_s=|\Sset|$ and scale $s_y=\max\{10^{-9},(n_s^{-1}\sum_{a\in\Sset}y_a^2)^{1/2}\}$, the generator objective is $F(\bm\theta)=\|\mathbf r(\bm\theta)\|^2$ with the stacked residual
\begin{equation}
\mathbf r(\bm\theta)=\begin{bmatrix}(P\Phi(\mathbf q(\bm\theta))-\mathbf y_{\Sset})/(s_y\sqrt{n_s})\\ \sqrt{\lambda/5}\,D_\sigma^{-1}(\bm\theta-\bm\theta^{(0)})\end{bmatrix},\qquad D_\sigma=\mathrm{diag}(\bm\sigma),
\end{equation}
so that $F=L(\mathbf q(\bm\theta))+\lambda R_G(\bm\theta)$, with the penalty block absent when $\lambda=0$. At an equilibrium with stored path flows $f_p$, the frozen-route operator is
\begin{equation}
R_{a,ij}=\sum_{p\in\mathcal P_{ij}:\,a\in p}\frac{f_p}{\sum_{p'\in\mathcal P_{ij}}f_{p'}},
\end{equation}
where $\mathcal P_{ij}$ is the stored route set of pair $(i,j)$. A pair with no route flow puts weight one on its cheapest stored route. By construction $R\mathbf q=\mathbf x$ up to solver drift. In a preliminary solver audit the maximum relative difference was $1.1\times10^{-16}$ at relative gaps $10^{-6}$ and $10^{-8}$. Holding $R$ fixed gives the approximate residual Jacobian
\begin{equation}
\tilde{\mathbf J}=\begin{bmatrix}PR\mathbf B/(s_y\sqrt{n_s})\\ \sqrt{\lambda/5}\,D_\sigma^{-1}\end{bmatrix}.
\end{equation}
It omits route reallocation and is not $P(\partial\Phi/\partial\mathbf q)\mathbf B$ \citep{TobinFriesz1988}. Because equilibrium path flows are not unique, two valid decompositions of the same equilibrium give different $R$ (Section~\ref{app-sec:numerics}).

\subsection{Damped Gauss–Newton steps, bounds and acceptance}

With $\mathbf g=\tilde{\mathbf J}^\top\mathbf r$ and $\mathbf H=\tilde{\mathbf J}^\top\tilde{\mathbf J}$, the free set $\mathcal F$ excludes coefficients at a bound whose gradient points outward (at the lower bound with $g_k>0$, or at the upper bound with $g_k<0$, bound tolerance $10^{-12}$) and the coefficients fixed by an ablation. The step solves
\begin{equation}
\big[\mathbf H_{\mathcal F\mathcal F}+\mu\,\mathrm{diag}(\mathrm{diag}(\mathbf H_{\mathcal F\mathcal F})+10^{-12})\big]\,\mathbf d=-\mathbf g_{\mathcal F},
\end{equation}
and the trial $\bm\theta+\mathbf d$ is clipped to the bounds. Each trial is evaluated by generating its OD matrix and solving equilibrium again, warm-started from the current route flows. It is accepted only if the fully evaluated objective decreases. Algorithm~\ref{alg:lm} gives the complete procedure.

\begin{algorithm}[htbp]
\caption{Calibration of the production–gravity generator (\texttt{estimators.fit\_generator}).}\label{alg:lm}
\begin{algorithmic}[1]
\State Start at $\bm\theta\gets\mathrm{clip}(\bm\theta^{(0)})$ with the equilibrium of $\mathbf q^{(0)}$
\State $\mu\gets10^{-3}$, evaluations $\gets1$, accepted $\gets0$
\While{accepted $<50$}
\State Form $R$ from the current route flows, $\mathbf B$ from Section~\ref{sec:jacobian}, $\tilde{\mathbf J}$, $\mathbf r$, $\mathbf g$, $\mathbf H$ and the free set $\mathcal F$
\If{$\|\mathbf g_{\mathcal F}\|<10^{-12}$} stop: stationary \EndIf
\For{up to 10 trials}
\State Solve the damped system for $\mathbf d$ (on failure set $\mu\gets4\mu$ and retry)
\State $\bm\theta'\gets\mathrm{clip}(\bm\theta+\mathbf d)$
\State Solve equilibrium for $\mathbf q(\bm\theta')$ at relative gap $10^{-5}$ with a warm start, and set evaluations $\gets$ evaluations $+1$
\If{$F(\bm\theta')<F(\bm\theta)$} accept, $\mu\gets\max(\mu/3,10^{-10})$, accepted $\gets$ accepted $+1$, exit the trial loop
\Else\ $\mu\gets4\mu$ \EndIf
\If{evaluations $\ge300$} exit the trial loop \EndIf
\EndFor
\If{no trial accepted} stop: no descent (or evaluation budget) \EndIf
\If{$\max_k|\theta'_k-\theta_k|<10^{-8}$} stop: small step \EndIf
\If{accepted $\ge10$ and the relative decrease over the last 10 accepted steps $<10^{-6}$} stop: stagnation \EndIf
\If{evaluations $\ge300$} stop: evaluation budget \EndIf
\EndWhile
\end{algorithmic}
\end{algorithm}

\subsection{Matrix estimators}\label{sec:ss}

The OD-factor and cellwise estimators use a successive-surrogate scheme with the same acceptance rule (Algorithm~\ref{alg:ss}). At the current iterate $\mathbf q_k$ with route operator $R_k$, the surrogate replaces $\Phi(\mathbf q)$ by the linear map $R_k\mathbf q+(\mathbf x_k-R_k\mathbf q_k)$, which is exact at $\mathbf q_k$, and adds the estimator's penalty. The surrogate is minimised by L-BFGS-B \citep{ByrdEtAl1995} with memory 20, at most 3,000 iterations and 6,000 function evaluations, and tolerances $10^{-15}$ (function) and $10^{-12}$ (projected gradient). The cellwise estimator works in the relative coordinates $v_{ij}=q_{ij}/q^{(0)}_{ij}$ with lower bound $10^{-12}$. The factor estimator works in $\bm\psi$, with the inner solve restricted to a box of half-width 0.5 around the current $\bm\psi$ and every trial step projected onto the factor and decay bounds. The trial sequence moves a fraction $\alpha\in\{1,0.5,0.25\}$ of the way to the surrogate minimiser and then re-solves the surrogate with a proximal term $\varrho\,P(\cdot)$ centred at the current iterate, for $\varrho$ equal to the current surrogate value times $10^{-1},10^{0},\ldots,10^{4}$. Here $P$ is the estimator's own penalty measured from the current iterate. Every trial is evaluated after a full equilibrium solve.

\begin{algorithm}[htbp]
\caption{Successive-surrogate calibration of the matrix estimators (\texttt{fit\_odme}, \texttt{fit\_factors\_ss}).}\label{alg:ss}
\begin{algorithmic}[1]
\State Start at the prior matrix and its equilibrium, with evaluations $\gets1$ and accepted $\gets0$
\While{accepted $<300$}
\State Form $R_k$ and minimise the surrogate objective by L-BFGS-B to obtain the candidate $\mathbf q^\star$ (or $\bm\psi^\star$)
\If{the relative change $\|\mathbf q^\star-\mathbf q_k\|_1/\|\mathbf q_k\|_1<10^{-9}$} stop: stationary \EndIf
\For{each trial in $\{\alpha=1,0.5,0.25\}$, then proximal re-solves with $\varrho=f_k\cdot10^{e}$, $e=-1,\ldots,4$}
\State Evaluate the trial by a full equilibrium solve, and set evaluations $\gets$ evaluations $+1$
\If{the objective decreases} accept, accepted $\gets$ accepted $+1$, exit the trial loop \EndIf
\If{evaluations $\ge1{,}200$} exit the trial loop \EndIf
\EndFor
\If{no trial accepted} stop: no descent (or evaluation budget) \EndIf
\If{stagnation as in Algorithm~\ref{alg:lm}} stop \EndIf
\If{the relative surrogate decrease $<10^{-9}$} stop: surrogate stationary \EndIf
\EndWhile
\end{algorithmic}
\end{algorithm}

\subsection{Cross-validation and refitting}

The calibration links of each dataset are assigned to three folds by their position in the world's link permutation, modulo 3. For each candidate weight, each fold is held out in turn. The estimator is fitted to the other two folds, and the validation loss is the mean squared validation residual divided by the squared count scale $s_y$ of the training folds. The weight with the lowest mean validation loss is selected, with ties resolved towards the larger weight, and $\lambda=\infty$ (the prior matrix, without fitting) is the largest. If the largest finite weight is selected, a weight ten times larger is evaluated once. The selected weight is refitted on all calibration links. Only noisy calibration counts enter this procedure.

\section{Numerical optimisation and convergence}\label{app-sec:numerics}

\subsection{Equilibrium solver}

Equilibrium is solved by path-based gradient projection with column generation \citep{JayakrishnanEtAl1994}. In each sweep, for each origin, a shortest-path tree is computed at current costs. The new shortest route of each OD pair is added to its route set if it is not already there, and flow is shifted from every other used route towards the cheapest one by a Newton step, based on the cost difference divided by the sum of derivatives over the links not shared by the two routes, truncated at the route's flow. Unused non-shortest routes are dropped. A warm start rescales stored route flows to the new demand, so the warm state is feasible. If it already satisfies the tolerance it is returned unchanged. The tolerances are relative gaps of $10^{-5}$ for all solves within estimation (at most 5,000 sweeps) and $10^{-6}$ for truth, evaluation and projection (at most 20,000 sweeps). In the canonical records, 22,640 base-year evaluation solves and 143,960 scenario-projection solves all reached the tolerance (largest gaps $9.9999\times10^{-7}$ and $9.99997\times10^{-7}$).

\subsection{Path decomposition}

Equilibrium link flows are unique, but path flows are not. In that audit, one demand was solved from a cold start and from a warm start at 1.3 times the demand. The link flows agreed to $3.7\times10^{-6}$ (relative), but the route sets differed (5,624 against 5,761 routes in EMA world 2000). The frozen-route count Jacobians differed by 1.4–8.1\% in relative Frobenius norm on EMA and 3.3–3.4\% on Chicago, the intercept column being identical because $R\mathbf q=\mathbf x$ for every decomposition. The Gauss–Newton step for a fixed non-uniform residual differed by 0.8–22.0\%. The proposed direction therefore depends on the solver history. Accepted iterates remain valid, because acceptance uses the re-solved objective, but the stopping point can depend on the history. Path-flow uniqueness criteria, such as entropy-maximising path flows, were not implemented.

\subsection{Finite-difference identifiability and stationarity diagnostics}

At the prior, at the truth (matched truth only) and at each stored generator fit, the count Jacobian was computed by central differences of the equilibrium map,
\begin{equation}
\mathbf J^{\mathrm{FD}}_{:,k}=\frac{P\Phi(\mathbf q(\bm\theta+h\sigma_k\mathbf e_k))-P\Phi(\mathbf q(\bm\theta-h\sigma_k\mathbf e_k))}{2h},\qquad h=10^{-3},
\end{equation}
which is the derivative with respect to the prior-scaled coordinates $\theta_k/\sigma_k$. The perturbed equilibria were solved warm from the point at relative gaps of $10^{-6}$ and $10^{-8}$ (EMA) or $10^{-7}$ (Chicago). Rows were divided by the declared count standard deviation $0.05\max\{y_a,0.01\bar y\}$. The effective rank counts singular values above $10^{-3}$ times the largest, and the condition index is the ratio of the largest to the smallest singular value. The frozen-route Jacobian was formed in the same coordinates and compared by relative Frobenius difference. Table~\ref{stab:d1d2} reports the results. Of 11,920 finite-difference Jacobians, 11,911 have effective rank five. The nine exceptions are all on EMA. Six are at fits under the pair-interaction truth ($\zeta=0$, $\rho=0.7$), three of them at the tight tolerance (rank four), and one each at a fit under the $(0.5,5)$ decay contrast, at a fit under the spatial production effect with $\phi=0.5$ and at the truth point of a matched dataset with capacity error, all at the looser tolerance. Full rank shows local sensitivity in all five directions. It is not a proof of global identification.

For stationarity, the gradient of the full regularised objective was computed by central differences at each stored fit, one-sided inward at an active bound, and projected. The Gauss–Newton predicted decrease is $\frac12\mathbf g_{\mathcal F}^\top\mathbf H_{\mathcal F\mathcal F}^{-1}\mathbf g_{\mathcal F}$ with $\mathbf H=2\mathbf J^{\top}\mathbf J+(2\lambda/5)D_\sigma^{-2}$ built from the finite-difference count Jacobian. The frozen-route gradient was compared with the finite-difference gradient. Their relative difference has median 1.02 on EMA and 1.05 on Chicago, because the frozen-route gradient is close to zero at fits that stop as surrogate-stationary, while the equilibrium gradient is not.

\begin{table}[htbp]\footnotesize
\tbl{Finite-difference identifiability (upper block) and stationarity (lower block) of the generator by network, evaluation point and equilibrium tolerance. Relative difference: frozen-route against finite-difference Jacobian, relative Frobenius norm. Predicted decrease: Gauss–Newton predicted relative decrease of the full objective at the stored fit.}
{\input{tables/tabS_d1d2.tex}}
\label{stab:d1d2}
\end{table}

\subsection{Polishing on the equilibrium objective}\label{sec:polish}

Starting from every stored generator fit in the post-confirmation programme (2,400 fits), a projected Gauss–Newton method minimised the full objective with the finite-difference Jacobian above. Each of at most 15 iterations solved the stacked least-squares problem on the free set, then halved the step up to eight times until the objective, evaluated after a full equilibrium solve, decreased. It stopped when no trial decreased the objective or the relative decrease fell below $10^{-6}$. Solves used relative gap $10^{-8}$ (EMA) or $10^{-7}$ (Chicago), and the recovery metrics after polishing were evaluated at $10^{-6}$, as in the main results. The EMA OD-factor fits with finite $\lambda$ (2,041 fits) were polished in the same way in $\bm\psi$, with central differences of step $10^{-4}$ and projection of every trial onto the factor and decay bounds. Each Gauss–Newton step then needs 296 equilibrium solves. On Chicago it would need 1,548, so the Chicago factor fits were not polished. The cellwise fits were not polished.

\begin{figure}[htbp]
\includegraphics[width=\linewidth]{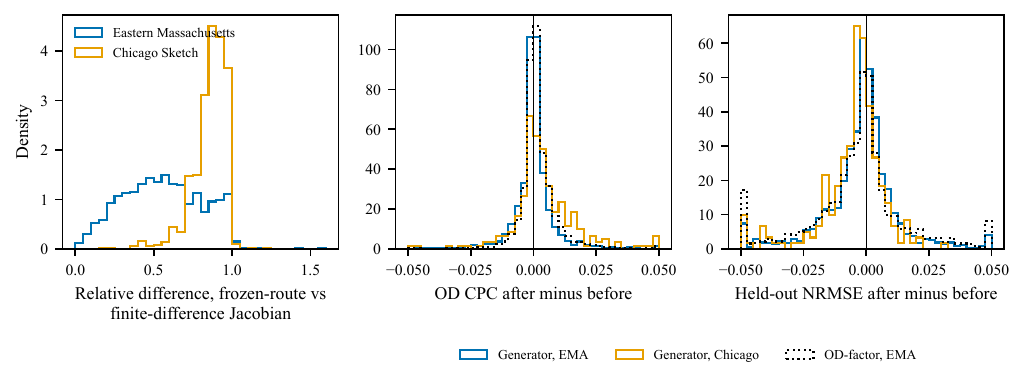}
\caption{Left: relative difference between the frozen-route and the finite-difference count Jacobian over all diagnostic points at the tight equilibrium tolerance. Centre and right: change in OD CPC and in held-out NRMSE when stored fits are polished on the equilibrium objective (values outside $\pm0.05$ are shown at the boundary).}
\label{sfig:numerics}
\end{figure}

Tables~\ref{stab:polish} and \ref{stab:polishchanges} and Figure~\ref{sfig:numerics} report the results. Seven of 240 paired contrasts change classification. Polishing only the generator while comparing with unpolished OD-factor fits inflates the generator's advantage: across 54 EMA strata in which both were polished, the mean difference between the asymmetric and the symmetric contrast is $+0.0017$ in OD CPC (maximum 0.0069) and $+0.0043$ in held-out NRMSE (maximum 0.0165). The main results therefore use the stored fits for all estimators.

\begin{table}[htbp]\footnotesize
\tbl{Polishing effect by network, estimator and truth family: range of per-stratum medians of the realised relative objective decrease and of the changes in OD CPC and held-out NRMSE (after minus before).}
{\input{tables/tabS_polish.tex}}
\label{stab:polish}
\end{table}

\begin{table}[htbp]\footnotesize
\tbl{Paired contrasts whose classification changes after polishing (mean paired difference and classification). On EMA, contrasts with the OD-factor model use polished fits of both estimators. Contrasts with cellwise adjustment use its stored fits.}
{\input{tables/tabS_polish_changes.tex}}
\label{stab:polishchanges}
\end{table}

\subsection{Multistart re-optimisation}

For EMA worlds 2000–2004, matched and two-purpose truth, first prior and 10\% coverage, the generator was refitted at the selected weight from five starts $\mathrm{clip}(\bm\theta^{(0)}+0.5\,\bm\sigma\odot\mathbf s_m)$ with fixed sign patterns $\mathbf s_m$ (all positive, all negative and three mixed patterns). Table~\ref{stab:multistart} reports the results. The best start lowers the stored objective by up to 10.4\% of its value, and in one dataset the stored fit is 0.2\% better than every start. The OD CPC differs by at most 0.031 across the starts and the stored fit.

\begin{table}[htbp]\footnotesize
\tbl{Five-start re-optimisation of the generator (EMA, first prior, 10\% coverage).}
{\input{tables/tabS_multistart.tex}}
\label{stab:multistart}
\end{table}

\subsection{Protocol-development gate for the matrix estimators}

Before the confirmation programme was frozen, a numerical gate was run on development worlds that are not part of the reported results: EMA and Chicago world 1000, first prior, 10\% coverage, matched and two-purpose truth. The generator, the OD-factor and the cellwise fits at their cross-validated weights were further optimised by two finite-difference methods on the equilibrium objective, a line-search polish and a function-first L-BFGS method. For the matrix estimators, the finite-difference gradient was taken along a restricted set of directions (the level, row and column patterns and the decay). A fit was classified as material if either method changed OD CPC or held-out NRMSE by at least 0.01. No binding fit was material. The largest changes were 0.0035 in OD CPC and 0.0062 in held-out NRMSE. A further diagnostic on the confirmation worlds refitted the OD-factor estimator at a penalty weight of $10^{-4}$, below its grid. This raised its held-out accuracy on Chicago but changed none of the 12 confirmation classifications of the generator against the factor model. These checks cover only a few datasets and restricted directions. They do not establish equilibrium stationarity of the matrix estimators in general.

\subsection{Stopping status and selected weights}

Tables~\ref{stab:status} and \ref{stab:lambda} list the stopping status and the cross-validated weights of every selected fit in the canonical records. Two fits stopped on the evaluation budget, both on EMA (one cellwise, one OD-factor). The generator selected $\lambda=0$ in 1,082 of 2,180 EMA datasets and 353 of 420 Chicago datasets, so its prior often acts only as the starting point. For cellwise adjustment, the inner L-BFGS-B solve reached its iteration limit in 81\% of outer iterations on Chicago and 32\% on EMA. Its outer iterations nevertheless stop through the acceptance test.

\begin{table}[htbp]\scriptsize
\tbl{Stopping status of the selected fits (canonical records).}
{\input{tables/tabS_status.tex}}
\label{stab:status}
\end{table}

\begin{table}[htbp]\scriptsize
\tbl{Cross-validated penalty weights of the selected fits ($\infty$: prior matrix).}
{\input{tables/tabS_lambda.tex}}
\label{stab:lambda}
\end{table}

\subsection{Unresolved numerical limitations}

The stored fits of every estimator are stationary for their frozen-route surrogates, not for the equilibrium objective. The polishing results bound the consequence for the generator on both networks and for the OD-factor estimator on EMA. They do not cover the cellwise estimator or the Chicago OD-factor estimator. The frozen-route proposals depend on the path decomposition. Exact equilibrium derivatives were not computed.

\section{Synthetic data generation}\label{sec:synthetic}

\subsection{Random streams}

Every random quantity of a world is drawn from its own stream, a NumPy generator seeded with \texttt{SeedSequence([world, crc32(network), stream])}, where \texttt{world} is the world seed (2000–2019) and \texttt{crc32} the 31-bit CRC of the network name. The streams are: population 0, attractiveness 1, covariate fields 2–4, priors 10 and 11, link permutation 20, count noise 21, population and employment cluster anchors 30 and 31, OD-interaction field 50, spatial production field 51 and capacity perturbation 52.

\subsection{Inputs, impedance and exposure}

Population and attractiveness are $N_i=1000\exp(0.45\,\varepsilon^N_i)$ and $A_i=500\exp(0.65\,\varepsilon^A_i)$ with independent standard normal $\varepsilon$. Let $d_{ij}=(c^{\mathrm{ff}}_{ij}+c^{\mathrm{ff}}_{ji})/2$ be the symmetric free-flow shortest-path time, $h_d$ the lower quartile of its positive finite values and $\mathbf W$ the kernel with entries proportional to $\exp(-d_{ij}/h_d)$, unit diagonal and rows normalised to one. For each covariate $k$, with $\mathbf e^{(k)}\sim N(\mathbf 0,\mathbf I)$ and $\mathrm{std}(\cdot)$ denoting standardisation to zero mean and unit population variance,
\begin{gather}
\mathbf v^{(k)}=\mathrm{std}\big(0.5\,\mathrm{std}(\mathbf W\mathbf e^{(k)})+0.5\,\mathbf e^{(k)}\big),\\
\mathrm{share}_{ik}=\ell_k+(u_k-\ell_k)\frac{1}{1+e^{-v^{(k)}_i}},\qquad z_{ik}=\frac{\mathrm{share}_{ik}-m_k}{s_k},
\end{gather}
with ranges $[\ell_k,u_k]$ of $[0.20,0.95]$, $[0.40,0.80]$ and $[0.30,0.80]$ and the base-year mean $m_k$ and standard deviation $s_k$. The support $\Omega$ contains every ordered pair of distinct zones with finite free-flow time in both directions, which is all pairs on both networks. The impedance is $c_{ij}=c^{\mathrm{ff}}_{ij}/\mathrm{median}_\Omega(c^{\mathrm{ff}})$. The exposure $\kappa$ is found by a bracketed secant search in $\log\kappa$ on the matched demand, solving equilibrium at gap $10^{-5}$, until the network travel-time inflation $I=\sum_at_ax_a/\sum_at^0_ax_a$ satisfies $|\log((I-1)/0.25)|<0.02$. The realised inflation lies between 1.245 and 1.255. $\kappa$ ranges from 0.48 to 1.34 on EMA and from 0.46 to 1.05 on Chicago.

\subsection{Priors, observations and supply perturbation}

For prior $m\in\{0,1\}$, $\bm\theta^{(0)}=\mathrm{clip}\big((\bm\gamma^*+0.20\,\bm\varepsilon_\gamma,\ \log2+0.25\,\varepsilon_\eta)\big)$ with independent standard normal draws from stream $10+m$, clipped to the coefficient bounds. The biased prior adds $(0,0.30,0,0,0.30)$ before clipping, and the prior matrix and the factor decay scale follow the biased prior. The eligible links (all links on EMA, all non-connector links on Chicago) are permuted. The first $\lceil0.2n_e\rceil$ form the held-out set, and the calibration set at coverage $c$ is the next $\lceil cn_e\rceil$ links. The calibration sets are therefore nested and disjoint from the held-out set. Counts are $y_a=\max\{0,x^*_a(1+0.05\,\epsilon_a)\}$, where $x^*_a$ is the equilibrium flow of the base-year truth at gap $10^{-6}$ and $\epsilon_a$ is one standard normal draw per link and world. Under capacity error, the estimators (including their cross-validation, prior equilibrium and route operators) use $C'_a=C_a\exp(0.15\,\xi_a-0.15^2/2)$ with $\xi_a$ standard normal, while truth, counts and evaluation use $C_a$.

\subsection{Truths}\label{sec:truthsS}

With $\bm\gamma^*=(0,0.35,0.20,0.25)$, $b^*=2$ and $O^*_i=\kappa N_i\exp(\gamma^*_0+\sum_k\gamma^*_kz_{ik})$, let $\mathcal G(\mathbf O,\mathbf A,b)$ denote the accessibility-corrected distribution of Equation~\eqref{seq:gen}. The truths are as follows.
\begin{itemize}
\item Matched: $\mathbf q^*=\mathcal G(\mathbf O^*,\mathbf A,b^*)$.
\item Two-purpose: $\mathbf q^*=\mathcal G(\mathbf w\odot\mathbf O^*,\mathbf A,b^{(1)})+\mathcal G((\mathbf 1-\mathbf w)\odot\mathbf O^*,\mathbf A^{(2)},b^{(2)})$ with $w_i=0.3+0.4/(1+e^{-z_{i3}})$ and $A^{(2)}_j\propto N_j^{0.6}A_j^{0.4}$. The code rescales $\mathbf A^{(2)}$ to the mean of $\mathbf A$, which does not affect the row-normalised shares. $(b^{(1)},b^{(2)})$ is $(1,3)$ for the two-purpose truth and $(0.5,5)$ or $(0.25,8)$ for the decay-contrast truths.
\item Mixture: $0.5\times$matched $+\,0.5\times$two-purpose $(1,3)$.
\item Destination-choice heterogeneity: $q^*_{ij}=O^*_i\exp(U_{ij})/\sum_{k:(i,k)\in\Omega}\exp(U_{ik})$ with $U_{ij}=\log A_j-b_ic_{ij}+\rho\,\omega_{ij}$ and $b_i=b^*\exp(\zeta z_{i1})$, without accessibility correction. The interaction field is built from $\mathbf U\in\mathbb R^{|\Zset|\times|\Zset|}$ with independent standard normal entries: $\bm\Omega=\mathbf W\mathbf U\mathbf W^\top$ is standardised over $\Omega$, centred within each origin over its supported destinations, rescaled to unit standard deviation over $\Omega$ and set to zero off $\Omega$. Labels hz$\zeta$a$\rho$ in the code denote $(\zeta,\rho)\in\{(0,0),(0.5,0),(1,0),(0,0.3),(0,0.7),(1,0.7)\}$.
\item Production interaction: the two-purpose $(1,3)$ truth with $O^*_i$ multiplied by $\exp\{\delta(h_i-\bar h)\}$, $h_i=\tanh(z_{i1})\tanh(z_{i3})$, $\bar h$ the base-year mean of $h_i$, $\delta\in\{0.25,0.5,1\}$. The confirmation programme's production-misspecification diagnostic is the case $\delta=0.25$ on worlds 2000–2009.
\item Spatial production effect: the two-purpose $(1,3)$ truth with $O^*_i$ multiplied by $\exp\{\phi(\chi_i\tanh(z_{i1})-\bar h^\chi)\}$, $\bm\chi=\mathrm{std}(\mathbf W\mathbf e^\chi)$ with $\mathbf e^\chi\sim N(\mathbf 0,\mathbf I)$, $\bar h^\chi$ the base-year mean of $\chi_i\tanh(z_{i1})$, $\phi\in\{0.5,1\}$.
\end{itemize}
Under a scenario (Section~\ref{sec:projection}) each truth follows the same law with the scenario inputs $(\mathbf N,\mathbf A,\mathbf z)$, while $\bm\Omega$, $\bm\chi$, $\bar h$ and $\bar h^\chi$ keep their base-year values. Because $b_i$ depends on $z_{i1}$, the composition scenario also changes the origin-specific decay of the destination-choice truths. Table~\ref{stab:deviation} gives the realised deviation of each truth from the matched truth.

\begin{table}[htbp]\footnotesize
\tbl{Realised OD common part of trips between each truth and the matched truth of the same world (base year).}
{\input{tables/tabS_deviation.tex}}
\label{stab:deviation}
\end{table}

\subsection{Verification of the stored data}

All of the above was implemented independently of the experiment code (script \texttt{verify\_equations.py}) and compared with the stored arrays of the 40 worlds. Population, attractiveness, shares, standardised covariates, impedances, priors (with and without bias), count noise and capacity perturbations agree to within a relative error of $2.1\times10^{-15}$. The held-out sets, nested calibration sets and scenario clusters are identical. All 1,520 stored truth arrays (base and two scenarios for every truth, and the noisy counts) agree to within $4.4\times10^{-15}$.

\section{Experimental configurations}\label{sec:configs}

\begin{table}[htbp]\footnotesize
\tbl{Datasets. Each dataset is one network, world, truth, prior and coverage. G: generator. G-s and G-d: slope-fixed and decay-fixed ablations. F: OD-factor. C: cellwise (relative entropy). C2: cellwise (squared deviation).}
{\begin{tabular}{>{\raggedright\arraybackslash}p{0.15\linewidth}>{\raggedright\arraybackslash}p{0.10\linewidth}>{\raggedright\arraybackslash}p{0.27\linewidth}>{\raggedright\arraybackslash}p{0.12\linewidth}>{\raggedright\arraybackslash}p{0.18\linewidth}r}\toprule
Programme & Network & Truths & Coverage (\%) & Estimators & Datasets\\\midrule
Confirmation & EMA, Chicago & Matched, two-purpose & 5, 10, 20 & G, F, C, G-s at 10\%, C2 on worlds 2000–2004 at 10\% & 480\\
Confirmation & EMA, Chicago & Production interaction $\delta=0.25$, worlds 2000–2009 & 10 & G, F, C & 40\\
Post-confirmation & EMA & Matched, two-purpose & 5, 10, 20, 40 & G, F, C, G-s, G-d, C2 on 2000–2004 at 10\% & 320\\
Post-confirmation & EMA & 14 misspecified truths & 10, 20, 40 & G, F, C, G-s, G-d & 1,680\\
Post-confirmation & EMA & Biased prior, capacity error (matched, two-purpose) & 10 & G, F, C, G-s, G-d & 160\\
Post-confirmation & Chicago & $\zeta=1$, $\rho=0.7$ & 10, 20 & G, F, C, G-s, G-d & 80\\
Post-confirmation & Chicago & Capacity error (matched, two-purpose) & 10 & G, F, C, G-s, G-d & 80\\
Post-confirmation & Chicago & Matched, two-purpose & 10 & G, G-s, G-d & 80\\\bottomrule
\end{tabular}}
\label{stab:datasets}
\end{table}

Table~\ref{stab:datasets} lists the datasets, each with 20 worlds and two priors unless stated. The post-confirmation EMA anchor datasets at 5–20\% and the Chicago generator ablations at 10\% repeat confirmation datasets. Their shared records (1,780 base-year evaluations) are identical, and the analysis uses the confirmation records for those strata. The canonical record set has 2,180 EMA and 420 Chicago datasets, and no dataset failed. Table~\ref{stab:settings} lists the settings shared by both programmes.

\begin{table}[htbp]\footnotesize
\tbl{Estimation and evaluation settings.}
{\begin{tabular}{p{0.32\linewidth}>{\raggedright\arraybackslash}p{0.61\linewidth}}\toprule
Item & Setting\\\midrule
Worlds, priors & 20 worlds (seeds 2000–2019) per network, two priors per world\\
Count noise & Multiplicative, standard deviation 0.05, one draw per link and world\\
Held-out links & 20\% of eligible links, fixed per world\\
Penalty grids & Generator and ablations $\{0,0.01,0.1,1,10\}$, matrix estimators $\{0.001,0.01,0.1,1,10,\infty\}$, one tenfold extension above the largest finite value\\
Cross-validation & Three folds within the calibration links, validation count loss, ties towards the larger weight\\
Generator bounds & $|\gamma_0|\le2$, $|\gamma_k|\le1.5$, $0.2\le b\le8$\\
Factor bounds & $|\tau|,|u_i|,|v_j|\le5$, $0.2\le b^{(0)}+\Delta b\le8$\\
Budgets & Generator family 50 accepted steps or 300 evaluations, matrix estimators 300 or 1,200\\
Equilibrium tolerances & Relative gap $10^{-5}$ in estimation (5,000 sweeps), $10^{-6}$ for truth, evaluation and projection (20,000 sweeps)\\
Primary coverage & 10\% (confirmation programme)\\
Inference & World-level paired differences after averaging the two priors, 95\% percentile bootstrap over worlds with $B=10{,}000$ (seed 20261028 for confirmation strata, 20261031 otherwise), paired-$t$ intervals stored alongside\\\bottomrule
\end{tabular}}
\label{stab:settings}
\end{table}

The confirmation programme followed a written protocol, frozen with its configuration and code before its worlds were generated. Its primary comparisons were the generator against the cellwise and OD-factor estimators at 10\% coverage, on both networks and both truths, for OD CPC, held-out NRMSE and the held-out flow change error in the composition scenario: 24 estimates in all. All 24 favoured the generator, and the regenerated values in these appendices reproduce the frozen report to within $3.3\times10^{-16}$. The post-confirmation programme was written down after the EMA confirmation results were known and before any of its own outcomes. Its results are exploratory.

\section{Additional results and robustness}\label{sec:additional}

\subsection{All strata}

Table~\ref{stab:allcontrasts} gives the four main paired comparisons for every stratum, and Table~\ref{stab:medians} the corresponding method medians, including calibration NRMSE.

\begin{landscape}
\begingroup\footnotesize
\input{tables/tabS_allcontrasts.tex}
\endgroup
\end{landscape}

\begin{landscape}
\begingroup\footnotesize\setlength{\tabcolsep}{4pt}
\input{tables/tabS_medians.tex}
\endgroup
\end{landscape}

\subsection{Biased prior and capacity error}

With the biased prior on EMA at 10\% coverage, the generator's OD CPC advantage over cellwise adjustment increases from 0.123 to 0.185 under matched truth and from 0.075 to 0.133 under the two-purpose truth (Table~\ref{stab:allcontrasts}). Under matched truth the bias lowers the median OD CPC of the shared prior matrix from 0.829 to 0.748. The generator's median falls only from 0.963 to 0.956, while the matrix estimators, which remain close to the prior matrix, fall from 0.864 to 0.789 (OD-factor) and from 0.837 to 0.763 (cellwise) (Table~\ref{stab:medians}). Capacity error in the estimators' assignment reduces the advantage over cellwise adjustment on EMA from 0.123 to 0.094 (matched) and from 0.075 to 0.060 (two-purpose). Every interval remains above zero. On Chicago the corresponding values are 0.133 and 0.096, close to the values without capacity error (0.141 and 0.090). Under capacity error, cellwise adjustment no longer has the lowest calibration error on Chicago matched truth, because its calibration fit was obtained on the perturbed network.

\subsection{Ablations and secondary cellwise variant}

Table~\ref{stab:ablation} compares the generator with its ablations and its prior at 10\% coverage. Under matched and two-purpose truth, fitting the slopes and fitting the decay both improve OD recovery: the generator's OD CPC exceeds that of the slope-fixed ablation by 0.060–0.118 and that of the decay-fixed ablation by 0.031–0.071 on the two networks. Over all 25 strata at 10\% coverage, fitting the slopes improves OD CPC in 21 (inconclusive under the conditions with $\rho=0.7$ and under $\phi=1$) and held-out NRMSE in 24. Fitting the decay improves OD CPC in 13 and is inconclusive in the other 12, which include every destination-choice condition, the stronger production misspecifications and the $(0.25,8)$ decay contrast. The squared-deviation cellwise variant, run on five worlds, recovers OD demand less accurately than the generator in every one of those worlds on both networks.

\begin{landscape}
\begingroup\footnotesize
\input{tables/tabS_ablation.tex}
\endgroup
\end{landscape}

\subsection{Count fit along the regularisation path}

\begin{figure}[htbp]
\includegraphics[width=\linewidth]{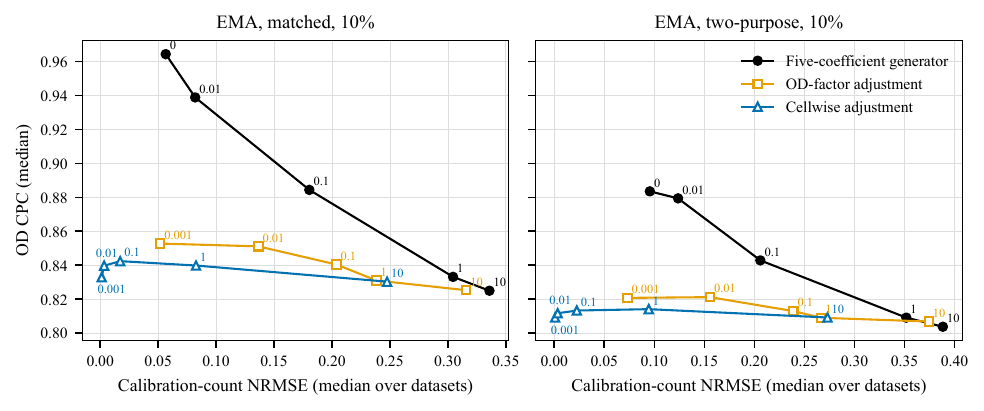}
\caption{Median calibration NRMSE against median OD CPC along each estimator's grid of penalty weights (labels), EMA, 10\% coverage, 40 datasets per truth.}
\label{sfig:path}
\end{figure}

On EMA at 10\% coverage under matched and two-purpose truth, the generator, the OD-factor and the cellwise estimators were also refitted at every finite grid weight (Figure~\ref{sfig:path}, Table~\ref{stab:path}). For the cellwise estimator the median Spearman correlation between calibration NRMSE and OD CPC along the path is 0.00, with a median OD CPC range of 0.008. For the generator it is $-1.00$ and $-0.95$, and for the OD-factor estimator $-1.00$ and $-0.80$. Calibration NRMSE and held-out NRMSE move together for all three estimators. These path fits are a diagnostic. They use the truth only for evaluation, not for selection.

\begin{table}[htbp]\footnotesize
\tbl{Regularisation-path diagnostic: median per-dataset Spearman correlations between calibration NRMSE and OD CPC or held-out NRMSE across the grid weights, and median OD CPC range along the path (EMA, 10\%).}
{\input{tables/tabS_lambdapath.tex}}
\label{stab:path}
\end{table}

\subsection{Distance to the nearest calibration link}

Each held-out link was assigned its free-flow distance to the nearest calibration link (the minimum symmetric node-to-node free-flow time over the endpoint pairs). Table~\ref{stab:d7} reports, per stratum, the median per-dataset Spearman correlation between that distance and the generator's absolute-error advantage over each comparator.

\begingroup\footnotesize
\input{tables/tabS_d7.tex}
\endgroup

\section{Scenario projection}\label{sec:projection}

\subsection{Scenarios and projection rules}

Each world defines two clusters of $\mathrm{round}(|\Zset|/4)$ zones, the zones nearest (in $d_{ij}$) to a randomly drawn anchor zone. They are selected before any fitting. The population and attractiveness scenario multiplies $N_i$ by 1.25 in the first cluster and $A_j$ by 1.25 in the second. The composition scenario raises the car-availability share by 0.10 and the employed-adult share by 0.05 in the first cluster, capped at the upper bounds of their ranges, with population and attractiveness unchanged. Scenario covariates are standardised with the base-year mean and standard deviation. Supply, support, $\kappa$ and impedance are unchanged, and no scenario counts are used. The future truth follows the law of Section~\ref{sec:truthsS} at the scenario inputs.

The generator is projected by re-evaluating Equation~\eqref{seq:gen} at the scenario inputs with its fitted coefficients (GEN). A base matrix $\hat{\mathbf q}_0$ of a matrix estimator is projected by
\begin{itemize}
\item The prior-ratio pivot (F3), $\hat q_{1,ij}=\hat q_{0,ij}\,q_{ij}(\bm\theta^{(0)},X_1)/q_{ij}(\bm\theta^{(0)},X_0)$, the main comparator rule of the confirmation protocol.
\item Row growth (F1), which scales each row to $\sum_j\hat q_{0,ij}\,O_i(\bm\theta^{(0)},X_1)/O_i(\bm\theta^{(0)},X_0)$ after multiplying column $j$ by $A_{1j}/A_{0j}$.
\item Furness (F2), which balances $\hat{\mathbf q}_0$ to the F1 origin targets and to destination targets $D_j\propto\sum_i\hat q_{0,ij}A_{1j}/A_{0j}$ rescaled to the same total, by iterative proportional fitting \citep{DemingStephan1940} with tolerance $10^{-10}$.
\item The hybrid (F3H), the pivot with the fitted generator's ratio $q_{ij}(\hat{\bm\theta},X_1)/q_{ij}(\hat{\bm\theta},X_0)$ in place of the prior's, which deliberately gives the matrix estimators the generator's fitted response.
\end{itemize}
The same-base comparison projects the generator's own base matrix by F3, so that the two projections share the base and differ only in the response rule. In the composition scenario F1 and F3 coincide for the matrix estimators. In the population and attractiveness scenario the generator's own projection equals its row-growth projection, because the accessibility denominator does not depend on $\mathbf N$ or $\mathbf A$. The largest difference in OD CPC between these two projections over 14,400 fits is $4.4\times10^{-16}$. The change error is
\begin{equation}
E_{\Delta x}=\frac{\|(\hat{\mathbf x}_1-\hat{\mathbf x}_0)-(\mathbf x^*_1-\mathbf x^*_0)\|_{2,\Hset}}{\|\mathbf x^*_1-\mathbf x^*_0\|_{2,\Hset}},
\end{equation}
computed on the held-out links, with each estimator's projected and base flows solved at gap $10^{-6}$.

\subsection{Results}

Table~\ref{stab:projection} reports the composition scenario at 10\% coverage. Under matched truth, the generator's change error is lower than that of the prior-pivoted cellwise projection by 0.473 on EMA and 0.445 on Chicago, and lower than that of the pivoted factor projection by 0.428 and 0.371. These are the projection endpoints of the confirmation programme. Most of this advantage remains when the generator's own base is projected by the prior ratio (same base: 0.399 and 0.347), so it comes from the fitted response rather than from the more accurate base matrix. When the matrix bases are projected with the generator's fitted response (hybrid), the advantage on EMA shrinks to 0.209 over cellwise and 0.121 over factor adjustment. Under the strongest destination-choice heterogeneity on Chicago, the advantage over the pivoted projections is large (1.300 and 1.160), but against the hybrid projections it is only 0.039 and 0.042, with intervals spanning zero. There, the advantage comes from the projection rule rather than from a better base estimate. Under the strongest spatial production effect on EMA, every projection contrast is inconclusive. The hybrid was not computed for the confirmation programme's Chicago records (matched and two-purpose truth), because it requires new projection equilibria. These projections are controlled scenarios with prescribed input changes. They are not observed temporal changes and do not validate forecasting accuracy.

\begin{landscape}
\begingroup\footnotesize
\input{tables/tabS_projection.tex}
\endgroup
\end{landscape}

\section{Reproducibility}\label{sec:repro}

\subsection{Code}
The experiments use a Python implementation in the package \texttt{generator\_pilot}. The confirmation programme ran a frozen copy of the code with fingerprint \texttt{29c105a2f5df8872}, and the post-confirmation programme ran a later copy with fingerprint \texttt{b7c8cd09bf6ac118}. The fingerprint is the first 16 hexadecimal digits of the SHA-256 hash over the package's Python and JSON files, and every record stores it. Both copies are preserved with file-level SHA-256 lists. The later copy differs from the frozen one only by additions (new truths, random streams, modifiers, ablation, diagnostics and projection rules), and it reproduces every shared confirmation record exactly. The configuration files have SHA-256 hashes \texttt{1c56c329\ldots} (confirmation) and \texttt{1d589278\ldots} (post-confirmation).

\subsection{Environment}
Runs used Python 3.11.15, NumPy 1.26.4, SciPy 1.17.1 and Numba 0.65.1 on Intel Xeon Platinum 8368Q nodes of the Cambridge Service for Data Driven Discovery, one thread per process.

\subsection{Data}
The networks are the Transportation Networks for Research files for Eastern Massachusetts and Chicago Sketch \citep{TNTP}, with SHA-256 hashes \texttt{c806a953\ldots} and \texttt{4396bff6\ldots}. All other data are synthetic and are regenerated from the world seeds by the equations of Section~\ref{sec:synthetic}.

\subsection{Outputs and analysis}
Each dataset's record (\texttt{record.json}) stores the configuration, cross-validation tables, fit histories, stopping status, equilibrium diagnostics and all evaluation metrics. The tables and figures of the article and these appendices are regenerated from these records by four scripts: \texttt{extract\_records.py} reads every record, \texttt{make\_results.py} computes the paired contrasts, medians and diagnostic summaries, and \texttt{make\_tables.py} and \texttt{make\_figures.py} write the tables and figures. \texttt{verify\_equations.py} performs the check of Section~\ref{sec:synthetic}. The regenerated contrasts reproduce the frozen confirmation report and the post-confirmation analysis tables exactly for every shared stratum.

\subsection{Availability}
The synthetic data, experiment configurations, records and analysis scripts are available from the corresponding author upon reasonable request.

\subsection{Known limitations}
Path-flow decompositions are not unique, so frozen-route proposals depend on solver history (Section~\ref{app-sec:numerics}). The stored fits are surrogate-stationary. The cellwise and the Chicago OD-factor fits were not polished. The total computing cost of the post-confirmation programme was not recorded, because the scheduler's accounting database was unavailable when it finished.

\bibliographystyle{unsrtnat}
\bibliography{references}
\end{document}

%% file: tables/tab_primary.tex
\begin{tabular}{llcccc}
\toprule
\multicolumn{6}{l}{\textit{Panel A. Median over worlds of the world mean}}\\
Network & Truth & Estimator & OD CPC & Held-out NRMSE & Calibration NRMSE\\
\midrule
EMA & Matched & Prior & 0.829 & 0.363 & 0.372\\
 &  & Generator & 0.963 & 0.060 & 0.057\\
 &  & OD-factor & 0.864 & 0.201 & 0.060\\
 &  & Cellwise & 0.837 & 0.281 & 0.010\\
\addlinespace[2pt]
EMA & Two-purpose $(1,3)$ & Prior & 0.797 & 0.397 & 0.403\\
 &  & Generator & 0.887 & 0.135 & 0.114\\
 &  & OD-factor & 0.817 & 0.250 & 0.103\\
 &  & Cellwise & 0.802 & 0.311 & 0.022\\
\addlinespace[2pt]
Chicago & Matched & Prior & 0.839 & 0.330 & 0.367\\
 &  & Generator & 0.981 & 0.026 & 0.073\\
 &  & OD-factor & 0.888 & 0.129 & 0.125\\
 &  & Cellwise & 0.831 & 0.151 & 0.016\\
\addlinespace[2pt]
Chicago & Two-purpose $(1,3)$ & Prior & 0.808 & 0.324 & 0.356\\
 &  & Generator & 0.892 & 0.065 & 0.094\\
 &  & OD-factor & 0.846 & 0.146 & 0.148\\
 &  & Cellwise & 0.806 & 0.164 & 0.017\\
\addlinespace[2pt]
\midrule
\multicolumn{6}{l}{\textit{Panel B. Mean paired difference, generator versus comparator (95\% interval, W)}}\\
Network & Truth & Comparator & OD CPC & Held-out NRMSE & Calibration NRMSE\\
\midrule
EMA & Matched & OD-factor & \begin{tabular}[t]{@{}c@{}}\textbf{+0.099} (20)\\[-1pt]{\scriptsize[+0.084, +0.115]}\end{tabular} & \begin{tabular}[t]{@{}c@{}}\textbf{+0.138} (20)\\[-1pt]{\scriptsize[+0.111, +0.168]}\end{tabular} & \begin{tabular}[t]{@{}c@{}}\textbf{+0.028} (11)\\[-1pt]{\scriptsize[+0.007, +0.052]}\end{tabular}\\
 &  & Cellwise & \begin{tabular}[t]{@{}c@{}}\textbf{+0.123} (20)\\[-1pt]{\scriptsize[+0.107, +0.139]}\end{tabular} & \begin{tabular}[t]{@{}c@{}}\textbf{+0.217} (20)\\[-1pt]{\scriptsize[+0.182, +0.255]}\end{tabular} & \begin{tabular}[t]{@{}c@{}}\textbf{$-$0.028} (3)\\[-1pt]{\scriptsize[$-$0.047, $-$0.004]}\end{tabular}\\
\addlinespace[2pt]
EMA & Two-purpose $(1,3)$ & OD-factor & \begin{tabular}[t]{@{}c@{}}\textbf{+0.064} (20)\\[-1pt]{\scriptsize[+0.051, +0.077]}\end{tabular} & \begin{tabular}[t]{@{}c@{}}\textbf{+0.116} (20)\\[-1pt]{\scriptsize[+0.097, +0.135]}\end{tabular} & \begin{tabular}[t]{@{}c@{}}+0.007 (9)\\[-1pt]{\scriptsize[$-$0.022, +0.038]}\end{tabular}\\
 &  & Cellwise & \begin{tabular}[t]{@{}c@{}}\textbf{+0.075} (19)\\[-1pt]{\scriptsize[+0.061, +0.090]}\end{tabular} & \begin{tabular}[t]{@{}c@{}}\textbf{+0.167} (19)\\[-1pt]{\scriptsize[+0.131, +0.201]}\end{tabular} & \begin{tabular}[t]{@{}c@{}}\textbf{$-$0.069} (3)\\[-1pt]{\scriptsize[$-$0.096, $-$0.036]}\end{tabular}\\
\addlinespace[2pt]
Chicago & Matched & OD-factor & \begin{tabular}[t]{@{}c@{}}\textbf{+0.093} (20)\\[-1pt]{\scriptsize[+0.078, +0.108]}\end{tabular} & \begin{tabular}[t]{@{}c@{}}\textbf{+0.111} (20)\\[-1pt]{\scriptsize[+0.090, +0.132]}\end{tabular} & \begin{tabular}[t]{@{}c@{}}\textbf{+0.058} (20)\\[-1pt]{\scriptsize[+0.046, +0.072]}\end{tabular}\\
 &  & Cellwise & \begin{tabular}[t]{@{}c@{}}\textbf{+0.141} (20)\\[-1pt]{\scriptsize[+0.123, +0.160]}\end{tabular} & \begin{tabular}[t]{@{}c@{}}\textbf{+0.135} (20)\\[-1pt]{\scriptsize[+0.113, +0.160]}\end{tabular} & \begin{tabular}[t]{@{}c@{}}\textbf{$-$0.056} (0)\\[-1pt]{\scriptsize[$-$0.062, $-$0.049]}\end{tabular}\\
\addlinespace[2pt]
Chicago & Two-purpose $(1,3)$ & OD-factor & \begin{tabular}[t]{@{}c@{}}\textbf{+0.048} (20)\\[-1pt]{\scriptsize[+0.037, +0.059]}\end{tabular} & \begin{tabular}[t]{@{}c@{}}\textbf{+0.093} (20)\\[-1pt]{\scriptsize[+0.074, +0.113]}\end{tabular} & \begin{tabular}[t]{@{}c@{}}\textbf{+0.058} (20)\\[-1pt]{\scriptsize[+0.044, +0.071]}\end{tabular}\\
 &  & Cellwise & \begin{tabular}[t]{@{}c@{}}\textbf{+0.090} (20)\\[-1pt]{\scriptsize[+0.074, +0.106]}\end{tabular} & \begin{tabular}[t]{@{}c@{}}\textbf{+0.108} (20)\\[-1pt]{\scriptsize[+0.088, +0.130]}\end{tabular} & \begin{tabular}[t]{@{}c@{}}\textbf{$-$0.073} (0)\\[-1pt]{\scriptsize[$-$0.081, $-$0.065]}\end{tabular}\\
\addlinespace[2pt]
\bottomrule
\end{tabular}

%% file: tables/tab_misspec.tex
\begin{tabular}{llcccccc}
\toprule
 & & & & \multicolumn{2}{c}{OD CPC} & \multicolumn{2}{c}{Held-out NRMSE}\\
\cmidrule(lr){5-6}\cmidrule(lr){7-8}
Network & Truth & Cov. & Dev. & vs OD-factor & vs cellwise & vs OD-factor & vs cellwise\\
\midrule
EMA & Two-purpose $(0.25,8)$ & 10\% & 0.755 & \begin{tabular}[t]{@{}c@{}}\textbf{+0.035} (20)\\[-1pt]{\scriptsize[+0.026, +0.044]}\end{tabular} & \begin{tabular}[t]{@{}c@{}}\textbf{+0.030} (19)\\[-1pt]{\scriptsize[+0.020, +0.039]}\end{tabular} & \begin{tabular}[t]{@{}c@{}}\textbf{+0.125} (20)\\[-1pt]{\scriptsize[+0.105, +0.145]}\end{tabular} & \begin{tabular}[t]{@{}c@{}}\textbf{+0.155} (19)\\[-1pt]{\scriptsize[+0.119, +0.194]}\end{tabular}\\
EMA & Two-purpose $(0.25,8)$ & 40\% & 0.755 & \begin{tabular}[t]{@{}c@{}}\textbf{+0.027} (19)\\[-1pt]{\scriptsize[+0.019, +0.036]}\end{tabular} & \begin{tabular}[t]{@{}c@{}}\textbf{+0.018} (17)\\[-1pt]{\scriptsize[+0.010, +0.027]}\end{tabular} & \begin{tabular}[t]{@{}c@{}}\textbf{+0.095} (20)\\[-1pt]{\scriptsize[+0.072, +0.120]}\end{tabular} & \begin{tabular}[t]{@{}c@{}}\textbf{+0.076} (19)\\[-1pt]{\scriptsize[+0.050, +0.102]}\end{tabular}\\
\addlinespace[3pt]
EMA & $\zeta=0$, $\rho=0$ & 10\% & 0.856 & \begin{tabular}[t]{@{}c@{}}+0.019 (18)\\[-1pt]{\scriptsize[$-$0.007, +0.037]}\end{tabular} & \begin{tabular}[t]{@{}c@{}}+0.018 (16)\\[-1pt]{\scriptsize[$-$0.009, +0.038]}\end{tabular} & \begin{tabular}[t]{@{}c@{}}\textbf{+0.057} (19)\\[-1pt]{\scriptsize[+0.038, +0.078]}\end{tabular} & \begin{tabular}[t]{@{}c@{}}\textbf{+0.070} (19)\\[-1pt]{\scriptsize[+0.046, +0.096]}\end{tabular}\\
EMA & $\zeta=0$, $\rho=0$ & 40\% & 0.856 & \begin{tabular}[t]{@{}c@{}}\textbf{+0.021} (14)\\[-1pt]{\scriptsize[+0.006, +0.035]}\end{tabular} & \begin{tabular}[t]{@{}c@{}}+0.012 (15)\\[-1pt]{\scriptsize[$-$0.004, +0.028]}\end{tabular} & \begin{tabular}[t]{@{}c@{}}\textbf{+0.065} (17)\\[-1pt]{\scriptsize[+0.042, +0.087]}\end{tabular} & \begin{tabular}[t]{@{}c@{}}\textbf{$-$0.021} (7)\\[-1pt]{\scriptsize[$-$0.041, $-$0.000]}\end{tabular}\\
EMA & $\zeta=0$, $\rho=0.7$ & 40\% & 0.765 & \begin{tabular}[t]{@{}c@{}}+0.013 (14)\\[-1pt]{\scriptsize[$-$0.004, +0.030]}\end{tabular} & \begin{tabular}[t]{@{}c@{}}\textbf{$-$0.035} (5)\\[-1pt]{\scriptsize[$-$0.054, $-$0.017]}\end{tabular} & \begin{tabular}[t]{@{}c@{}}\textbf{+0.039} (13)\\[-1pt]{\scriptsize[+0.003, +0.074]}\end{tabular} & \begin{tabular}[t]{@{}c@{}}\textbf{$-$0.183} (1)\\[-1pt]{\scriptsize[$-$0.226, $-$0.138]}\end{tabular}\\
EMA & $\zeta=1$, $\rho=0$ & 40\% & 0.681 & \begin{tabular}[t]{@{}c@{}}\textbf{$-$0.008} (8)\\[-1pt]{\scriptsize[$-$0.016, $-$0.001]}\end{tabular} & \begin{tabular}[t]{@{}c@{}}\textbf{$-$0.016} (6)\\[-1pt]{\scriptsize[$-$0.026, $-$0.006]}\end{tabular} & \begin{tabular}[t]{@{}c@{}}\textbf{+0.055} (14)\\[-1pt]{\scriptsize[+0.019, +0.092]}\end{tabular} & \begin{tabular}[t]{@{}c@{}}\textbf{$-$0.035} (6)\\[-1pt]{\scriptsize[$-$0.064, $-$0.009]}\end{tabular}\\
EMA & $\zeta=1$, $\rho=0.7$ & 10\% & 0.619 & \begin{tabular}[t]{@{}c@{}}+0.001 (11)\\[-1pt]{\scriptsize[$-$0.013, +0.016]}\end{tabular} & \begin{tabular}[t]{@{}c@{}}\textbf{$-$0.019} (6)\\[-1pt]{\scriptsize[$-$0.035, $-$0.003]}\end{tabular} & \begin{tabular}[t]{@{}c@{}}\textbf{+0.089} (18)\\[-1pt]{\scriptsize[+0.045, +0.142]}\end{tabular} & \begin{tabular}[t]{@{}c@{}}+0.017 (9)\\[-1pt]{\scriptsize[$-$0.036, +0.082]}\end{tabular}\\
EMA & $\zeta=1$, $\rho=0.7$ & 20\% & 0.619 & \begin{tabular}[t]{@{}c@{}}$-$0.004 (11)\\[-1pt]{\scriptsize[$-$0.016, +0.008]}\end{tabular} & \begin{tabular}[t]{@{}c@{}}\textbf{$-$0.034} (3)\\[-1pt]{\scriptsize[$-$0.047, $-$0.021]}\end{tabular} & \begin{tabular}[t]{@{}c@{}}\textbf{+0.046} (12)\\[-1pt]{\scriptsize[+0.006, +0.086]}\end{tabular} & \begin{tabular}[t]{@{}c@{}}\textbf{$-$0.098} (3)\\[-1pt]{\scriptsize[$-$0.147, $-$0.050]}\end{tabular}\\
EMA & $\zeta=1$, $\rho=0.7$ & 40\% & 0.619 & \begin{tabular}[t]{@{}c@{}}\textbf{$-$0.014} (6)\\[-1pt]{\scriptsize[$-$0.029, $-$0.000]}\end{tabular} & \begin{tabular}[t]{@{}c@{}}\textbf{$-$0.049} (2)\\[-1pt]{\scriptsize[$-$0.067, $-$0.033]}\end{tabular} & \begin{tabular}[t]{@{}c@{}}\textbf{+0.048} (16)\\[-1pt]{\scriptsize[+0.021, +0.079]}\end{tabular} & \begin{tabular}[t]{@{}c@{}}\textbf{$-$0.177} (1)\\[-1pt]{\scriptsize[$-$0.218, $-$0.131]}\end{tabular}\\
Chicago & $\zeta=1$, $\rho=0.7$ & 10\% & 0.626 & \begin{tabular}[t]{@{}c@{}}\textbf{$-$0.041} (3)\\[-1pt]{\scriptsize[$-$0.063, $-$0.021]}\end{tabular} & \begin{tabular}[t]{@{}c@{}}\textbf{$-$0.052} (4)\\[-1pt]{\scriptsize[$-$0.075, $-$0.028]}\end{tabular} & \begin{tabular}[t]{@{}c@{}}\textbf{+0.069} (20)\\[-1pt]{\scriptsize[+0.048, +0.093]}\end{tabular} & \begin{tabular}[t]{@{}c@{}}\textbf{$-$0.150} (1)\\[-1pt]{\scriptsize[$-$0.178, $-$0.119]}\end{tabular}\\
Chicago & $\zeta=1$, $\rho=0.7$ & 20\% & 0.626 & \begin{tabular}[t]{@{}c@{}}\textbf{$-$0.041} (4)\\[-1pt]{\scriptsize[$-$0.065, $-$0.018]}\end{tabular} & \begin{tabular}[t]{@{}c@{}}\textbf{$-$0.064} (3)\\[-1pt]{\scriptsize[$-$0.089, $-$0.040]}\end{tabular} & \begin{tabular}[t]{@{}c@{}}\textbf{+0.055} (19)\\[-1pt]{\scriptsize[+0.034, +0.078]}\end{tabular} & \begin{tabular}[t]{@{}c@{}}\textbf{$-$0.250} (0)\\[-1pt]{\scriptsize[$-$0.277, $-$0.222]}\end{tabular}\\
\addlinespace[3pt]
EMA & Interaction $\delta=1$ & 40\% & 0.792 & \begin{tabular}[t]{@{}c@{}}\textbf{+0.015} (14)\\[-1pt]{\scriptsize[+0.003, +0.029]}\end{tabular} & \begin{tabular}[t]{@{}c@{}}\textbf{+0.052} (20)\\[-1pt]{\scriptsize[+0.039, +0.068]}\end{tabular} & \begin{tabular}[t]{@{}c@{}}\textbf{+0.035} (13)\\[-1pt]{\scriptsize[+0.008, +0.069]}\end{tabular} & \begin{tabular}[t]{@{}c@{}}\textbf{+0.061} (16)\\[-1pt]{\scriptsize[+0.030, +0.096]}\end{tabular}\\
\addlinespace[3pt]
EMA & Spatial $\phi=0.5$ & 40\% & 0.833 & \begin{tabular}[t]{@{}c@{}}$-$0.002 (11)\\[-1pt]{\scriptsize[$-$0.019, +0.014]}\end{tabular} & \begin{tabular}[t]{@{}c@{}}\textbf{+0.025} (17)\\[-1pt]{\scriptsize[+0.013, +0.037]}\end{tabular} & \begin{tabular}[t]{@{}c@{}}+0.004 (11)\\[-1pt]{\scriptsize[$-$0.041, +0.046]}\end{tabular} & \begin{tabular}[t]{@{}c@{}}+0.009 (11)\\[-1pt]{\scriptsize[$-$0.028, +0.047]}\end{tabular}\\
EMA & Spatial $\phi=1$ & 10\% & 0.715 & \begin{tabular}[t]{@{}c@{}}\textbf{$-$0.043} (6)\\[-1pt]{\scriptsize[$-$0.082, $-$0.014]}\end{tabular} & \begin{tabular}[t]{@{}c@{}}$-$0.015 (9)\\[-1pt]{\scriptsize[$-$0.043, +0.009]}\end{tabular} & \begin{tabular}[t]{@{}c@{}}$-$0.035 (10)\\[-1pt]{\scriptsize[$-$0.127, +0.052]}\end{tabular} & \begin{tabular}[t]{@{}c@{}}\textbf{+0.066} (16)\\[-1pt]{\scriptsize[+0.010, +0.124]}\end{tabular}\\
EMA & Spatial $\phi=1$ & 20\% & 0.715 & \begin{tabular}[t]{@{}c@{}}\textbf{$-$0.054} (6)\\[-1pt]{\scriptsize[$-$0.107, $-$0.008]}\end{tabular} & \begin{tabular}[t]{@{}c@{}}$-$0.014 (11)\\[-1pt]{\scriptsize[$-$0.052, +0.020]}\end{tabular} & \begin{tabular}[t]{@{}c@{}}$-$0.075 (8)\\[-1pt]{\scriptsize[$-$0.193, +0.025]}\end{tabular} & \begin{tabular}[t]{@{}c@{}}$-$0.012 (11)\\[-1pt]{\scriptsize[$-$0.102, +0.065]}\end{tabular}\\
EMA & Spatial $\phi=1$ & 40\% & 0.715 & \begin{tabular}[t]{@{}c@{}}\textbf{$-$0.081} (2)\\[-1pt]{\scriptsize[$-$0.122, $-$0.045]}\end{tabular} & \begin{tabular}[t]{@{}c@{}}$-$0.006 (7)\\[-1pt]{\scriptsize[$-$0.031, +0.025]}\end{tabular} & \begin{tabular}[t]{@{}c@{}}\textbf{$-$0.117} (4)\\[-1pt]{\scriptsize[$-$0.222, $-$0.027]}\end{tabular} & \begin{tabular}[t]{@{}c@{}}\textbf{$-$0.089} (6)\\[-1pt]{\scriptsize[$-$0.180, $-$0.013]}\end{tabular}\\
\bottomrule
\end{tabular}

%% file: tables/tab_numerics.tex
\begin{tabular}{>{\raggedright\arraybackslash}p{0.52\linewidth}cc}
\toprule
Diagnostic & EMA & Chicago\\
\midrule
\multicolumn{3}{l}{\textit{Count sensitivities (all diagnostic points)}}\\
Finite-difference Jacobians evaluated & 10,400 & 1,520\\
\quad with effective rank five & 10,391 & 1,520\\
Condition index, median (10th, 90th percentile)\textsuperscript{a} & 6.1 (3.4, 48.4) & 8.3 (4.6, 30.7)\\
Relative difference, frozen-route vs finite-difference Jacobian\textsuperscript{a} & 0.55 (0.23, 0.92) & 0.89 (0.73, 0.97)\\
\multicolumn{3}{l}{\textit{Stationarity of stored generator fits}}\\
Predicted relative objective decrease, median (share $>1\%$)\textsuperscript{a} & 2.0\% (70\%) & 2.5\% (76\%)\\
Fits with a coefficient at a bound & 30 of 2,160 & 4 of 240\\
\multicolumn{3}{l}{\textit{Polishing on the equilibrium objective (range of stratum medians)}}\\
Realised objective decrease & 0.6–7.3\% & 1.1–5.9\%\\
Change in OD CPC & $-$0.0027 to +0.0049 & $-$0.0002 to +0.0091\\
Change in held-out NRMSE & $-$0.0115 to +0.0028 & $-$0.0098 to $-$0.0001\\
Contrast classifications changed\textsuperscript{b} & 7 of 216 & 0 of 24\\
\multicolumn{3}{l}{\textit{Five-start re-optimisation (EMA, 10\%, 10 datasets)}}\\
Objective decrease at the best start (share of stored) & $-$0.2 to 10.4\% & –\\
Largest OD CPC range across starts and stored fit & 0.031 & –\\
\bottomrule
\end{tabular}

%% file: tables/tabS_d1d2.tex
\begin{tabular}{lllrcccc}
\toprule
Network & Point & UE gap & $n$ & Eff.\ rank 5 & Cond.\ index (median) & Rel.\ difference (median) & Rel.\ difference (90th pct.)\\
\midrule
Chicago & fit & 1e-07 & 320 & 320 & 8.0 & 0.892 & 0.975\\
Chicago & fit & 1e-06 & 320 & 320 & 7.8 & 0.872 & 0.975\\
Chicago & prior & 1e-07 & 320 & 320 & 9.3 & 0.857 & 0.969\\
Chicago & prior & 1e-06 & 320 & 320 & 9.3 & 0.827 & 0.966\\
Chicago & truth & 1e-07 & 120 & 120 & 7.5 & 0.899 & 0.966\\
Chicago & truth & 1e-06 & 120 & 120 & 7.4 & 0.881 & 0.972\\
EMA & fit & 1e-08 & 2,420 & 2,417 & 6.4 & 0.606 & 0.943\\
EMA & fit & 1e-06 & 2,420 & 2,415 & 6.3 & 0.599 & 0.946\\
EMA & prior & 1e-08 & 2,420 & 2,420 & 6.0 & 0.489 & 0.864\\
EMA & prior & 1e-06 & 2,420 & 2,420 & 6.5 & 0.514 & 0.897\\
EMA & truth & 1e-08 & 360 & 360 & 5.3 & 0.593 & 0.967\\
EMA & truth & 1e-06 & 360 & 359 & 5.3 & 0.603 & 0.972\\
\midrule
Network & Set & UE gap & $n$ & At bound & Pred.\ decrease (median) & Share $>1\%$ & Gradient rel.\ difference (median)\\
\midrule
Chicago & confirmation & 1e-07 & 80 & 0 & 4.00\% & 92\% & 1.040\\
Chicago & confirmation & 1e-06 & 80 & 0 & 3.85\% & 89\% & 1.032\\
Chicago & post-confirmation & 1e-07 & 240 & 4 & 2.51\% & 76\% & 1.047\\
Chicago & post-confirmation & 1e-06 & 240 & 4 & 2.31\% & 73\% & 1.047\\
EMA & confirmation & 1e-08 & 260 & 0 & 4.16\% & 82\% & 1.032\\
EMA & confirmation & 1e-06 & 260 & 0 & 4.47\% & 82\% & 1.029\\
EMA & post-confirmation & 1e-08 & 2,160 & 30 & 2.01\% & 70\% & 1.017\\
EMA & post-confirmation & 1e-06 & 2,160 & 30 & 1.99\% & 68\% & 1.013\\
\bottomrule
\end{tabular}

%% file: tables/tabS_polish.tex
\begin{tabular}{lllrccc}
\toprule
Network & Estimator & Family & Strata & Realised decrease & $\Delta$ OD CPC & $\Delta$ held-out NRMSE\\
\midrule
Chicago & Generator & anchors & 2 & 5.5–5.9\% & $-$0.0002 to +0.0091 & $-$0.0098 to $-$0.0047\\
Chicago & Generator & destination choice & 2 & 1.1–1.3\% & +0.0008 to +0.0009 & $-$0.0007 to $-$0.0001\\
Chicago & Generator & prior/supply & 2 & 3.7–5.4\% & +0.0023 to +0.0055 & $-$0.0032 to $-$0.0030\\
EMA & Generator & anchors & 8 & 2.1–7.0\% & $-$0.0005 to +0.0049 & $-$0.0115 to +0.0005\\
EMA & Generator & destination choice & 18 & 0.6–2.3\% & $-$0.0019 to +0.0002 & $-$0.0029 to +0.0017\\
EMA & Generator & prior/supply & 4 & 3.6–7.3\% & $-$0.0027 to +0.0020 & $-$0.0045 to $-$0.0000\\
EMA & Generator & production & 15 & 0.8–3.8\% & $-$0.0003 to +0.0020 & $-$0.0025 to +0.0028\\
EMA & Generator & two-purpose severity & 9 & 1.8–5.3\% & $-$0.0008 to +0.0010 & $-$0.0047 to +0.0000\\
EMA & OD-factor & anchors & 8 & 4.2–7.2\% & +0.0000 to +0.0014 & $-$0.0046 to $-$0.0000\\
EMA & OD-factor & destination choice & 18 & 2.2–5.7\% & $-$0.0004 to +0.0015 & $-$0.0044 to +0.0017\\
EMA & OD-factor & prior/supply & 4 & 4.8–9.3\% & +0.0004 to +0.0018 & $-$0.0053 to +0.0025\\
EMA & OD-factor & production & 15 & 4.1–9.2\% & +0.0001 to +0.0040 & $-$0.0098 to $-$0.0001\\
EMA & OD-factor & two-purpose severity & 9 & 2.7–6.0\% & +0.0000 to +0.0011 & $-$0.0037 to $-$0.0003\\
\bottomrule
\end{tabular}

%% file: tables/tabS_polish_changes.tex
\begin{tabular}{llcllcc}
\toprule
Network & Truth & Cov. & Comparator & Metric & Stored fits & After polishing\\
\midrule
EMA & $\zeta=0$, $\rho=0$ & 40\% & OD-factor & OD CPC & +0.021 (generator) & +0.016 (spans 0)\\
EMA & $\zeta=0$, $\rho=0.3$ & 10\% & cellwise & OD CPC & +0.020 (generator) & +0.015 (spans 0)\\
EMA & $\zeta=0$, $\rho=0.7$ & 10\% & cellwise & OD CPC & $-$0.018 (spans 0) & $-$0.027 (comparator)\\
EMA & $\zeta=0$, $\rho=0.7$ & 20\% & cellwise & OD CPC & $-$0.019 (spans 0) & $-$0.022 (comparator)\\
EMA & Interaction $\delta=1$ & 40\% & OD-factor & OD CPC & +0.015 (generator) & +0.010 (spans 0)\\
EMA & Interaction $\delta=1$ & 40\% & OD-factor & held-out NRMSE & +0.035 (generator) & +0.025 (spans 0)\\
EMA & Spatial $\phi=0.5$ & 10\% & OD-factor & OD CPC & +0.013 (generator) & +0.010 (spans 0)\\
\bottomrule
\end{tabular}

%% file: tables/tabS_multistart.tex
\begin{tabular}{llcccccc}
\toprule
World & Truth & Stored objective & Best start & Worst start & Stored OD CPC & Min.\ OD CPC & Max.\ OD CPC\\
\midrule
2000 & Matched & 0.006612 & 0.006291 & 0.006601 & 0.9326 & 0.9309 & 0.9383\\
2001 & Matched & 0.001011 & 0.001000 & 0.001041 & 0.9585 & 0.9540 & 0.9611\\
2002 & Matched & 0.011505 & 0.011529 & 0.011899 & 0.9306 & 0.9247 & 0.9302\\
2003 & Matched & 0.000614 & 0.000551 & 0.000652 & 0.9667 & 0.9664 & 0.9732\\
2004 & Matched & 0.003390 & 0.003325 & 0.003446 & 0.9713 & 0.9700 & 0.9756\\
2000 & Two-purpose $(1,3)$ & 0.011244 & 0.010968 & 0.011246 & 0.8542 & 0.8539 & 0.8574\\
2001 & Two-purpose $(1,3)$ & 0.015261 & 0.014291 & 0.015926 & 0.8432 & 0.8191 & 0.8499\\
2002 & Two-purpose $(1,3)$ & 0.018637 & 0.018591 & 0.019172 & 0.8792 & 0.8721 & 0.8789\\
2003 & Two-purpose $(1,3)$ & 0.009772 & 0.009463 & 0.009650 & 0.8879 & 0.8852 & 0.8878\\
2004 & Two-purpose $(1,3)$ & 0.006588 & 0.006526 & 0.006633 & 0.9017 & 0.9013 & 0.9020\\
\bottomrule
\end{tabular}

%% file: tables/tabS_status.tex
\begin{tabular}{llrrrrrrrr}
\toprule
Network & Estimator & no descent & small step & stagnation & stationary & surrogate stationary & max accepted & eval budget & unchanged Q0\\
\midrule
Chicago & Generator & 234 & 174 & 12 & 0 & 0 & 0 & 0 & 0\\
Chicago & Slopes fixed & 129 & 89 & 21 & 1 & 0 & 0 & 0 & 0\\
Chicago & Decay fixed & 105 & 121 & 14 & 0 & 0 & 0 & 0 & 0\\
Chicago & Cellwise (KL) & 411 & 0 & 0 & 6 & 0 & 0 & 0 & 3\\
Chicago & Cellwise (squared) & 20 & 0 & 0 & 0 & 0 & 0 & 0 & 0\\
Chicago & OD-factor & 278 & 0 & 26 & 29 & 79 & 0 & 0 & 8\\
EMA & Generator & 895 & 1102 & 179 & 0 & 0 & 4 & 0 & 0\\
EMA & Slopes fixed & 589 & 1259 & 306 & 3 & 0 & 3 & 0 & 0\\
EMA & Decay fixed & 816 & 1153 & 188 & 0 & 0 & 3 & 0 & 0\\
EMA & Cellwise (KL) & 1651 & 0 & 0 & 430 & 2 & 0 & 1 & 96\\
EMA & Cellwise (squared) & 16 & 0 & 0 & 3 & 1 & 0 & 0 & 0\\
EMA & OD-factor & 1321 & 0 & 178 & 189 & 370 & 0 & 1 & 121\\
\bottomrule
\end{tabular}

%% file: tables/tabS_lambda.tex
\begin{tabular}{llrrrrrrrr}
\toprule
Network & Estimator & 0 & 0.001 & 0.01 & 0.1 & 1 & 10 & 100 & $\infty$\\
\midrule
Chicago & Generator & 353 & 0 & 62 & 1 & 1 & 0 & 3 & 0\\
Chicago & Slopes fixed & 146 & 0 & 58 & 21 & 4 & 2 & 9 & 0\\
Chicago & Decay fixed & 165 & 0 & 66 & 3 & 1 & 0 & 5 & 0\\
Chicago & Cellwise (KL) & 0 & 185 & 122 & 93 & 17 & 0 & 0 & 3\\
Chicago & Cellwise (squared) & 0 & 9 & 1 & 8 & 2 & 0 & 0 & 0\\
Chicago & OD-factor & 0 & 393 & 11 & 6 & 1 & 1 & 0 & 8\\
EMA & Generator & 1082 & 0 & 804 & 203 & 26 & 7 & 58 & 0\\
EMA & Slopes fixed & 749 & 0 & 724 & 365 & 105 & 27 & 190 & 0\\
EMA & Decay fixed & 904 & 0 & 868 & 244 & 32 & 15 & 97 & 0\\
EMA & Cellwise (KL) & 0 & 1007 & 541 & 362 & 119 & 51 & 4 & 96\\
EMA & Cellwise (squared) & 0 & 8 & 2 & 3 & 1 & 6 & 0 & 0\\
EMA & OD-factor & 0 & 1360 & 390 & 158 & 121 & 28 & 2 & 121\\
\bottomrule
\end{tabular}

%% file: tables/tabS_deviation.tex
\begin{tabular}{llccc}
\toprule
Network & Truth & Median & Minimum & Maximum\\
\midrule
EMA & Two-purpose $(1,3)$ & 0.895 & 0.870 & 0.912\\
EMA & Mixture $\alpha=0.5$ & 0.947 & 0.935 & 0.956\\
EMA & Two-purpose $(0.5,5)$ & 0.842 & 0.822 & 0.860\\
EMA & Two-purpose $(0.25,8)$ & 0.755 & 0.737 & 0.774\\
EMA & $\zeta=0$, $\rho=0$ & 0.856 & 0.825 & 0.877\\
EMA & $\zeta=0.5$, $\rho=0$ & 0.791 & 0.761 & 0.813\\
EMA & $\zeta=1$, $\rho=0$ & 0.681 & 0.637 & 0.716\\
EMA & $\zeta=0$, $\rho=0.3$ & 0.832 & 0.760 & 0.888\\
EMA & $\zeta=0$, $\rho=0.7$ & 0.765 & 0.646 & 0.806\\
EMA & $\zeta=1$, $\rho=0.7$ & 0.619 & 0.561 & 0.675\\
EMA & Interaction $\delta=0.25$ & 0.886 & 0.864 & 0.902\\
EMA & Interaction $\delta=0.5$ & 0.862 & 0.848 & 0.880\\
EMA & Interaction $\delta=1$ & 0.792 & 0.773 & 0.821\\
EMA & Spatial $\phi=0.5$ & 0.833 & 0.774 & 0.866\\
EMA & Spatial $\phi=1$ & 0.715 & 0.436 & 0.794\\
EMA & Interaction $\delta=0.25$\textsuperscript{a} & 0.889 & 0.864 & 0.902\\
Chicago & Two-purpose $(1,3)$ & 0.892 & 0.883 & 0.903\\
Chicago & $\zeta=1$, $\rho=0.7$ & 0.626 & 0.583 & 0.667\\
Chicago & Interaction $\delta=0.25$\textsuperscript{a} & 0.885 & 0.880 & 0.891\\
\bottomrule
\end{tabular}

%% file: tables/tabS_allcontrasts.tex
\begin{longtable}{llcccccc}
\caption{Mean paired differences for every evaluated stratum (generator minus comparator for OD CPC, comparator minus generator for held-out NRMSE). Each cell gives the mean and W, the number of worlds favouring the generator, with the 95\% world-bootstrap interval beneath. Bold: interval excludes zero. Dev.: median OD CPC between the truth and the matched truth in the same world.}\label{stab:allcontrasts}\\
\toprule
 & & & & \multicolumn{2}{c}{OD CPC} & \multicolumn{2}{c}{Held-out NRMSE}\\
\cmidrule(lr){5-6}\cmidrule(lr){7-8}
Network & Truth & Cov. & Dev. & vs OD-factor & vs cellwise & vs OD-factor & vs cellwise\\
\midrule
\endfirsthead
\toprule
 & & & & \multicolumn{2}{c}{OD CPC} & \multicolumn{2}{c}{Held-out NRMSE}\\
\cmidrule(lr){5-6}\cmidrule(lr){7-8}
Network & Truth & Cov. & Dev. & vs OD-factor & vs cellwise & vs OD-factor & vs cellwise\\
\midrule
\endhead
\bottomrule
\multicolumn{8}{l}{\footnotesize \textsuperscript{a} Confirmation-programme diagnostic, worlds 2000–2009 only.}\\
\endfoot
EMA & Matched & 5\% & 1.000 & \begin{tabular}[t]{@{}c@{}}\textbf{+0.070} (19)\\[-1pt]{\scriptsize[+0.052, +0.089]}\end{tabular} & \begin{tabular}[t]{@{}c@{}}\textbf{+0.093} (20)\\[-1pt]{\scriptsize[+0.073, +0.114]}\end{tabular} & \begin{tabular}[t]{@{}c@{}}\textbf{+0.106} (19)\\[-1pt]{\scriptsize[+0.072, +0.146]}\end{tabular} & \begin{tabular}[t]{@{}c@{}}\textbf{+0.196} (20)\\[-1pt]{\scriptsize[+0.153, +0.241]}\end{tabular}\\
 &  & 10\% &  & \begin{tabular}[t]{@{}c@{}}\textbf{+0.099} (20)\\[-1pt]{\scriptsize[+0.084, +0.115]}\end{tabular} & \begin{tabular}[t]{@{}c@{}}\textbf{+0.123} (20)\\[-1pt]{\scriptsize[+0.107, +0.139]}\end{tabular} & \begin{tabular}[t]{@{}c@{}}\textbf{+0.138} (20)\\[-1pt]{\scriptsize[+0.111, +0.168]}\end{tabular} & \begin{tabular}[t]{@{}c@{}}\textbf{+0.217} (20)\\[-1pt]{\scriptsize[+0.182, +0.255]}\end{tabular}\\
 &  & 20\% &  & \begin{tabular}[t]{@{}c@{}}\textbf{+0.101} (20)\\[-1pt]{\scriptsize[+0.085, +0.117]}\end{tabular} & \begin{tabular}[t]{@{}c@{}}\textbf{+0.125} (20)\\[-1pt]{\scriptsize[+0.110, +0.140]}\end{tabular} & \begin{tabular}[t]{@{}c@{}}\textbf{+0.142} (20)\\[-1pt]{\scriptsize[+0.114, +0.170]}\end{tabular} & \begin{tabular}[t]{@{}c@{}}\textbf{+0.185} (20)\\[-1pt]{\scriptsize[+0.150, +0.223]}\end{tabular}\\
 &  & 40\% &  & \begin{tabular}[t]{@{}c@{}}\textbf{+0.093} (20)\\[-1pt]{\scriptsize[+0.078, +0.110]}\end{tabular} & \begin{tabular}[t]{@{}c@{}}\textbf{+0.122} (20)\\[-1pt]{\scriptsize[+0.109, +0.134]}\end{tabular} & \begin{tabular}[t]{@{}c@{}}\textbf{+0.122} (20)\\[-1pt]{\scriptsize[+0.095, +0.149]}\end{tabular} & \begin{tabular}[t]{@{}c@{}}\textbf{+0.142} (20)\\[-1pt]{\scriptsize[+0.117, +0.168]}\end{tabular}\\
\addlinespace[2pt]
EMA & Two-purpose $(1,3)$ & 5\% & 0.895 & \begin{tabular}[t]{@{}c@{}}\textbf{+0.044} (20)\\[-1pt]{\scriptsize[+0.030, +0.058]}\end{tabular} & \begin{tabular}[t]{@{}c@{}}\textbf{+0.057} (20)\\[-1pt]{\scriptsize[+0.042, +0.074]}\end{tabular} & \begin{tabular}[t]{@{}c@{}}\textbf{+0.089} (19)\\[-1pt]{\scriptsize[+0.054, +0.124]}\end{tabular} & \begin{tabular}[t]{@{}c@{}}\textbf{+0.150} (19)\\[-1pt]{\scriptsize[+0.106, +0.193]}\end{tabular}\\
 &  & 10\% &  & \begin{tabular}[t]{@{}c@{}}\textbf{+0.064} (20)\\[-1pt]{\scriptsize[+0.051, +0.077]}\end{tabular} & \begin{tabular}[t]{@{}c@{}}\textbf{+0.075} (19)\\[-1pt]{\scriptsize[+0.061, +0.090]}\end{tabular} & \begin{tabular}[t]{@{}c@{}}\textbf{+0.116} (20)\\[-1pt]{\scriptsize[+0.097, +0.135]}\end{tabular} & \begin{tabular}[t]{@{}c@{}}\textbf{+0.167} (19)\\[-1pt]{\scriptsize[+0.131, +0.201]}\end{tabular}\\
 &  & 20\% &  & \begin{tabular}[t]{@{}c@{}}\textbf{+0.060} (20)\\[-1pt]{\scriptsize[+0.046, +0.075]}\end{tabular} & \begin{tabular}[t]{@{}c@{}}\textbf{+0.072} (20)\\[-1pt]{\scriptsize[+0.059, +0.085]}\end{tabular} & \begin{tabular}[t]{@{}c@{}}\textbf{+0.112} (19)\\[-1pt]{\scriptsize[+0.086, +0.138]}\end{tabular} & \begin{tabular}[t]{@{}c@{}}\textbf{+0.142} (20)\\[-1pt]{\scriptsize[+0.114, +0.172]}\end{tabular}\\
 &  & 40\% &  & \begin{tabular}[t]{@{}c@{}}\textbf{+0.057} (20)\\[-1pt]{\scriptsize[+0.044, +0.070]}\end{tabular} & \begin{tabular}[t]{@{}c@{}}\textbf{+0.068} (20)\\[-1pt]{\scriptsize[+0.058, +0.079]}\end{tabular} & \begin{tabular}[t]{@{}c@{}}\textbf{+0.091} (19)\\[-1pt]{\scriptsize[+0.067, +0.114]}\end{tabular} & \begin{tabular}[t]{@{}c@{}}\textbf{+0.091} (20)\\[-1pt]{\scriptsize[+0.069, +0.113]}\end{tabular}\\
\addlinespace[2pt]
EMA & Mixture $\alpha=0.5$ & 10\% & 0.947 & \begin{tabular}[t]{@{}c@{}}\textbf{+0.084} (20)\\[-1pt]{\scriptsize[+0.069, +0.100]}\end{tabular} & \begin{tabular}[t]{@{}c@{}}\textbf{+0.103} (20)\\[-1pt]{\scriptsize[+0.087, +0.120]}\end{tabular} & \begin{tabular}[t]{@{}c@{}}\textbf{+0.135} (20)\\[-1pt]{\scriptsize[+0.110, +0.163]}\end{tabular} & \begin{tabular}[t]{@{}c@{}}\textbf{+0.201} (20)\\[-1pt]{\scriptsize[+0.164, +0.239]}\end{tabular}\\
 &  & 20\% &  & \begin{tabular}[t]{@{}c@{}}\textbf{+0.081} (20)\\[-1pt]{\scriptsize[+0.065, +0.097]}\end{tabular} & \begin{tabular}[t]{@{}c@{}}\textbf{+0.099} (20)\\[-1pt]{\scriptsize[+0.084, +0.113]}\end{tabular} & \begin{tabular}[t]{@{}c@{}}\textbf{+0.129} (20)\\[-1pt]{\scriptsize[+0.102, +0.156]}\end{tabular} & \begin{tabular}[t]{@{}c@{}}\textbf{+0.166} (20)\\[-1pt]{\scriptsize[+0.134, +0.200]}\end{tabular}\\
 &  & 40\% &  & \begin{tabular}[t]{@{}c@{}}\textbf{+0.075} (20)\\[-1pt]{\scriptsize[+0.061, +0.091]}\end{tabular} & \begin{tabular}[t]{@{}c@{}}\textbf{+0.097} (20)\\[-1pt]{\scriptsize[+0.086, +0.109]}\end{tabular} & \begin{tabular}[t]{@{}c@{}}\textbf{+0.110} (20)\\[-1pt]{\scriptsize[+0.084, +0.136]}\end{tabular} & \begin{tabular}[t]{@{}c@{}}\textbf{+0.119} (20)\\[-1pt]{\scriptsize[+0.096, +0.144]}\end{tabular}\\
\addlinespace[2pt]
EMA & Two-purpose $(0.5,5)$ & 10\% & 0.842 & \begin{tabular}[t]{@{}c@{}}\textbf{+0.045} (19)\\[-1pt]{\scriptsize[+0.034, +0.057]}\end{tabular} & \begin{tabular}[t]{@{}c@{}}\textbf{+0.046} (19)\\[-1pt]{\scriptsize[+0.034, +0.058]}\end{tabular} & \begin{tabular}[t]{@{}c@{}}\textbf{+0.116} (20)\\[-1pt]{\scriptsize[+0.096, +0.138]}\end{tabular} & \begin{tabular}[t]{@{}c@{}}\textbf{+0.157} (19)\\[-1pt]{\scriptsize[+0.122, +0.192]}\end{tabular}\\
 &  & 20\% &  & \begin{tabular}[t]{@{}c@{}}\textbf{+0.039} (18)\\[-1pt]{\scriptsize[+0.027, +0.051]}\end{tabular} & \begin{tabular}[t]{@{}c@{}}\textbf{+0.040} (19)\\[-1pt]{\scriptsize[+0.029, +0.051]}\end{tabular} & \begin{tabular}[t]{@{}c@{}}\textbf{+0.113} (20)\\[-1pt]{\scriptsize[+0.086, +0.138]}\end{tabular} & \begin{tabular}[t]{@{}c@{}}\textbf{+0.133} (19)\\[-1pt]{\scriptsize[+0.101, +0.165]}\end{tabular}\\
 &  & 40\% &  & \begin{tabular}[t]{@{}c@{}}\textbf{+0.038} (19)\\[-1pt]{\scriptsize[+0.028, +0.048]}\end{tabular} & \begin{tabular}[t]{@{}c@{}}\textbf{+0.036} (19)\\[-1pt]{\scriptsize[+0.028, +0.046]}\end{tabular} & \begin{tabular}[t]{@{}c@{}}\textbf{+0.090} (20)\\[-1pt]{\scriptsize[+0.066, +0.115]}\end{tabular} & \begin{tabular}[t]{@{}c@{}}\textbf{+0.082} (20)\\[-1pt]{\scriptsize[+0.058, +0.107]}\end{tabular}\\
\addlinespace[2pt]
EMA & Two-purpose $(0.25,8)$ & 10\% & 0.755 & \begin{tabular}[t]{@{}c@{}}\textbf{+0.035} (20)\\[-1pt]{\scriptsize[+0.026, +0.044]}\end{tabular} & \begin{tabular}[t]{@{}c@{}}\textbf{+0.030} (19)\\[-1pt]{\scriptsize[+0.020, +0.039]}\end{tabular} & \begin{tabular}[t]{@{}c@{}}\textbf{+0.125} (20)\\[-1pt]{\scriptsize[+0.105, +0.145]}\end{tabular} & \begin{tabular}[t]{@{}c@{}}\textbf{+0.155} (19)\\[-1pt]{\scriptsize[+0.119, +0.194]}\end{tabular}\\
 &  & 20\% &  & \begin{tabular}[t]{@{}c@{}}\textbf{+0.030} (19)\\[-1pt]{\scriptsize[+0.022, +0.039]}\end{tabular} & \begin{tabular}[t]{@{}c@{}}\textbf{+0.025} (17)\\[-1pt]{\scriptsize[+0.017, +0.034]}\end{tabular} & \begin{tabular}[t]{@{}c@{}}\textbf{+0.112} (20)\\[-1pt]{\scriptsize[+0.088, +0.136]}\end{tabular} & \begin{tabular}[t]{@{}c@{}}\textbf{+0.117} (19)\\[-1pt]{\scriptsize[+0.080, +0.153]}\end{tabular}\\
 &  & 40\% &  & \begin{tabular}[t]{@{}c@{}}\textbf{+0.027} (19)\\[-1pt]{\scriptsize[+0.019, +0.036]}\end{tabular} & \begin{tabular}[t]{@{}c@{}}\textbf{+0.018} (17)\\[-1pt]{\scriptsize[+0.010, +0.027]}\end{tabular} & \begin{tabular}[t]{@{}c@{}}\textbf{+0.095} (20)\\[-1pt]{\scriptsize[+0.072, +0.120]}\end{tabular} & \begin{tabular}[t]{@{}c@{}}\textbf{+0.076} (19)\\[-1pt]{\scriptsize[+0.050, +0.102]}\end{tabular}\\
\addlinespace[2pt]
EMA & $\zeta=0$, $\rho=0$ & 10\% & 0.856 & \begin{tabular}[t]{@{}c@{}}+0.019 (18)\\[-1pt]{\scriptsize[$-$0.007, +0.037]}\end{tabular} & \begin{tabular}[t]{@{}c@{}}+0.018 (16)\\[-1pt]{\scriptsize[$-$0.009, +0.038]}\end{tabular} & \begin{tabular}[t]{@{}c@{}}\textbf{+0.057} (19)\\[-1pt]{\scriptsize[+0.038, +0.078]}\end{tabular} & \begin{tabular}[t]{@{}c@{}}\textbf{+0.070} (19)\\[-1pt]{\scriptsize[+0.046, +0.096]}\end{tabular}\\
 &  & 20\% &  & \begin{tabular}[t]{@{}c@{}}\textbf{+0.039} (18)\\[-1pt]{\scriptsize[+0.024, +0.054]}\end{tabular} & \begin{tabular}[t]{@{}c@{}}\textbf{+0.032} (18)\\[-1pt]{\scriptsize[+0.019, +0.045]}\end{tabular} & \begin{tabular}[t]{@{}c@{}}\textbf{+0.069} (19)\\[-1pt]{\scriptsize[+0.047, +0.094]}\end{tabular} & \begin{tabular}[t]{@{}c@{}}\textbf{+0.039} (15)\\[-1pt]{\scriptsize[+0.012, +0.069]}\end{tabular}\\
 &  & 40\% &  & \begin{tabular}[t]{@{}c@{}}\textbf{+0.021} (14)\\[-1pt]{\scriptsize[+0.006, +0.035]}\end{tabular} & \begin{tabular}[t]{@{}c@{}}+0.012 (15)\\[-1pt]{\scriptsize[$-$0.004, +0.028]}\end{tabular} & \begin{tabular}[t]{@{}c@{}}\textbf{+0.065} (17)\\[-1pt]{\scriptsize[+0.042, +0.087]}\end{tabular} & \begin{tabular}[t]{@{}c@{}}\textbf{$-$0.021} (7)\\[-1pt]{\scriptsize[$-$0.041, $-$0.000]}\end{tabular}\\
\addlinespace[2pt]
EMA & $\zeta=0.5$, $\rho=0$ & 10\% & 0.791 & \begin{tabular}[t]{@{}c@{}}+0.002 (12)\\[-1pt]{\scriptsize[$-$0.022, +0.019]}\end{tabular} & \begin{tabular}[t]{@{}c@{}}+0.009 (13)\\[-1pt]{\scriptsize[$-$0.016, +0.028]}\end{tabular} & \begin{tabular}[t]{@{}c@{}}\textbf{+0.037} (14)\\[-1pt]{\scriptsize[+0.012, +0.062]}\end{tabular} & \begin{tabular}[t]{@{}c@{}}\textbf{+0.064} (14)\\[-1pt]{\scriptsize[+0.026, +0.105]}\end{tabular}\\
 &  & 20\% &  & \begin{tabular}[t]{@{}c@{}}\textbf{+0.013} (14)\\[-1pt]{\scriptsize[+0.003, +0.024]}\end{tabular} & \begin{tabular}[t]{@{}c@{}}\textbf{+0.014} (13)\\[-1pt]{\scriptsize[+0.002, +0.026]}\end{tabular} & \begin{tabular}[t]{@{}c@{}}\textbf{+0.042} (15)\\[-1pt]{\scriptsize[+0.015, +0.071]}\end{tabular} & \begin{tabular}[t]{@{}c@{}}+0.010 (9)\\[-1pt]{\scriptsize[$-$0.021, +0.044]}\end{tabular}\\
 &  & 40\% &  & \begin{tabular}[t]{@{}c@{}}+0.002 (13)\\[-1pt]{\scriptsize[$-$0.006, +0.010]}\end{tabular} & \begin{tabular}[t]{@{}c@{}}+0.001 (9)\\[-1pt]{\scriptsize[$-$0.011, +0.013]}\end{tabular} & \begin{tabular}[t]{@{}c@{}}\textbf{+0.045} (15)\\[-1pt]{\scriptsize[+0.015, +0.078]}\end{tabular} & \begin{tabular}[t]{@{}c@{}}\textbf{$-$0.045} (5)\\[-1pt]{\scriptsize[$-$0.069, $-$0.021]}\end{tabular}\\
\addlinespace[2pt]
EMA & $\zeta=1$, $\rho=0$ & 10\% & 0.681 & \begin{tabular}[t]{@{}c@{}}+0.001 (12)\\[-1pt]{\scriptsize[$-$0.014, +0.014]}\end{tabular} & \begin{tabular}[t]{@{}c@{}}$-$0.002 (10)\\[-1pt]{\scriptsize[$-$0.018, +0.012]}\end{tabular} & \begin{tabular}[t]{@{}c@{}}\textbf{+0.065} (14)\\[-1pt]{\scriptsize[+0.034, +0.097]}\end{tabular} & \begin{tabular}[t]{@{}c@{}}\textbf{+0.097} (14)\\[-1pt]{\scriptsize[+0.051, +0.147]}\end{tabular}\\
 &  & 20\% &  & \begin{tabular}[t]{@{}c@{}}+0.000 (11)\\[-1pt]{\scriptsize[$-$0.007, +0.007]}\end{tabular} & \begin{tabular}[t]{@{}c@{}}$-$0.004 (7)\\[-1pt]{\scriptsize[$-$0.013, +0.006]}\end{tabular} & \begin{tabular}[t]{@{}c@{}}\textbf{+0.045} (14)\\[-1pt]{\scriptsize[+0.015, +0.078]}\end{tabular} & \begin{tabular}[t]{@{}c@{}}+0.013 (11)\\[-1pt]{\scriptsize[$-$0.022, +0.046]}\end{tabular}\\
 &  & 40\% &  & \begin{tabular}[t]{@{}c@{}}\textbf{$-$0.008} (8)\\[-1pt]{\scriptsize[$-$0.016, $-$0.001]}\end{tabular} & \begin{tabular}[t]{@{}c@{}}\textbf{$-$0.016} (6)\\[-1pt]{\scriptsize[$-$0.026, $-$0.006]}\end{tabular} & \begin{tabular}[t]{@{}c@{}}\textbf{+0.055} (14)\\[-1pt]{\scriptsize[+0.019, +0.092]}\end{tabular} & \begin{tabular}[t]{@{}c@{}}\textbf{$-$0.035} (6)\\[-1pt]{\scriptsize[$-$0.064, $-$0.009]}\end{tabular}\\
\addlinespace[2pt]
EMA & $\zeta=0$, $\rho=0.3$ & 10\% & 0.832 & \begin{tabular}[t]{@{}c@{}}\textbf{+0.027} (19)\\[-1pt]{\scriptsize[+0.014, +0.038]}\end{tabular} & \begin{tabular}[t]{@{}c@{}}\textbf{+0.020} (16)\\[-1pt]{\scriptsize[+0.000, +0.038]}\end{tabular} & \begin{tabular}[t]{@{}c@{}}\textbf{+0.073} (17)\\[-1pt]{\scriptsize[+0.049, +0.098]}\end{tabular} & \begin{tabular}[t]{@{}c@{}}\textbf{+0.053} (15)\\[-1pt]{\scriptsize[+0.021, +0.091]}\end{tabular}\\
 &  & 20\% &  & \begin{tabular}[t]{@{}c@{}}\textbf{+0.029} (17)\\[-1pt]{\scriptsize[+0.007, +0.051]}\end{tabular} & \begin{tabular}[t]{@{}c@{}}+0.011 (12)\\[-1pt]{\scriptsize[$-$0.008, +0.030]}\end{tabular} & \begin{tabular}[t]{@{}c@{}}\textbf{+0.068} (17)\\[-1pt]{\scriptsize[+0.035, +0.104]}\end{tabular} & \begin{tabular}[t]{@{}c@{}}$-$0.001 (9)\\[-1pt]{\scriptsize[$-$0.036, +0.037]}\end{tabular}\\
 &  & 40\% &  & \begin{tabular}[t]{@{}c@{}}\textbf{+0.024} (16)\\[-1pt]{\scriptsize[+0.009, +0.039]}\end{tabular} & \begin{tabular}[t]{@{}c@{}}+0.006 (10)\\[-1pt]{\scriptsize[$-$0.014, +0.024]}\end{tabular} & \begin{tabular}[t]{@{}c@{}}\textbf{+0.066} (16)\\[-1pt]{\scriptsize[+0.035, +0.098]}\end{tabular} & \begin{tabular}[t]{@{}c@{}}\textbf{$-$0.060} (5)\\[-1pt]{\scriptsize[$-$0.094, $-$0.022]}\end{tabular}\\
\addlinespace[2pt]
EMA & $\zeta=0$, $\rho=0.7$ & 10\% & 0.765 & \begin{tabular}[t]{@{}c@{}}+0.016 (16)\\[-1pt]{\scriptsize[$-$0.008, +0.036]}\end{tabular} & \begin{tabular}[t]{@{}c@{}}$-$0.018 (9)\\[-1pt]{\scriptsize[$-$0.046, +0.004]}\end{tabular} & \begin{tabular}[t]{@{}c@{}}\textbf{+0.073} (18)\\[-1pt]{\scriptsize[+0.044, +0.106]}\end{tabular} & \begin{tabular}[t]{@{}c@{}}$-$0.012 (7)\\[-1pt]{\scriptsize[$-$0.054, +0.033]}\end{tabular}\\
 &  & 20\% &  & \begin{tabular}[t]{@{}c@{}}\textbf{+0.019} (17)\\[-1pt]{\scriptsize[+0.003, +0.035]}\end{tabular} & \begin{tabular}[t]{@{}c@{}}$-$0.019 (6)\\[-1pt]{\scriptsize[$-$0.042, +0.003]}\end{tabular} & \begin{tabular}[t]{@{}c@{}}\textbf{+0.075} (19)\\[-1pt]{\scriptsize[+0.031, +0.129]}\end{tabular} & \begin{tabular}[t]{@{}c@{}}\textbf{$-$0.091} (2)\\[-1pt]{\scriptsize[$-$0.133, $-$0.042]}\end{tabular}\\
 &  & 40\% &  & \begin{tabular}[t]{@{}c@{}}+0.013 (14)\\[-1pt]{\scriptsize[$-$0.004, +0.030]}\end{tabular} & \begin{tabular}[t]{@{}c@{}}\textbf{$-$0.035} (5)\\[-1pt]{\scriptsize[$-$0.054, $-$0.017]}\end{tabular} & \begin{tabular}[t]{@{}c@{}}\textbf{+0.039} (13)\\[-1pt]{\scriptsize[+0.003, +0.074]}\end{tabular} & \begin{tabular}[t]{@{}c@{}}\textbf{$-$0.183} (1)\\[-1pt]{\scriptsize[$-$0.226, $-$0.138]}\end{tabular}\\
\addlinespace[2pt]
EMA & $\zeta=1$, $\rho=0.7$ & 10\% & 0.619 & \begin{tabular}[t]{@{}c@{}}+0.001 (11)\\[-1pt]{\scriptsize[$-$0.013, +0.016]}\end{tabular} & \begin{tabular}[t]{@{}c@{}}\textbf{$-$0.019} (6)\\[-1pt]{\scriptsize[$-$0.035, $-$0.003]}\end{tabular} & \begin{tabular}[t]{@{}c@{}}\textbf{+0.089} (18)\\[-1pt]{\scriptsize[+0.045, +0.142]}\end{tabular} & \begin{tabular}[t]{@{}c@{}}+0.017 (9)\\[-1pt]{\scriptsize[$-$0.036, +0.082]}\end{tabular}\\
 &  & 20\% &  & \begin{tabular}[t]{@{}c@{}}$-$0.004 (11)\\[-1pt]{\scriptsize[$-$0.016, +0.008]}\end{tabular} & \begin{tabular}[t]{@{}c@{}}\textbf{$-$0.034} (3)\\[-1pt]{\scriptsize[$-$0.047, $-$0.021]}\end{tabular} & \begin{tabular}[t]{@{}c@{}}\textbf{+0.046} (12)\\[-1pt]{\scriptsize[+0.006, +0.086]}\end{tabular} & \begin{tabular}[t]{@{}c@{}}\textbf{$-$0.098} (3)\\[-1pt]{\scriptsize[$-$0.147, $-$0.050]}\end{tabular}\\
 &  & 40\% &  & \begin{tabular}[t]{@{}c@{}}\textbf{$-$0.014} (6)\\[-1pt]{\scriptsize[$-$0.029, $-$0.000]}\end{tabular} & \begin{tabular}[t]{@{}c@{}}\textbf{$-$0.049} (2)\\[-1pt]{\scriptsize[$-$0.067, $-$0.033]}\end{tabular} & \begin{tabular}[t]{@{}c@{}}\textbf{+0.048} (16)\\[-1pt]{\scriptsize[+0.021, +0.079]}\end{tabular} & \begin{tabular}[t]{@{}c@{}}\textbf{$-$0.177} (1)\\[-1pt]{\scriptsize[$-$0.218, $-$0.131]}\end{tabular}\\
\addlinespace[2pt]
EMA & Interaction $\delta=0.25$ & 10\% & 0.886 & \begin{tabular}[t]{@{}c@{}}\textbf{+0.061} (19)\\[-1pt]{\scriptsize[+0.047, +0.075]}\end{tabular} & \begin{tabular}[t]{@{}c@{}}\textbf{+0.073} (19)\\[-1pt]{\scriptsize[+0.056, +0.088]}\end{tabular} & \begin{tabular}[t]{@{}c@{}}\textbf{+0.112} (20)\\[-1pt]{\scriptsize[+0.091, +0.132]}\end{tabular} & \begin{tabular}[t]{@{}c@{}}\textbf{+0.166} (19)\\[-1pt]{\scriptsize[+0.128, +0.202]}\end{tabular}\\
 &  & 20\% &  & \begin{tabular}[t]{@{}c@{}}\textbf{+0.058} (20)\\[-1pt]{\scriptsize[+0.043, +0.073]}\end{tabular} & \begin{tabular}[t]{@{}c@{}}\textbf{+0.070} (20)\\[-1pt]{\scriptsize[+0.057, +0.083]}\end{tabular} & \begin{tabular}[t]{@{}c@{}}\textbf{+0.110} (19)\\[-1pt]{\scriptsize[+0.081, +0.139]}\end{tabular} & \begin{tabular}[t]{@{}c@{}}\textbf{+0.138} (19)\\[-1pt]{\scriptsize[+0.109, +0.168]}\end{tabular}\\
 &  & 40\% &  & \begin{tabular}[t]{@{}c@{}}\textbf{+0.052} (20)\\[-1pt]{\scriptsize[+0.039, +0.066]}\end{tabular} & \begin{tabular}[t]{@{}c@{}}\textbf{+0.069} (20)\\[-1pt]{\scriptsize[+0.059, +0.081]}\end{tabular} & \begin{tabular}[t]{@{}c@{}}\textbf{+0.086} (19)\\[-1pt]{\scriptsize[+0.060, +0.111]}\end{tabular} & \begin{tabular}[t]{@{}c@{}}\textbf{+0.088} (20)\\[-1pt]{\scriptsize[+0.064, +0.113]}\end{tabular}\\
\addlinespace[2pt]
EMA & Interaction $\delta=0.5$ & 10\% & 0.862 & \begin{tabular}[t]{@{}c@{}}\textbf{+0.054} (19)\\[-1pt]{\scriptsize[+0.039, +0.069]}\end{tabular} & \begin{tabular}[t]{@{}c@{}}\textbf{+0.065} (19)\\[-1pt]{\scriptsize[+0.048, +0.081]}\end{tabular} & \begin{tabular}[t]{@{}c@{}}\textbf{+0.106} (19)\\[-1pt]{\scriptsize[+0.080, +0.133]}\end{tabular} & \begin{tabular}[t]{@{}c@{}}\textbf{+0.156} (19)\\[-1pt]{\scriptsize[+0.120, +0.190]}\end{tabular}\\
 &  & 20\% &  & \begin{tabular}[t]{@{}c@{}}\textbf{+0.049} (20)\\[-1pt]{\scriptsize[+0.035, +0.064]}\end{tabular} & \begin{tabular}[t]{@{}c@{}}\textbf{+0.064} (20)\\[-1pt]{\scriptsize[+0.052, +0.078]}\end{tabular} & \begin{tabular}[t]{@{}c@{}}\textbf{+0.095} (18)\\[-1pt]{\scriptsize[+0.065, +0.129]}\end{tabular} & \begin{tabular}[t]{@{}c@{}}\textbf{+0.126} (19)\\[-1pt]{\scriptsize[+0.095, +0.156]}\end{tabular}\\
 &  & 40\% &  & \begin{tabular}[t]{@{}c@{}}\textbf{+0.038} (20)\\[-1pt]{\scriptsize[+0.028, +0.051]}\end{tabular} & \begin{tabular}[t]{@{}c@{}}\textbf{+0.063} (20)\\[-1pt]{\scriptsize[+0.051, +0.076]}\end{tabular} & \begin{tabular}[t]{@{}c@{}}\textbf{+0.070} (19)\\[-1pt]{\scriptsize[+0.046, +0.097]}\end{tabular} & \begin{tabular}[t]{@{}c@{}}\textbf{+0.081} (20)\\[-1pt]{\scriptsize[+0.056, +0.108]}\end{tabular}\\
\addlinespace[2pt]
EMA & Interaction $\delta=1$ & 10\% & 0.792 & \begin{tabular}[t]{@{}c@{}}\textbf{+0.042} (17)\\[-1pt]{\scriptsize[+0.025, +0.060]}\end{tabular} & \begin{tabular}[t]{@{}c@{}}\textbf{+0.056} (19)\\[-1pt]{\scriptsize[+0.038, +0.076]}\end{tabular} & \begin{tabular}[t]{@{}c@{}}\textbf{+0.084} (16)\\[-1pt]{\scriptsize[+0.045, +0.125]}\end{tabular} & \begin{tabular}[t]{@{}c@{}}\textbf{+0.130} (19)\\[-1pt]{\scriptsize[+0.090, +0.175]}\end{tabular}\\
 &  & 20\% &  & \begin{tabular}[t]{@{}c@{}}\textbf{+0.035} (18)\\[-1pt]{\scriptsize[+0.020, +0.052]}\end{tabular} & \begin{tabular}[t]{@{}c@{}}\textbf{+0.060} (20)\\[-1pt]{\scriptsize[+0.045, +0.077]}\end{tabular} & \begin{tabular}[t]{@{}c@{}}\textbf{+0.070} (19)\\[-1pt]{\scriptsize[+0.034, +0.112]}\end{tabular} & \begin{tabular}[t]{@{}c@{}}\textbf{+0.116} (18)\\[-1pt]{\scriptsize[+0.075, +0.161]}\end{tabular}\\
 &  & 40\% &  & \begin{tabular}[t]{@{}c@{}}\textbf{+0.015} (14)\\[-1pt]{\scriptsize[+0.003, +0.029]}\end{tabular} & \begin{tabular}[t]{@{}c@{}}\textbf{+0.052} (20)\\[-1pt]{\scriptsize[+0.039, +0.068]}\end{tabular} & \begin{tabular}[t]{@{}c@{}}\textbf{+0.035} (13)\\[-1pt]{\scriptsize[+0.008, +0.069]}\end{tabular} & \begin{tabular}[t]{@{}c@{}}\textbf{+0.061} (16)\\[-1pt]{\scriptsize[+0.030, +0.096]}\end{tabular}\\
\addlinespace[2pt]
EMA & Spatial $\phi=0.5$ & 10\% & 0.833 & \begin{tabular}[t]{@{}c@{}}\textbf{+0.013} (14)\\[-1pt]{\scriptsize[+0.000, +0.025]}\end{tabular} & \begin{tabular}[t]{@{}c@{}}\textbf{+0.025} (18)\\[-1pt]{\scriptsize[+0.012, +0.038]}\end{tabular} & \begin{tabular}[t]{@{}c@{}}+0.027 (12)\\[-1pt]{\scriptsize[$-$0.002, +0.055]}\end{tabular} & \begin{tabular}[t]{@{}c@{}}\textbf{+0.081} (16)\\[-1pt]{\scriptsize[+0.047, +0.116]}\end{tabular}\\
 &  & 20\% &  & \begin{tabular}[t]{@{}c@{}}+0.009 (13)\\[-1pt]{\scriptsize[$-$0.008, +0.025]}\end{tabular} & \begin{tabular}[t]{@{}c@{}}\textbf{+0.029} (15)\\[-1pt]{\scriptsize[+0.015, +0.042]}\end{tabular} & \begin{tabular}[t]{@{}c@{}}+0.017 (13)\\[-1pt]{\scriptsize[$-$0.030, +0.060]}\end{tabular} & \begin{tabular}[t]{@{}c@{}}\textbf{+0.065} (16)\\[-1pt]{\scriptsize[+0.025, +0.104]}\end{tabular}\\
 &  & 40\% &  & \begin{tabular}[t]{@{}c@{}}$-$0.002 (11)\\[-1pt]{\scriptsize[$-$0.019, +0.014]}\end{tabular} & \begin{tabular}[t]{@{}c@{}}\textbf{+0.025} (17)\\[-1pt]{\scriptsize[+0.013, +0.037]}\end{tabular} & \begin{tabular}[t]{@{}c@{}}+0.004 (11)\\[-1pt]{\scriptsize[$-$0.041, +0.046]}\end{tabular} & \begin{tabular}[t]{@{}c@{}}+0.009 (11)\\[-1pt]{\scriptsize[$-$0.028, +0.047]}\end{tabular}\\
\addlinespace[2pt]
EMA & Spatial $\phi=1$ & 10\% & 0.715 & \begin{tabular}[t]{@{}c@{}}\textbf{$-$0.043} (6)\\[-1pt]{\scriptsize[$-$0.082, $-$0.014]}\end{tabular} & \begin{tabular}[t]{@{}c@{}}$-$0.015 (9)\\[-1pt]{\scriptsize[$-$0.043, +0.009]}\end{tabular} & \begin{tabular}[t]{@{}c@{}}$-$0.035 (10)\\[-1pt]{\scriptsize[$-$0.127, +0.052]}\end{tabular} & \begin{tabular}[t]{@{}c@{}}\textbf{+0.066} (16)\\[-1pt]{\scriptsize[+0.010, +0.124]}\end{tabular}\\
 &  & 20\% &  & \begin{tabular}[t]{@{}c@{}}\textbf{$-$0.054} (6)\\[-1pt]{\scriptsize[$-$0.107, $-$0.008]}\end{tabular} & \begin{tabular}[t]{@{}c@{}}$-$0.014 (11)\\[-1pt]{\scriptsize[$-$0.052, +0.020]}\end{tabular} & \begin{tabular}[t]{@{}c@{}}$-$0.075 (8)\\[-1pt]{\scriptsize[$-$0.193, +0.025]}\end{tabular} & \begin{tabular}[t]{@{}c@{}}$-$0.012 (11)\\[-1pt]{\scriptsize[$-$0.102, +0.065]}\end{tabular}\\
 &  & 40\% &  & \begin{tabular}[t]{@{}c@{}}\textbf{$-$0.081} (2)\\[-1pt]{\scriptsize[$-$0.122, $-$0.045]}\end{tabular} & \begin{tabular}[t]{@{}c@{}}$-$0.006 (7)\\[-1pt]{\scriptsize[$-$0.031, +0.025]}\end{tabular} & \begin{tabular}[t]{@{}c@{}}\textbf{$-$0.117} (4)\\[-1pt]{\scriptsize[$-$0.222, $-$0.027]}\end{tabular} & \begin{tabular}[t]{@{}c@{}}\textbf{$-$0.089} (6)\\[-1pt]{\scriptsize[$-$0.180, $-$0.013]}\end{tabular}\\
\addlinespace[2pt]
EMA & Interaction $\delta=0.25$\textsuperscript{a} & 10\% & 0.889 & \begin{tabular}[t]{@{}c@{}}\textbf{+0.060} (10)\\[-1pt]{\scriptsize[+0.043, +0.081]}\end{tabular} & \begin{tabular}[t]{@{}c@{}}\textbf{+0.072} (10)\\[-1pt]{\scriptsize[+0.054, +0.094]}\end{tabular} & \begin{tabular}[t]{@{}c@{}}\textbf{+0.115} (10)\\[-1pt]{\scriptsize[+0.086, +0.142]}\end{tabular} & \begin{tabular}[t]{@{}c@{}}\textbf{+0.164} (9)\\[-1pt]{\scriptsize[+0.105, +0.220]}\end{tabular}\\
\addlinespace[2pt]
EMA & Matched, biased prior & 10\% & 1.000 & \begin{tabular}[t]{@{}c@{}}\textbf{+0.154} (20)\\[-1pt]{\scriptsize[+0.134, +0.173]}\end{tabular} & \begin{tabular}[t]{@{}c@{}}\textbf{+0.185} (20)\\[-1pt]{\scriptsize[+0.166, +0.202]}\end{tabular} & \begin{tabular}[t]{@{}c@{}}\textbf{+0.241} (20)\\[-1pt]{\scriptsize[+0.200, +0.282]}\end{tabular} & \begin{tabular}[t]{@{}c@{}}\textbf{+0.322} (20)\\[-1pt]{\scriptsize[+0.276, +0.371]}\end{tabular}\\
\addlinespace[2pt]
EMA & Two-purpose, biased prior & 10\% & 0.895 & \begin{tabular}[t]{@{}c@{}}\textbf{+0.111} (20)\\[-1pt]{\scriptsize[+0.091, +0.130]}\end{tabular} & \begin{tabular}[t]{@{}c@{}}\textbf{+0.133} (20)\\[-1pt]{\scriptsize[+0.111, +0.152]}\end{tabular} & \begin{tabular}[t]{@{}c@{}}\textbf{+0.223} (20)\\[-1pt]{\scriptsize[+0.185, +0.261]}\end{tabular} & \begin{tabular}[t]{@{}c@{}}\textbf{+0.276} (20)\\[-1pt]{\scriptsize[+0.225, +0.323]}\end{tabular}\\
\addlinespace[2pt]
EMA & Matched, capacity error & 10\% & 1.000 & \begin{tabular}[t]{@{}c@{}}\textbf{+0.076} (19)\\[-1pt]{\scriptsize[+0.055, +0.097]}\end{tabular} & \begin{tabular}[t]{@{}c@{}}\textbf{+0.094} (20)\\[-1pt]{\scriptsize[+0.073, +0.117]}\end{tabular} & \begin{tabular}[t]{@{}c@{}}\textbf{+0.118} (19)\\[-1pt]{\scriptsize[+0.080, +0.159]}\end{tabular} & \begin{tabular}[t]{@{}c@{}}\textbf{+0.188} (20)\\[-1pt]{\scriptsize[+0.143, +0.235]}\end{tabular}\\
\addlinespace[2pt]
EMA & Two-purpose, capacity error & 10\% & 0.895 & \begin{tabular}[t]{@{}c@{}}\textbf{+0.048} (18)\\[-1pt]{\scriptsize[+0.024, +0.068]}\end{tabular} & \begin{tabular}[t]{@{}c@{}}\textbf{+0.060} (18)\\[-1pt]{\scriptsize[+0.034, +0.082]}\end{tabular} & \begin{tabular}[t]{@{}c@{}}\textbf{+0.095} (19)\\[-1pt]{\scriptsize[+0.068, +0.124]}\end{tabular} & \begin{tabular}[t]{@{}c@{}}\textbf{+0.150} (20)\\[-1pt]{\scriptsize[+0.113, +0.188]}\end{tabular}\\
\addlinespace[2pt]
Chicago & Matched & 5\% & 1.000 & \begin{tabular}[t]{@{}c@{}}\textbf{+0.086} (20)\\[-1pt]{\scriptsize[+0.071, +0.101]}\end{tabular} & \begin{tabular}[t]{@{}c@{}}\textbf{+0.125} (20)\\[-1pt]{\scriptsize[+0.107, +0.144]}\end{tabular} & \begin{tabular}[t]{@{}c@{}}\textbf{+0.104} (20)\\[-1pt]{\scriptsize[+0.083, +0.125]}\end{tabular} & \begin{tabular}[t]{@{}c@{}}\textbf{+0.165} (20)\\[-1pt]{\scriptsize[+0.130, +0.200]}\end{tabular}\\
 &  & 10\% &  & \begin{tabular}[t]{@{}c@{}}\textbf{+0.093} (20)\\[-1pt]{\scriptsize[+0.078, +0.108]}\end{tabular} & \begin{tabular}[t]{@{}c@{}}\textbf{+0.141} (20)\\[-1pt]{\scriptsize[+0.123, +0.160]}\end{tabular} & \begin{tabular}[t]{@{}c@{}}\textbf{+0.111} (20)\\[-1pt]{\scriptsize[+0.090, +0.132]}\end{tabular} & \begin{tabular}[t]{@{}c@{}}\textbf{+0.135} (20)\\[-1pt]{\scriptsize[+0.113, +0.160]}\end{tabular}\\
 &  & 20\% &  & \begin{tabular}[t]{@{}c@{}}\textbf{+0.095} (20)\\[-1pt]{\scriptsize[+0.081, +0.110]}\end{tabular} & \begin{tabular}[t]{@{}c@{}}\textbf{+0.141} (20)\\[-1pt]{\scriptsize[+0.123, +0.159]}\end{tabular} & \begin{tabular}[t]{@{}c@{}}\textbf{+0.112} (20)\\[-1pt]{\scriptsize[+0.095, +0.130]}\end{tabular} & \begin{tabular}[t]{@{}c@{}}\textbf{+0.104} (20)\\[-1pt]{\scriptsize[+0.093, +0.117]}\end{tabular}\\
\addlinespace[2pt]
Chicago & Two-purpose $(1,3)$ & 5\% & 0.892 & \begin{tabular}[t]{@{}c@{}}\textbf{+0.047} (20)\\[-1pt]{\scriptsize[+0.035, +0.059]}\end{tabular} & \begin{tabular}[t]{@{}c@{}}\textbf{+0.083} (20)\\[-1pt]{\scriptsize[+0.067, +0.099]}\end{tabular} & \begin{tabular}[t]{@{}c@{}}\textbf{+0.094} (20)\\[-1pt]{\scriptsize[+0.073, +0.117]}\end{tabular} & \begin{tabular}[t]{@{}c@{}}\textbf{+0.148} (20)\\[-1pt]{\scriptsize[+0.117, +0.181]}\end{tabular}\\
 &  & 10\% &  & \begin{tabular}[t]{@{}c@{}}\textbf{+0.048} (20)\\[-1pt]{\scriptsize[+0.037, +0.059]}\end{tabular} & \begin{tabular}[t]{@{}c@{}}\textbf{+0.090} (20)\\[-1pt]{\scriptsize[+0.074, +0.106]}\end{tabular} & \begin{tabular}[t]{@{}c@{}}\textbf{+0.093} (20)\\[-1pt]{\scriptsize[+0.074, +0.113]}\end{tabular} & \begin{tabular}[t]{@{}c@{}}\textbf{+0.108} (20)\\[-1pt]{\scriptsize[+0.088, +0.130]}\end{tabular}\\
 &  & 20\% &  & \begin{tabular}[t]{@{}c@{}}\textbf{+0.047} (20)\\[-1pt]{\scriptsize[+0.037, +0.057]}\end{tabular} & \begin{tabular}[t]{@{}c@{}}\textbf{+0.088} (20)\\[-1pt]{\scriptsize[+0.073, +0.104]}\end{tabular} & \begin{tabular}[t]{@{}c@{}}\textbf{+0.090} (20)\\[-1pt]{\scriptsize[+0.073, +0.107]}\end{tabular} & \begin{tabular}[t]{@{}c@{}}\textbf{+0.072} (20)\\[-1pt]{\scriptsize[+0.061, +0.084]}\end{tabular}\\
\addlinespace[2pt]
Chicago & $\zeta=1$, $\rho=0.7$ & 10\% & 0.626 & \begin{tabular}[t]{@{}c@{}}\textbf{$-$0.041} (3)\\[-1pt]{\scriptsize[$-$0.063, $-$0.021]}\end{tabular} & \begin{tabular}[t]{@{}c@{}}\textbf{$-$0.052} (4)\\[-1pt]{\scriptsize[$-$0.075, $-$0.028]}\end{tabular} & \begin{tabular}[t]{@{}c@{}}\textbf{+0.069} (20)\\[-1pt]{\scriptsize[+0.048, +0.093]}\end{tabular} & \begin{tabular}[t]{@{}c@{}}\textbf{$-$0.150} (1)\\[-1pt]{\scriptsize[$-$0.178, $-$0.119]}\end{tabular}\\
 &  & 20\% &  & \begin{tabular}[t]{@{}c@{}}\textbf{$-$0.041} (4)\\[-1pt]{\scriptsize[$-$0.065, $-$0.018]}\end{tabular} & \begin{tabular}[t]{@{}c@{}}\textbf{$-$0.064} (3)\\[-1pt]{\scriptsize[$-$0.089, $-$0.040]}\end{tabular} & \begin{tabular}[t]{@{}c@{}}\textbf{+0.055} (19)\\[-1pt]{\scriptsize[+0.034, +0.078]}\end{tabular} & \begin{tabular}[t]{@{}c@{}}\textbf{$-$0.250} (0)\\[-1pt]{\scriptsize[$-$0.277, $-$0.222]}\end{tabular}\\
\addlinespace[2pt]
Chicago & Interaction $\delta=0.25$\textsuperscript{a} & 10\% & 0.885 & \begin{tabular}[t]{@{}c@{}}\textbf{+0.047} (10)\\[-1pt]{\scriptsize[+0.034, +0.060]}\end{tabular} & \begin{tabular}[t]{@{}c@{}}\textbf{+0.097} (10)\\[-1pt]{\scriptsize[+0.082, +0.114]}\end{tabular} & \begin{tabular}[t]{@{}c@{}}\textbf{+0.095} (10)\\[-1pt]{\scriptsize[+0.071, +0.124]}\end{tabular} & \begin{tabular}[t]{@{}c@{}}\textbf{+0.117} (10)\\[-1pt]{\scriptsize[+0.091, +0.146]}\end{tabular}\\
\addlinespace[2pt]
Chicago & Matched, capacity error & 10\% & 1.000 & \begin{tabular}[t]{@{}c@{}}\textbf{+0.081} (20)\\[-1pt]{\scriptsize[+0.070, +0.093]}\end{tabular} & \begin{tabular}[t]{@{}c@{}}\textbf{+0.133} (20)\\[-1pt]{\scriptsize[+0.112, +0.153]}\end{tabular} & \begin{tabular}[t]{@{}c@{}}\textbf{+0.093} (20)\\[-1pt]{\scriptsize[+0.075, +0.111]}\end{tabular} & \begin{tabular}[t]{@{}c@{}}\textbf{+0.124} (20)\\[-1pt]{\scriptsize[+0.102, +0.148]}\end{tabular}\\
\addlinespace[2pt]
Chicago & Two-purpose, capacity error & 10\% & 0.892 & \begin{tabular}[t]{@{}c@{}}\textbf{+0.047} (20)\\[-1pt]{\scriptsize[+0.038, +0.057]}\end{tabular} & \begin{tabular}[t]{@{}c@{}}\textbf{+0.096} (20)\\[-1pt]{\scriptsize[+0.077, +0.114]}\end{tabular} & \begin{tabular}[t]{@{}c@{}}\textbf{+0.085} (20)\\[-1pt]{\scriptsize[+0.066, +0.105]}\end{tabular} & \begin{tabular}[t]{@{}c@{}}\textbf{+0.110} (20)\\[-1pt]{\scriptsize[+0.089, +0.133]}\end{tabular}\\
\end{longtable}

%% file: tables/tabS_medians.tex
\begin{longtable}{llccccccccccccc}
\caption{Median over worlds of the world-mean metric for each estimator: prior generator (P), five-coefficient generator (G), OD-factor adjustment (F) and cellwise adjustment (C). \textsuperscript{a}Confirmation-programme diagnostic, worlds 2000–2009 only.}\label{stab:medians}\\
\toprule
 & & & \multicolumn{4}{c}{OD CPC} & \multicolumn{4}{c}{Held-out NRMSE} & \multicolumn{4}{c}{Calibration NRMSE}\\
\cmidrule(lr){4-7}\cmidrule(lr){8-11}\cmidrule(lr){12-15}
Network & Truth & Cov. & P & G & F & C & P & G & F & C & P & G & F & C\\
\midrule
\endfirsthead
\toprule
Network & Truth & Cov. & P & G & F & C & P & G & F & C & P & G & F & C\\
\midrule
\endhead
\bottomrule
\endfoot
EMA & Matched & 5\% & 0.829 & 0.929 & 0.856 & 0.832 & 0.363 & 0.121 & 0.224 & 0.307 & 0.371 & 0.068 & 0.111 & 0.008\\
 &  & 10\% & 0.829 & 0.963 & 0.864 & 0.837 & 0.363 & 0.060 & 0.201 & 0.281 & 0.372 & 0.057 & 0.060 & 0.010\\
 &  & 20\% & 0.829 & 0.975 & 0.875 & 0.847 & 0.363 & 0.040 & 0.175 & 0.219 & 0.382 & 0.065 & 0.088 & 0.022\\
 &  & 40\% & 0.829 & 0.979 & 0.892 & 0.851 & 0.363 & 0.033 & 0.148 & 0.168 & 0.383 & 0.074 & 0.101 & 0.029\\
\addlinespace[2pt]
EMA & Two-purpose $(1,3)$ & 5\% & 0.797 & 0.859 & 0.808 & 0.799 & 0.397 & 0.175 & 0.258 & 0.352 & 0.372 & 0.101 & 0.141 & 0.008\\
 &  & 10\% & 0.797 & 0.887 & 0.817 & 0.802 & 0.397 & 0.135 & 0.250 & 0.311 & 0.403 & 0.114 & 0.103 & 0.022\\
 &  & 20\% & 0.797 & 0.888 & 0.825 & 0.812 & 0.397 & 0.119 & 0.223 & 0.255 & 0.426 & 0.113 & 0.119 & 0.028\\
 &  & 40\% & 0.797 & 0.896 & 0.840 & 0.823 & 0.397 & 0.109 & 0.206 & 0.186 & 0.406 & 0.119 & 0.126 & 0.028\\
\addlinespace[2pt]
EMA & Mixture $\alpha=0.5$ & 10\% & 0.821 & 0.934 & 0.852 & 0.827 & 0.378 & 0.088 & 0.208 & 0.293 & 0.380 & 0.077 & 0.081 & 0.016\\
 &  & 20\% & 0.821 & 0.940 & 0.851 & 0.836 & 0.378 & 0.067 & 0.191 & 0.225 & 0.394 & 0.083 & 0.107 & 0.027\\
 &  & 40\% & 0.821 & 0.942 & 0.871 & 0.843 & 0.378 & 0.062 & 0.170 & 0.171 & 0.390 & 0.087 & 0.109 & 0.027\\
\addlinespace[2pt]
EMA & Two-purpose $(0.5,5)$ & 10\% & 0.771 & 0.826 & 0.786 & 0.776 & 0.402 & 0.148 & 0.256 & 0.318 & 0.424 & 0.125 & 0.097 & 0.012\\
 &  & 20\% & 0.771 & 0.827 & 0.791 & 0.783 & 0.402 & 0.137 & 0.247 & 0.252 & 0.429 & 0.138 & 0.146 & 0.026\\
 &  & 40\% & 0.771 & 0.834 & 0.798 & 0.794 & 0.402 & 0.119 & 0.207 & 0.190 & 0.432 & 0.137 & 0.142 & 0.025\\
\addlinespace[2pt]
EMA & Two-purpose $(0.25,8)$ & 10\% & 0.709 & 0.746 & 0.715 & 0.718 & 0.409 & 0.173 & 0.298 & 0.339 & 0.455 & 0.131 & 0.141 & 0.014\\
 &  & 20\% & 0.709 & 0.749 & 0.719 & 0.722 & 0.409 & 0.168 & 0.275 & 0.277 & 0.471 & 0.157 & 0.166 & 0.014\\
 &  & 40\% & 0.709 & 0.751 & 0.726 & 0.733 & 0.409 & 0.152 & 0.231 & 0.206 & 0.456 & 0.156 & 0.172 & 0.018\\
\addlinespace[2pt]
EMA & $\zeta=0$, $\rho=0$ & 10\% & 0.769 & 0.806 & 0.786 & 0.784 & 0.463 & 0.293 & 0.372 & 0.371 & 0.426 & 0.215 & 0.210 & 0.020\\
 &  & 20\% & 0.769 & 0.836 & 0.795 & 0.794 & 0.463 & 0.268 & 0.334 & 0.292 & 0.458 & 0.247 & 0.219 & 0.021\\
 &  & 40\% & 0.769 & 0.838 & 0.809 & 0.814 & 0.463 & 0.261 & 0.323 & 0.224 & 0.477 & 0.246 & 0.222 & 0.022\\
\addlinespace[2pt]
EMA & $\zeta=0.5$, $\rho=0$ & 10\% & 0.728 & 0.767 & 0.744 & 0.738 & 0.481 & 0.336 & 0.368 & 0.389 & 0.492 & 0.251 & 0.174 & 0.011\\
 &  & 20\% & 0.728 & 0.775 & 0.751 & 0.754 & 0.481 & 0.305 & 0.355 & 0.325 & 0.497 & 0.282 & 0.221 & 0.010\\
 &  & 40\% & 0.728 & 0.776 & 0.768 & 0.768 & 0.481 & 0.294 & 0.343 & 0.249 & 0.493 & 0.294 & 0.197 & 0.027\\
\addlinespace[2pt]
EMA & $\zeta=1$, $\rho=0$ & 10\% & 0.638 & 0.650 & 0.650 & 0.648 & 0.591 & 0.398 & 0.447 & 0.477 & 0.644 & 0.300 & 0.228 & 0.010\\
 &  & 20\% & 0.638 & 0.662 & 0.663 & 0.664 & 0.591 & 0.367 & 0.406 & 0.397 & 0.634 & 0.346 & 0.265 & 0.013\\
 &  & 40\% & 0.638 & 0.666 & 0.666 & 0.678 & 0.591 & 0.348 & 0.389 & 0.317 & 0.629 & 0.354 & 0.250 & 0.018\\
\addlinespace[2pt]
EMA & $\zeta=0$, $\rho=0.3$ & 10\% & 0.749 & 0.797 & 0.764 & 0.770 & 0.516 & 0.375 & 0.452 & 0.405 & 0.471 & 0.256 & 0.228 & 0.022\\
 &  & 20\% & 0.749 & 0.800 & 0.772 & 0.789 & 0.516 & 0.372 & 0.413 & 0.343 & 0.525 & 0.304 & 0.271 & 0.042\\
 &  & 40\% & 0.749 & 0.809 & 0.789 & 0.800 & 0.516 & 0.348 & 0.395 & 0.241 & 0.563 & 0.291 & 0.288 & 0.026\\
\addlinespace[2pt]
EMA & $\zeta=0$, $\rho=0.7$ & 10\% & 0.693 & 0.715 & 0.686 & 0.711 & 0.656 & 0.578 & 0.658 & 0.553 & 0.632 & 0.396 & 0.364 & 0.023\\
 &  & 20\% & 0.693 & 0.722 & 0.692 & 0.729 & 0.656 & 0.551 & 0.616 & 0.446 & 0.692 & 0.478 & 0.423 & 0.056\\
 &  & 40\% & 0.693 & 0.718 & 0.714 & 0.760 & 0.656 & 0.507 & 0.563 & 0.318 & 0.708 & 0.444 & 0.481 & 0.040\\
\addlinespace[2pt]
EMA & $\zeta=1$, $\rho=0.7$ & 10\% & 0.590 & 0.589 & 0.587 & 0.605 & 0.729 & 0.604 & 0.672 & 0.614 & 0.696 & 0.511 & 0.346 & 0.065\\
 &  & 20\% & 0.590 & 0.589 & 0.592 & 0.625 & 0.729 & 0.537 & 0.616 & 0.454 & 0.734 & 0.536 & 0.498 & 0.017\\
 &  & 40\% & 0.590 & 0.605 & 0.608 & 0.642 & 0.729 & 0.527 & 0.581 & 0.377 & 0.762 & 0.514 & 0.473 & 0.023\\
\addlinespace[2pt]
EMA & Interaction $\delta=0.25$ & 10\% & 0.793 & 0.875 & 0.813 & 0.802 & 0.393 & 0.133 & 0.236 & 0.320 & 0.393 & 0.121 & 0.118 & 0.009\\
 &  & 20\% & 0.793 & 0.879 & 0.828 & 0.808 & 0.393 & 0.126 & 0.236 & 0.258 & 0.431 & 0.122 & 0.115 & 0.036\\
 &  & 40\% & 0.793 & 0.885 & 0.839 & 0.818 & 0.393 & 0.113 & 0.204 & 0.182 & 0.427 & 0.123 & 0.128 & 0.026\\
\addlinespace[2pt]
EMA & Interaction $\delta=0.5$ & 10\% & 0.783 & 0.851 & 0.800 & 0.791 & 0.424 & 0.176 & 0.273 & 0.339 & 0.399 & 0.123 & 0.125 & 0.016\\
 &  & 20\% & 0.783 & 0.859 & 0.814 & 0.799 & 0.424 & 0.164 & 0.230 & 0.278 & 0.444 & 0.140 & 0.118 & 0.041\\
 &  & 40\% & 0.783 & 0.865 & 0.831 & 0.805 & 0.424 & 0.144 & 0.202 & 0.199 & 0.446 & 0.137 & 0.132 & 0.026\\
\addlinespace[2pt]
EMA & Interaction $\delta=1$ & 10\% & 0.737 & 0.796 & 0.754 & 0.749 & 0.490 & 0.229 & 0.343 & 0.400 & 0.481 & 0.151 & 0.133 & 0.045\\
 &  & 20\% & 0.737 & 0.812 & 0.778 & 0.750 & 0.490 & 0.207 & 0.275 & 0.324 & 0.467 & 0.172 & 0.130 & 0.048\\
 &  & 40\% & 0.737 & 0.820 & 0.804 & 0.768 & 0.490 & 0.205 & 0.220 & 0.255 & 0.481 & 0.176 & 0.138 & 0.042\\
\addlinespace[2pt]
EMA & Spatial $\phi=0.5$ & 10\% & 0.773 & 0.803 & 0.794 & 0.783 & 0.437 & 0.251 & 0.274 & 0.345 & 0.435 & 0.173 & 0.124 & 0.004\\
 &  & 20\% & 0.773 & 0.825 & 0.810 & 0.788 & 0.437 & 0.211 & 0.236 & 0.275 & 0.465 & 0.185 & 0.134 & 0.027\\
 &  & 40\% & 0.773 & 0.831 & 0.828 & 0.801 & 0.437 & 0.200 & 0.209 & 0.204 & 0.456 & 0.194 & 0.132 & 0.036\\
\addlinespace[2pt]
EMA & Spatial $\phi=1$ & 10\% & 0.673 & 0.678 & 0.714 & 0.704 & 0.563 & 0.451 & 0.359 & 0.498 & 0.587 & 0.279 & 0.179 & 0.033\\
 &  & 20\% & 0.673 & 0.696 & 0.747 & 0.713 & 0.563 & 0.357 & 0.307 & 0.392 & 0.589 & 0.322 & 0.172 & 0.043\\
 &  & 40\% & 0.673 & 0.731 & 0.776 & 0.740 & 0.563 & 0.363 & 0.253 & 0.295 & 0.620 & 0.352 & 0.167 & 0.063\\
\addlinespace[2pt]
EMA & Interaction $\delta=0.25$\textsuperscript{a} & 10\% & 0.793 & 0.873 & 0.817 & 0.802 & 0.424 & 0.138 & 0.251 & 0.320 & 0.393 & 0.119 & 0.091 & 0.004\\
\addlinespace[2pt]
EMA & Matched, biased prior & 10\% & 0.748 & 0.956 & 0.789 & 0.763 & 0.465 & 0.067 & 0.320 & 0.385 & 0.466 & 0.064 & 0.127 & 0.007\\
\addlinespace[2pt]
EMA & Two-purpose, biased prior & 10\% & 0.725 & 0.880 & 0.752 & 0.734 & 0.511 & 0.135 & 0.358 & 0.424 & 0.524 & 0.104 & 0.160 & 0.010\\
\addlinespace[2pt]
EMA & Matched, capacity error & 10\% & 0.829 & 0.931 & 0.856 & 0.836 & 0.363 & 0.091 & 0.213 & 0.286 & 0.372 & 0.083 & 0.101 & 0.080\\
\addlinespace[2pt]
EMA & Two-purpose, capacity error & 10\% & 0.797 & 0.881 & 0.818 & 0.800 & 0.397 & 0.156 & 0.249 & 0.310 & 0.403 & 0.136 & 0.131 & 0.093\\
\addlinespace[2pt]
Chicago & Matched & 5\% & 0.839 & 0.966 & 0.881 & 0.839 & 0.330 & 0.050 & 0.135 & 0.185 & 0.372 & 0.075 & 0.127 & 0.009\\
 &  & 10\% & 0.839 & 0.981 & 0.888 & 0.831 & 0.330 & 0.026 & 0.129 & 0.151 & 0.367 & 0.073 & 0.125 & 0.016\\
 &  & 20\% & 0.839 & 0.987 & 0.892 & 0.846 & 0.330 & 0.021 & 0.125 & 0.121 & 0.354 & 0.073 & 0.128 & 0.033\\
\addlinespace[2pt]
Chicago & Two-purpose $(1,3)$ & 5\% & 0.808 & 0.886 & 0.843 & 0.803 & 0.324 & 0.078 & 0.157 & 0.212 & 0.372 & 0.095 & 0.142 & 0.009\\
 &  & 10\% & 0.808 & 0.892 & 0.846 & 0.806 & 0.324 & 0.065 & 0.146 & 0.164 & 0.356 & 0.094 & 0.148 & 0.017\\
 &  & 20\% & 0.808 & 0.894 & 0.852 & 0.804 & 0.324 & 0.063 & 0.142 & 0.137 & 0.346 & 0.093 & 0.153 & 0.036\\
\addlinespace[2pt]
Chicago & $\zeta=1$, $\rho=0.7$ & 10\% & 0.591 & 0.571 & 0.608 & 0.615 & 0.816 & 0.530 & 0.590 & 0.373 & 0.769 & 0.514 & 0.498 & 0.018\\
 &  & 20\% & 0.591 & 0.554 & 0.608 & 0.625 & 0.816 & 0.513 & 0.544 & 0.265 & 0.788 & 0.510 & 0.526 & 0.021\\
\addlinespace[2pt]
Chicago & Interaction $\delta=0.25$\textsuperscript{a} & 10\% & 0.800 & 0.883 & 0.839 & 0.785 & 0.517 & 0.070 & 0.157 & 0.175 & 0.474 & 0.098 & 0.152 & 0.019\\
\addlinespace[2pt]
Chicago & Matched, capacity error & 10\% & 0.839 & 0.963 & 0.873 & 0.825 & 0.330 & 0.049 & 0.140 & 0.178 & 0.367 & 0.088 & 0.136 & 0.105\\
\addlinespace[2pt]
Chicago & Two-purpose, capacity error & 10\% & 0.808 & 0.886 & 0.837 & 0.790 & 0.324 & 0.078 & 0.166 & 0.190 & 0.356 & 0.106 & 0.159 & 0.101\\
\end{longtable}

%% file: tables/tabS_ablation.tex
\begin{longtable}{llcccccc}
\caption{Generator against its ablations and the prior at 10\% coverage (and the squared-deviation cellwise variant on its five-world subset). Cells as in Table~\ref{stab:allcontrasts}.}\label{stab:ablation}\\
\toprule
 & & \multicolumn{3}{c}{OD CPC} & \multicolumn{3}{c}{Held-out NRMSE}\\
\cmidrule(lr){3-5}\cmidrule(lr){6-8}
Network & Truth & vs slopes fixed & vs decay fixed & vs prior & vs slopes fixed & vs decay fixed & vs prior\\
\midrule
\endfirsthead
\toprule
 & & \multicolumn{3}{c}{OD CPC} & \multicolumn{3}{c}{Held-out NRMSE}\\
\cmidrule(lr){3-5}\cmidrule(lr){6-8}
Network & Truth & vs slopes fixed & vs decay fixed & vs prior & vs slopes fixed & vs decay fixed & vs prior\\
\midrule
\endhead
\bottomrule
\endfoot
EMA & Matched & \begin{tabular}[t]{@{}c@{}}\textbf{+0.118} (20)\\[-1pt]{\scriptsize[+0.099, +0.140]}\end{tabular} & \begin{tabular}[t]{@{}c@{}}\textbf{+0.048} (17)\\[-1pt]{\scriptsize[+0.031, +0.065]}\end{tabular} & \begin{tabular}[t]{@{}c@{}}\textbf{+0.136} (20)\\[-1pt]{\scriptsize[+0.118, +0.155]}\end{tabular} & \begin{tabular}[t]{@{}c@{}}\textbf{+0.186} (20)\\[-1pt]{\scriptsize[+0.154, +0.220]}\end{tabular} & \begin{tabular}[t]{@{}c@{}}\textbf{+0.052} (19)\\[-1pt]{\scriptsize[+0.030, +0.076]}\end{tabular} & \begin{tabular}[t]{@{}c@{}}\textbf{+0.324} (20)\\[-1pt]{\scriptsize[+0.264, +0.395]}\end{tabular}\\
EMA & Two-purpose $(1,3)$ & \begin{tabular}[t]{@{}c@{}}\textbf{+0.077} (20)\\[-1pt]{\scriptsize[+0.062, +0.094]}\end{tabular} & \begin{tabular}[t]{@{}c@{}}\textbf{+0.031} (18)\\[-1pt]{\scriptsize[+0.018, +0.046]}\end{tabular} & \begin{tabular}[t]{@{}c@{}}\textbf{+0.087} (20)\\[-1pt]{\scriptsize[+0.071, +0.104]}\end{tabular} & \begin{tabular}[t]{@{}c@{}}\textbf{+0.141} (20)\\[-1pt]{\scriptsize[+0.118, +0.166]}\end{tabular} & \begin{tabular}[t]{@{}c@{}}\textbf{+0.061} (16)\\[-1pt]{\scriptsize[+0.038, +0.085]}\end{tabular} & \begin{tabular}[t]{@{}c@{}}\textbf{+0.270} (20)\\[-1pt]{\scriptsize[+0.214, +0.333]}\end{tabular}\\
EMA & Mixture $\alpha=0.5$ & \begin{tabular}[t]{@{}c@{}}\textbf{+0.102} (20)\\[-1pt]{\scriptsize[+0.082, +0.123]}\end{tabular} & \begin{tabular}[t]{@{}c@{}}\textbf{+0.041} (19)\\[-1pt]{\scriptsize[+0.026, +0.058]}\end{tabular} & \begin{tabular}[t]{@{}c@{}}\textbf{+0.115} (20)\\[-1pt]{\scriptsize[+0.098, +0.134]}\end{tabular} & \begin{tabular}[t]{@{}c@{}}\textbf{+0.169} (20)\\[-1pt]{\scriptsize[+0.141, +0.201]}\end{tabular} & \begin{tabular}[t]{@{}c@{}}\textbf{+0.057} (18)\\[-1pt]{\scriptsize[+0.033, +0.083]}\end{tabular} & \begin{tabular}[t]{@{}c@{}}\textbf{+0.305} (20)\\[-1pt]{\scriptsize[+0.245, +0.374]}\end{tabular}\\
EMA & Two-purpose $(0.5,5)$ & \begin{tabular}[t]{@{}c@{}}\textbf{+0.064} (20)\\[-1pt]{\scriptsize[+0.045, +0.086]}\end{tabular} & \begin{tabular}[t]{@{}c@{}}\textbf{+0.013} (13)\\[-1pt]{\scriptsize[+0.004, +0.022]}\end{tabular} & \begin{tabular}[t]{@{}c@{}}\textbf{+0.057} (19)\\[-1pt]{\scriptsize[+0.044, +0.070]}\end{tabular} & \begin{tabular}[t]{@{}c@{}}\textbf{+0.131} (20)\\[-1pt]{\scriptsize[+0.106, +0.158]}\end{tabular} & \begin{tabular}[t]{@{}c@{}}\textbf{+0.059} (17)\\[-1pt]{\scriptsize[+0.035, +0.084]}\end{tabular} & \begin{tabular}[t]{@{}c@{}}\textbf{+0.266} (20)\\[-1pt]{\scriptsize[+0.205, +0.334]}\end{tabular}\\
EMA & Two-purpose $(0.25,8)$ & \begin{tabular}[t]{@{}c@{}}\textbf{+0.049} (19)\\[-1pt]{\scriptsize[+0.034, +0.066]}\end{tabular} & \begin{tabular}[t]{@{}c@{}}+0.006 (11)\\[-1pt]{\scriptsize[$-$0.000, +0.013]}\end{tabular} & \begin{tabular}[t]{@{}c@{}}\textbf{+0.038} (19)\\[-1pt]{\scriptsize[+0.028, +0.050]}\end{tabular} & \begin{tabular}[t]{@{}c@{}}\textbf{+0.123} (20)\\[-1pt]{\scriptsize[+0.096, +0.151]}\end{tabular} & \begin{tabular}[t]{@{}c@{}}\textbf{+0.068} (18)\\[-1pt]{\scriptsize[+0.041, +0.096]}\end{tabular} & \begin{tabular}[t]{@{}c@{}}\textbf{+0.267} (20)\\[-1pt]{\scriptsize[+0.200, +0.345]}\end{tabular}\\
EMA & $\zeta=0$, $\rho=0$ & \begin{tabular}[t]{@{}c@{}}\textbf{+0.040} (17)\\[-1pt]{\scriptsize[+0.025, +0.057]}\end{tabular} & \begin{tabular}[t]{@{}c@{}}$-$0.003 (12)\\[-1pt]{\scriptsize[$-$0.026, +0.015]}\end{tabular} & \begin{tabular}[t]{@{}c@{}}\textbf{+0.034} (17)\\[-1pt]{\scriptsize[+0.008, +0.056]}\end{tabular} & \begin{tabular}[t]{@{}c@{}}\textbf{+0.050} (16)\\[-1pt]{\scriptsize[+0.024, +0.077]}\end{tabular} & \begin{tabular}[t]{@{}c@{}}\textbf{+0.026} (15)\\[-1pt]{\scriptsize[+0.013, +0.041]}\end{tabular} & \begin{tabular}[t]{@{}c@{}}\textbf{+0.165} (20)\\[-1pt]{\scriptsize[+0.112, +0.228]}\end{tabular}\\
EMA & $\zeta=0.5$, $\rho=0$ & \begin{tabular}[t]{@{}c@{}}\textbf{+0.041} (18)\\[-1pt]{\scriptsize[+0.025, +0.059]}\end{tabular} & \begin{tabular}[t]{@{}c@{}}$-$0.009 (13)\\[-1pt]{\scriptsize[$-$0.032, +0.006]}\end{tabular} & \begin{tabular}[t]{@{}c@{}}+0.023 (16)\\[-1pt]{\scriptsize[$-$0.002, +0.044]}\end{tabular} & \begin{tabular}[t]{@{}c@{}}\textbf{+0.068} (19)\\[-1pt]{\scriptsize[+0.035, +0.112]}\end{tabular} & \begin{tabular}[t]{@{}c@{}}+0.006 (8)\\[-1pt]{\scriptsize[$-$0.004, +0.020]}\end{tabular} & \begin{tabular}[t]{@{}c@{}}\textbf{+0.191} (20)\\[-1pt]{\scriptsize[+0.121, +0.275]}\end{tabular}\\
EMA & $\zeta=1$, $\rho=0$ & \begin{tabular}[t]{@{}c@{}}\textbf{+0.033} (18)\\[-1pt]{\scriptsize[+0.021, +0.045]}\end{tabular} & \begin{tabular}[t]{@{}c@{}}$-$0.005 (11)\\[-1pt]{\scriptsize[$-$0.021, +0.006]}\end{tabular} & \begin{tabular}[t]{@{}c@{}}+0.011 (15)\\[-1pt]{\scriptsize[$-$0.006, +0.025]}\end{tabular} & \begin{tabular}[t]{@{}c@{}}\textbf{+0.103} (18)\\[-1pt]{\scriptsize[+0.065, +0.148]}\end{tabular} & \begin{tabular}[t]{@{}c@{}}+0.008 (12)\\[-1pt]{\scriptsize[$-$0.006, +0.022]}\end{tabular} & \begin{tabular}[t]{@{}c@{}}\textbf{+0.253} (20)\\[-1pt]{\scriptsize[+0.167, +0.352]}\end{tabular}\\
EMA & $\zeta=0$, $\rho=0.3$ & \begin{tabular}[t]{@{}c@{}}\textbf{+0.043} (16)\\[-1pt]{\scriptsize[+0.018, +0.068]}\end{tabular} & \begin{tabular}[t]{@{}c@{}}+0.007 (14)\\[-1pt]{\scriptsize[$-$0.009, +0.020]}\end{tabular} & \begin{tabular}[t]{@{}c@{}}\textbf{+0.038} (18)\\[-1pt]{\scriptsize[+0.017, +0.058]}\end{tabular} & \begin{tabular}[t]{@{}c@{}}\textbf{+0.058} (15)\\[-1pt]{\scriptsize[+0.024, +0.094]}\end{tabular} & \begin{tabular}[t]{@{}c@{}}\textbf{+0.028} (14)\\[-1pt]{\scriptsize[+0.009, +0.049]}\end{tabular} & \begin{tabular}[t]{@{}c@{}}\textbf{+0.159} (19)\\[-1pt]{\scriptsize[+0.097, +0.234]}\end{tabular}\\
EMA & $\zeta=0$, $\rho=0.7$ & \begin{tabular}[t]{@{}c@{}}+0.008 (14)\\[-1pt]{\scriptsize[$-$0.021, +0.035]}\end{tabular} & \begin{tabular}[t]{@{}c@{}}$-$0.000 (11)\\[-1pt]{\scriptsize[$-$0.025, +0.017]}\end{tabular} & \begin{tabular}[t]{@{}c@{}}+0.003 (14)\\[-1pt]{\scriptsize[$-$0.025, +0.026]}\end{tabular} & \begin{tabular}[t]{@{}c@{}}+0.034 (13)\\[-1pt]{\scriptsize[$-$0.005, +0.072]}\end{tabular} & \begin{tabular}[t]{@{}c@{}}\textbf{+0.035} (11)\\[-1pt]{\scriptsize[+0.004, +0.073]}\end{tabular} & \begin{tabular}[t]{@{}c@{}}\textbf{+0.109} (17)\\[-1pt]{\scriptsize[+0.051, +0.178]}\end{tabular}\\
EMA & $\zeta=1$, $\rho=0.7$ & \begin{tabular}[t]{@{}c@{}}+0.017 (12)\\[-1pt]{\scriptsize[$-$0.001, +0.036]}\end{tabular} & \begin{tabular}[t]{@{}c@{}}$-$0.003 (12)\\[-1pt]{\scriptsize[$-$0.021, +0.013]}\end{tabular} & \begin{tabular}[t]{@{}c@{}}+0.001 (12)\\[-1pt]{\scriptsize[$-$0.016, +0.016]}\end{tabular} & \begin{tabular}[t]{@{}c@{}}\textbf{+0.061} (16)\\[-1pt]{\scriptsize[+0.026, +0.098]}\end{tabular} & \begin{tabular}[t]{@{}c@{}}+0.007 (9)\\[-1pt]{\scriptsize[$-$0.011, +0.029]}\end{tabular} & \begin{tabular}[t]{@{}c@{}}\textbf{+0.177} (18)\\[-1pt]{\scriptsize[+0.092, +0.273]}\end{tabular}\\
EMA & Interaction $\delta=0.25$ & \begin{tabular}[t]{@{}c@{}}\textbf{+0.074} (19)\\[-1pt]{\scriptsize[+0.058, +0.091]}\end{tabular} & \begin{tabular}[t]{@{}c@{}}\textbf{+0.027} (16)\\[-1pt]{\scriptsize[+0.013, +0.042]}\end{tabular} & \begin{tabular}[t]{@{}c@{}}\textbf{+0.084} (20)\\[-1pt]{\scriptsize[+0.067, +0.101]}\end{tabular} & \begin{tabular}[t]{@{}c@{}}\textbf{+0.142} (20)\\[-1pt]{\scriptsize[+0.116, +0.169]}\end{tabular} & \begin{tabular}[t]{@{}c@{}}\textbf{+0.053} (17)\\[-1pt]{\scriptsize[+0.030, +0.079]}\end{tabular} & \begin{tabular}[t]{@{}c@{}}\textbf{+0.268} (20)\\[-1pt]{\scriptsize[+0.213, +0.331]}\end{tabular}\\
EMA & Interaction $\delta=0.5$ & \begin{tabular}[t]{@{}c@{}}\textbf{+0.070} (19)\\[-1pt]{\scriptsize[+0.051, +0.089]}\end{tabular} & \begin{tabular}[t]{@{}c@{}}\textbf{+0.019} (14)\\[-1pt]{\scriptsize[+0.006, +0.033]}\end{tabular} & \begin{tabular}[t]{@{}c@{}}\textbf{+0.076} (19)\\[-1pt]{\scriptsize[+0.059, +0.093]}\end{tabular} & \begin{tabular}[t]{@{}c@{}}\textbf{+0.137} (20)\\[-1pt]{\scriptsize[+0.108, +0.169]}\end{tabular} & \begin{tabular}[t]{@{}c@{}}\textbf{+0.038} (13)\\[-1pt]{\scriptsize[+0.016, +0.060]}\end{tabular} & \begin{tabular}[t]{@{}c@{}}\textbf{+0.257} (20)\\[-1pt]{\scriptsize[+0.206, +0.311]}\end{tabular}\\
EMA & Interaction $\delta=1$ & \begin{tabular}[t]{@{}c@{}}\textbf{+0.064} (20)\\[-1pt]{\scriptsize[+0.045, +0.083]}\end{tabular} & \begin{tabular}[t]{@{}c@{}}+0.011 (13)\\[-1pt]{\scriptsize[$-$0.000, +0.025]}\end{tabular} & \begin{tabular}[t]{@{}c@{}}\textbf{+0.067} (20)\\[-1pt]{\scriptsize[+0.049, +0.087]}\end{tabular} & \begin{tabular}[t]{@{}c@{}}\textbf{+0.128} (18)\\[-1pt]{\scriptsize[+0.083, +0.176]}\end{tabular} & \begin{tabular}[t]{@{}c@{}}\textbf{+0.026} (15)\\[-1pt]{\scriptsize[+0.009, +0.046]}\end{tabular} & \begin{tabular}[t]{@{}c@{}}\textbf{+0.237} (20)\\[-1pt]{\scriptsize[+0.182, +0.294]}\end{tabular}\\
EMA & Spatial $\phi=0.5$ & \begin{tabular}[t]{@{}c@{}}\textbf{+0.030} (19)\\[-1pt]{\scriptsize[+0.021, +0.040]}\end{tabular} & \begin{tabular}[t]{@{}c@{}}+0.010 (12)\\[-1pt]{\scriptsize[$-$0.000, +0.021]}\end{tabular} & \begin{tabular}[t]{@{}c@{}}\textbf{+0.039} (20)\\[-1pt]{\scriptsize[+0.028, +0.050]}\end{tabular} & \begin{tabular}[t]{@{}c@{}}\textbf{+0.082} (16)\\[-1pt]{\scriptsize[+0.043, +0.123]}\end{tabular} & \begin{tabular}[t]{@{}c@{}}+0.011 (11)\\[-1pt]{\scriptsize[$-$0.018, +0.039]}\end{tabular} & \begin{tabular}[t]{@{}c@{}}\textbf{+0.169} (19)\\[-1pt]{\scriptsize[+0.114, +0.224]}\end{tabular}\\
EMA & Spatial $\phi=1$ & \begin{tabular}[t]{@{}c@{}}+0.022 (11)\\[-1pt]{\scriptsize[$-$0.021, +0.074]}\end{tabular} & \begin{tabular}[t]{@{}c@{}}$-$0.012 (9)\\[-1pt]{\scriptsize[$-$0.036, +0.004]}\end{tabular} & \begin{tabular}[t]{@{}c@{}}+0.012 (13)\\[-1pt]{\scriptsize[$-$0.016, +0.040]}\end{tabular} & \begin{tabular}[t]{@{}c@{}}\textbf{+0.089} (15)\\[-1pt]{\scriptsize[+0.031, +0.150]}\end{tabular} & \begin{tabular}[t]{@{}c@{}}+0.004 (13)\\[-1pt]{\scriptsize[$-$0.029, +0.037]}\end{tabular} & \begin{tabular}[t]{@{}c@{}}\textbf{+0.173} (19)\\[-1pt]{\scriptsize[+0.116, +0.229]}\end{tabular}\\
EMA & Matched, biased prior & \begin{tabular}[t]{@{}c@{}}\textbf{+0.147} (20)\\[-1pt]{\scriptsize[+0.120, +0.176]}\end{tabular} & \begin{tabular}[t]{@{}c@{}}\textbf{+0.130} (20)\\[-1pt]{\scriptsize[+0.101, +0.159]}\end{tabular} & \begin{tabular}[t]{@{}c@{}}\textbf{+0.200} (20)\\[-1pt]{\scriptsize[+0.179, +0.218]}\end{tabular} & \begin{tabular}[t]{@{}c@{}}\textbf{+0.236} (20)\\[-1pt]{\scriptsize[+0.194, +0.281]}\end{tabular} & \begin{tabular}[t]{@{}c@{}}\textbf{+0.160} (20)\\[-1pt]{\scriptsize[+0.113, +0.206]}\end{tabular} & \begin{tabular}[t]{@{}c@{}}\textbf{+0.445} (20)\\[-1pt]{\scriptsize[+0.363, +0.546]}\end{tabular}\\
EMA & Two-purpose, biased prior & \begin{tabular}[t]{@{}c@{}}\textbf{+0.103} (20)\\[-1pt]{\scriptsize[+0.077, +0.131]}\end{tabular} & \begin{tabular}[t]{@{}c@{}}\textbf{+0.108} (20)\\[-1pt]{\scriptsize[+0.083, +0.133]}\end{tabular} & \begin{tabular}[t]{@{}c@{}}\textbf{+0.146} (20)\\[-1pt]{\scriptsize[+0.124, +0.167]}\end{tabular} & \begin{tabular}[t]{@{}c@{}}\textbf{+0.181} (20)\\[-1pt]{\scriptsize[+0.145, +0.218]}\end{tabular} & \begin{tabular}[t]{@{}c@{}}\textbf{+0.174} (19)\\[-1pt]{\scriptsize[+0.122, +0.223]}\end{tabular} & \begin{tabular}[t]{@{}c@{}}\textbf{+0.398} (20)\\[-1pt]{\scriptsize[+0.320, +0.493]}\end{tabular}\\
EMA & Matched, capacity error & \begin{tabular}[t]{@{}c@{}}\textbf{+0.085} (19)\\[-1pt]{\scriptsize[+0.063, +0.107]}\end{tabular} & \begin{tabular}[t]{@{}c@{}}\textbf{+0.027} (14)\\[-1pt]{\scriptsize[+0.007, +0.050]}\end{tabular} & \begin{tabular}[t]{@{}c@{}}\textbf{+0.106} (20)\\[-1pt]{\scriptsize[+0.083, +0.130]}\end{tabular} & \begin{tabular}[t]{@{}c@{}}\textbf{+0.147} (19)\\[-1pt]{\scriptsize[+0.107, +0.194]}\end{tabular} & \begin{tabular}[t]{@{}c@{}}\textbf{+0.033} (12)\\[-1pt]{\scriptsize[+0.005, +0.065]}\end{tabular} & \begin{tabular}[t]{@{}c@{}}\textbf{+0.284} (20)\\[-1pt]{\scriptsize[+0.219, +0.357]}\end{tabular}\\
EMA & Two-purpose, capacity error & \begin{tabular}[t]{@{}c@{}}\textbf{+0.061} (19)\\[-1pt]{\scriptsize[+0.038, +0.081]}\end{tabular} & \begin{tabular}[t]{@{}c@{}}+0.016 (13)\\[-1pt]{\scriptsize[$-$0.004, +0.035]}\end{tabular} & \begin{tabular}[t]{@{}c@{}}\textbf{+0.069} (19)\\[-1pt]{\scriptsize[+0.043, +0.091]}\end{tabular} & \begin{tabular}[t]{@{}c@{}}\textbf{+0.114} (20)\\[-1pt]{\scriptsize[+0.084, +0.148]}\end{tabular} & \begin{tabular}[t]{@{}c@{}}\textbf{+0.036} (14)\\[-1pt]{\scriptsize[+0.016, +0.059]}\end{tabular} & \begin{tabular}[t]{@{}c@{}}\textbf{+0.244} (20)\\[-1pt]{\scriptsize[+0.184, +0.309]}\end{tabular}\\
Chicago & Matched & \begin{tabular}[t]{@{}c@{}}\textbf{+0.103} (20)\\[-1pt]{\scriptsize[+0.082, +0.124]}\end{tabular} & \begin{tabular}[t]{@{}c@{}}\textbf{+0.071} (20)\\[-1pt]{\scriptsize[+0.052, +0.090]}\end{tabular} & \begin{tabular}[t]{@{}c@{}}\textbf{+0.143} (20)\\[-1pt]{\scriptsize[+0.123, +0.164]}\end{tabular} & \begin{tabular}[t]{@{}c@{}}\textbf{+0.149} (20)\\[-1pt]{\scriptsize[+0.115, +0.186]}\end{tabular} & \begin{tabular}[t]{@{}c@{}}\textbf{+0.055} (20)\\[-1pt]{\scriptsize[+0.041, +0.071]}\end{tabular} & \begin{tabular}[t]{@{}c@{}}\textbf{+0.362} (20)\\[-1pt]{\scriptsize[+0.284, +0.440]}\end{tabular}\\
Chicago & Two-purpose $(1,3)$ & \begin{tabular}[t]{@{}c@{}}\textbf{+0.060} (20)\\[-1pt]{\scriptsize[+0.045, +0.075]}\end{tabular} & \begin{tabular}[t]{@{}c@{}}\textbf{+0.035} (17)\\[-1pt]{\scriptsize[+0.021, +0.051]}\end{tabular} & \begin{tabular}[t]{@{}c@{}}\textbf{+0.086} (20)\\[-1pt]{\scriptsize[+0.070, +0.103]}\end{tabular} & \begin{tabular}[t]{@{}c@{}}\textbf{+0.118} (20)\\[-1pt]{\scriptsize[+0.085, +0.152]}\end{tabular} & \begin{tabular}[t]{@{}c@{}}\textbf{+0.065} (19)\\[-1pt]{\scriptsize[+0.050, +0.080]}\end{tabular} & \begin{tabular}[t]{@{}c@{}}\textbf{+0.325} (20)\\[-1pt]{\scriptsize[+0.257, +0.392]}\end{tabular}\\
Chicago & $\zeta=1$, $\rho=0.7$ & \begin{tabular}[t]{@{}c@{}}$-$0.020 (5)\\[-1pt]{\scriptsize[$-$0.047, +0.008]}\end{tabular} & \begin{tabular}[t]{@{}c@{}}$-$0.012 (9)\\[-1pt]{\scriptsize[$-$0.037, +0.012]}\end{tabular} & \begin{tabular}[t]{@{}c@{}}\textbf{$-$0.027} (8)\\[-1pt]{\scriptsize[$-$0.051, $-$0.003]}\end{tabular} & \begin{tabular}[t]{@{}c@{}}\textbf{+0.134} (20)\\[-1pt]{\scriptsize[+0.103, +0.164]}\end{tabular} & \begin{tabular}[t]{@{}c@{}}\textbf{+0.035} (16)\\[-1pt]{\scriptsize[+0.013, +0.063]}\end{tabular} & \begin{tabular}[t]{@{}c@{}}\textbf{+0.344} (20)\\[-1pt]{\scriptsize[+0.266, +0.432]}\end{tabular}\\
Chicago & Matched, capacity error & \begin{tabular}[t]{@{}c@{}}\textbf{+0.085} (20)\\[-1pt]{\scriptsize[+0.067, +0.103]}\end{tabular} & \begin{tabular}[t]{@{}c@{}}\textbf{+0.050} (16)\\[-1pt]{\scriptsize[+0.029, +0.072]}\end{tabular} & \begin{tabular}[t]{@{}c@{}}\textbf{+0.120} (20)\\[-1pt]{\scriptsize[+0.098, +0.142]}\end{tabular} & \begin{tabular}[t]{@{}c@{}}\textbf{+0.128} (20)\\[-1pt]{\scriptsize[+0.095, +0.162]}\end{tabular} & \begin{tabular}[t]{@{}c@{}}\textbf{+0.036} (17)\\[-1pt]{\scriptsize[+0.020, +0.054]}\end{tabular} & \begin{tabular}[t]{@{}c@{}}\textbf{+0.334} (20)\\[-1pt]{\scriptsize[+0.257, +0.414]}\end{tabular}\\
Chicago & Two-purpose, capacity error & \begin{tabular}[t]{@{}c@{}}\textbf{+0.055} (20)\\[-1pt]{\scriptsize[+0.041, +0.068]}\end{tabular} & \begin{tabular}[t]{@{}c@{}}\textbf{+0.031} (16)\\[-1pt]{\scriptsize[+0.017, +0.047]}\end{tabular} & \begin{tabular}[t]{@{}c@{}}\textbf{+0.080} (19)\\[-1pt]{\scriptsize[+0.063, +0.097]}\end{tabular} & \begin{tabular}[t]{@{}c@{}}\textbf{+0.106} (19)\\[-1pt]{\scriptsize[+0.074, +0.141]}\end{tabular} & \begin{tabular}[t]{@{}c@{}}\textbf{+0.058} (18)\\[-1pt]{\scriptsize[+0.042, +0.074]}\end{tabular} & \begin{tabular}[t]{@{}c@{}}\textbf{+0.310} (20)\\[-1pt]{\scriptsize[+0.242, +0.380]}\end{tabular}\\
\midrule
\multicolumn{8}{l}{\textit{Squared-deviation cellwise variant (worlds 2000–2004, 10\%)}}\\
EMA & Matched & \multicolumn{3}{c}{OD CPC: \begin{tabular}[t]{@{}c@{}}\textbf{+0.140} (5)\\[-1pt]{\scriptsize[+0.107, +0.173]}\end{tabular}} & \multicolumn{3}{c}{NRMSE: \begin{tabular}[t]{@{}c@{}}\textbf{+0.259} (5)\\[-1pt]{\scriptsize[+0.157, +0.360]}\end{tabular}}\\
EMA & Two-purpose $(1,3)$ & \multicolumn{3}{c}{OD CPC: \begin{tabular}[t]{@{}c@{}}\textbf{+0.091} (5)\\[-1pt]{\scriptsize[+0.065, +0.125]}\end{tabular}} & \multicolumn{3}{c}{NRMSE: \begin{tabular}[t]{@{}c@{}}\textbf{+0.190} (5)\\[-1pt]{\scriptsize[+0.120, +0.265]}\end{tabular}}\\
Chicago & Matched & \multicolumn{3}{c}{OD CPC: \begin{tabular}[t]{@{}c@{}}\textbf{+0.164} (5)\\[-1pt]{\scriptsize[+0.133, +0.200]}\end{tabular}} & \multicolumn{3}{c}{NRMSE: \begin{tabular}[t]{@{}c@{}}\textbf{+0.141} (5)\\[-1pt]{\scriptsize[+0.117, +0.174]}\end{tabular}}\\
Chicago & Two-purpose $(1,3)$ & \multicolumn{3}{c}{OD CPC: \begin{tabular}[t]{@{}c@{}}\textbf{+0.112} (5)\\[-1pt]{\scriptsize[+0.086, +0.141]}\end{tabular}} & \multicolumn{3}{c}{NRMSE: \begin{tabular}[t]{@{}c@{}}\textbf{+0.114} (5)\\[-1pt]{\scriptsize[+0.086, +0.144]}\end{tabular}}\\
\end{longtable}

%% file: tables/tabS_lambdapath.tex
\begin{tabular}{llrccc}
\toprule
Truth & Estimator & Datasets & Spearman, calibration vs OD CPC & Spearman, calibration vs held-out & OD CPC range along path\\
\midrule
Matched & Generator & 40 & $-$1.00 & +1.00 & 0.138\\
Matched & Cellwise (KL) & 40 & +0.00 & +1.00 & 0.008\\
Matched & OD-factor & 40 & $-$1.00 & +0.95 & 0.030\\
Two-purpose $(1,3)$ & Generator & 40 & $-$0.95 & +1.00 & 0.086\\
Two-purpose $(1,3)$ & Cellwise (KL) & 40 & +0.00 & +1.00 & 0.008\\
Two-purpose $(1,3)$ & OD-factor & 40 & $-$0.80 & +0.90 & 0.025\\
\bottomrule
\end{tabular}

%% file: tables/tabS_d7.tex
\begin{longtable}{llccc}
\caption{Median per-dataset Spearman correlation between a held-out link’s free-flow distance to the nearest calibration link and the generator’s absolute-error advantage over the comparator. Positive values: the advantage grows with distance.}\label{stab:d7}\\
\toprule
Network & Truth & Cov. & vs OD-factor & vs cellwise\\
\midrule
\endfirsthead
\toprule
Network & Truth & Cov. & vs OD-factor & vs cellwise\\
\midrule
\endhead
\bottomrule
\endfoot
Chicago & $\zeta=1$, $\rho=0.7$ & 10\% & +0.000 & +0.054\\
Chicago & $\zeta=1$, $\rho=0.7$ & 20\% & $-$0.002 & +0.044\\
Chicago & Matched, capacity error & 10\% & +0.017 & +0.020\\
Chicago & Two-purpose, capacity error & 10\% & +0.008 & +0.025\\
EMA & $\zeta=0.5$, $\rho=0$ & 10\% & +0.057 & +0.014\\
EMA & $\zeta=0.5$, $\rho=0$ & 20\% & +0.013 & $-$0.008\\
EMA & $\zeta=0.5$, $\rho=0$ & 40\% & +0.134 & +0.007\\
EMA & $\zeta=0$, $\rho=0$ & 10\% & +0.058 & $-$0.003\\
EMA & $\zeta=0$, $\rho=0$ & 20\% & +0.070 & +0.037\\
EMA & $\zeta=0$, $\rho=0$ & 40\% & $-$0.002 & $-$0.065\\
EMA & $\zeta=0$, $\rho=0.3$ & 10\% & +0.016 & $-$0.003\\
EMA & $\zeta=0$, $\rho=0.3$ & 20\% & +0.031 & +0.036\\
EMA & $\zeta=0$, $\rho=0.3$ & 40\% & +0.046 & $-$0.023\\
EMA & $\zeta=0$, $\rho=0.7$ & 10\% & +0.010 & +0.091\\
EMA & $\zeta=0$, $\rho=0.7$ & 20\% & $-$0.010 & +0.023\\
EMA & $\zeta=0$, $\rho=0.7$ & 40\% & +0.020 & +0.030\\
EMA & $\zeta=1$, $\rho=0$ & 10\% & $-$0.013 & +0.005\\
EMA & $\zeta=1$, $\rho=0$ & 20\% & $-$0.003 & +0.008\\
EMA & $\zeta=1$, $\rho=0$ & 40\% & +0.187 & $-$0.010\\
EMA & $\zeta=1$, $\rho=0.7$ & 10\% & +0.028 & $-$0.024\\
EMA & $\zeta=1$, $\rho=0.7$ & 20\% & $-$0.033 & $-$0.057\\
EMA & $\zeta=1$, $\rho=0.7$ & 40\% & +0.020 & +0.097\\
EMA & Matched & 5\% & +0.054 & +0.065\\
EMA & Matched & 10\% & +0.033 & +0.102\\
EMA & Matched & 20\% & +0.059 & +0.076\\
EMA & Matched & 40\% & +0.080 & $-$0.028\\
EMA & Matched, capacity error & 10\% & +0.037 & +0.045\\
EMA & Matched, biased prior & 10\% & +0.072 & +0.069\\
EMA & Mixture $\alpha=0.5$ & 10\% & +0.004 & +0.058\\
EMA & Mixture $\alpha=0.5$ & 20\% & +0.104 & +0.026\\
EMA & Mixture $\alpha=0.5$ & 40\% & +0.116 & +0.025\\
EMA & Interaction $\delta=0.25$ & 10\% & +0.062 & +0.037\\
EMA & Interaction $\delta=0.25$ & 20\% & +0.071 & +0.030\\
EMA & Interaction $\delta=0.25$ & 40\% & +0.122 & +0.034\\
EMA & Interaction $\delta=0.5$ & 10\% & $-$0.041 & +0.030\\
EMA & Interaction $\delta=0.5$ & 20\% & +0.022 & +0.039\\
EMA & Interaction $\delta=0.5$ & 40\% & +0.129 & $-$0.031\\
EMA & Interaction $\delta=1$ & 10\% & $-$0.040 & +0.021\\
EMA & Interaction $\delta=1$ & 20\% & $-$0.018 & +0.028\\
EMA & Interaction $\delta=1$ & 40\% & +0.069 & +0.093\\
EMA & Spatial $\phi=0.5$ & 10\% & +0.045 & +0.008\\
EMA & Spatial $\phi=0.5$ & 20\% & +0.036 & +0.026\\
EMA & Spatial $\phi=0.5$ & 40\% & $-$0.043 & +0.088\\
EMA & Spatial $\phi=1$ & 10\% & +0.025 & +0.005\\
EMA & Spatial $\phi=1$ & 20\% & +0.010 & $-$0.010\\
EMA & Spatial $\phi=1$ & 40\% & +0.093 & +0.131\\
EMA & Two-purpose $(1,3)$ & 5\% & +0.061 & +0.069\\
EMA & Two-purpose $(1,3)$ & 10\% & +0.039 & +0.055\\
EMA & Two-purpose $(1,3)$ & 20\% & +0.055 & +0.012\\
EMA & Two-purpose $(1,3)$ & 40\% & +0.139 & +0.056\\
EMA & Two-purpose, capacity error & 10\% & $-$0.038 & +0.008\\
EMA & Two-purpose, biased prior & 10\% & +0.076 & +0.062\\
EMA & Two-purpose $(0.25,8)$ & 10\% & +0.049 & +0.040\\
EMA & Two-purpose $(0.25,8)$ & 20\% & +0.045 & +0.014\\
EMA & Two-purpose $(0.25,8)$ & 40\% & +0.160 & +0.061\\
EMA & Two-purpose $(0.5,5)$ & 10\% & +0.033 & +0.039\\
EMA & Two-purpose $(0.5,5)$ & 20\% & +0.057 & +0.010\\
EMA & Two-purpose $(0.5,5)$ & 40\% & +0.146 & +0.047\\
\end{longtable}

%% file: tables/tabS_projection.tex
\begin{longtable}{lllccccc}
\caption{Composition scenario: held-out link-flow change error of the generator re-evaluated at the scenario inputs, against matrix bases projected by the prior ratio (F3), the generator’s own base projected by the prior ratio (same base), and matrix bases projected by the fitted generator’s ratio (hybrid). Positive values favour the generator. Cells as in Table~\ref{stab:allcontrasts}. Coverage 10\%. Prog.: C confirmation, P post-confirmation programme. \textsuperscript{a}Confirmation-programme diagnostic, worlds 2000–2009 only.}\label{stab:projection}\\
\toprule
Network & Truth & Prog. & vs OD-factor, F3 & vs cellwise, F3 & vs same base & vs OD-factor, hybrid & vs cellwise, hybrid\\
\midrule
\endfirsthead
\toprule
Network & Truth & Prog. & vs OD-factor, F3 & vs cellwise, F3 & vs same base & vs OD-factor, hybrid & vs cellwise, hybrid\\
\midrule
\endhead
\bottomrule
\endfoot
EMA & Matched & C & \begin{tabular}[t]{@{}c@{}}\textbf{+0.428} (20)\\[-1pt]{\scriptsize[+0.311, +0.556]}\end{tabular} & \begin{tabular}[t]{@{}c@{}}\textbf{+0.473} (20)\\[-1pt]{\scriptsize[+0.343, +0.623]}\end{tabular} & \begin{tabular}[t]{@{}c@{}}\textbf{+0.399} (20)\\[-1pt]{\scriptsize[+0.267, +0.540]}\end{tabular} & \begin{tabular}[t]{@{}c@{}}\textbf{+0.121} (19)\\[-1pt]{\scriptsize[+0.078, +0.162]}\end{tabular} & \begin{tabular}[t]{@{}c@{}}\textbf{+0.209} (19)\\[-1pt]{\scriptsize[+0.153, +0.264]}\end{tabular}\\
EMA & Two-purpose $(1,3)$ & C & \begin{tabular}[t]{@{}c@{}}\textbf{+0.360} (20)\\[-1pt]{\scriptsize[+0.247, +0.480]}\end{tabular} & \begin{tabular}[t]{@{}c@{}}\textbf{+0.394} (19)\\[-1pt]{\scriptsize[+0.266, +0.537]}\end{tabular} & \begin{tabular}[t]{@{}c@{}}\textbf{+0.338} (18)\\[-1pt]{\scriptsize[+0.224, +0.465]}\end{tabular} & \begin{tabular}[t]{@{}c@{}}\textbf{+0.093} (18)\\[-1pt]{\scriptsize[+0.052, +0.134]}\end{tabular} & \begin{tabular}[t]{@{}c@{}}\textbf{+0.156} (19)\\[-1pt]{\scriptsize[+0.093, +0.218]}\end{tabular}\\
EMA & Mixture $\alpha=0.5$ & P & \begin{tabular}[t]{@{}c@{}}\textbf{+0.428} (20)\\[-1pt]{\scriptsize[+0.314, +0.550]}\end{tabular} & \begin{tabular}[t]{@{}c@{}}\textbf{+0.461} (20)\\[-1pt]{\scriptsize[+0.338, +0.603]}\end{tabular} & \begin{tabular}[t]{@{}c@{}}\textbf{+0.396} (20)\\[-1pt]{\scriptsize[+0.273, +0.535]}\end{tabular} & \begin{tabular}[t]{@{}c@{}}\textbf{+0.111} (18)\\[-1pt]{\scriptsize[+0.070, +0.150]}\end{tabular} & \begin{tabular}[t]{@{}c@{}}\textbf{+0.197} (20)\\[-1pt]{\scriptsize[+0.147, +0.248]}\end{tabular}\\
EMA & Two-purpose $(0.5,5)$ & P & \begin{tabular}[t]{@{}c@{}}\textbf{+0.327} (19)\\[-1pt]{\scriptsize[+0.217, +0.443]}\end{tabular} & \begin{tabular}[t]{@{}c@{}}\textbf{+0.362} (19)\\[-1pt]{\scriptsize[+0.242, +0.498]}\end{tabular} & \begin{tabular}[t]{@{}c@{}}\textbf{+0.290} (18)\\[-1pt]{\scriptsize[+0.185, +0.404]}\end{tabular} & \begin{tabular}[t]{@{}c@{}}\textbf{+0.076} (14)\\[-1pt]{\scriptsize[+0.012, +0.138]}\end{tabular} & \begin{tabular}[t]{@{}c@{}}\textbf{+0.152} (15)\\[-1pt]{\scriptsize[+0.083, +0.228]}\end{tabular}\\
EMA & Two-purpose $(0.25,8)$ & P & \begin{tabular}[t]{@{}c@{}}\textbf{+0.327} (19)\\[-1pt]{\scriptsize[+0.230, +0.427]}\end{tabular} & \begin{tabular}[t]{@{}c@{}}\textbf{+0.357} (20)\\[-1pt]{\scriptsize[+0.248, +0.478]}\end{tabular} & \begin{tabular}[t]{@{}c@{}}\textbf{+0.271} (19)\\[-1pt]{\scriptsize[+0.176, +0.372]}\end{tabular} & \begin{tabular}[t]{@{}c@{}}\textbf{+0.097} (15)\\[-1pt]{\scriptsize[+0.034, +0.157]}\end{tabular} & \begin{tabular}[t]{@{}c@{}}\textbf{+0.176} (17)\\[-1pt]{\scriptsize[+0.094, +0.266]}\end{tabular}\\
EMA & $\zeta=0$, $\rho=0$ & P & \begin{tabular}[t]{@{}c@{}}\textbf{+0.200} (15)\\[-1pt]{\scriptsize[+0.060, +0.338]}\end{tabular} & \begin{tabular}[t]{@{}c@{}}\textbf{+0.199} (16)\\[-1pt]{\scriptsize[+0.071, +0.337]}\end{tabular} & \begin{tabular}[t]{@{}c@{}}\textbf{+0.160} (16)\\[-1pt]{\scriptsize[+0.032, +0.281]}\end{tabular} & \begin{tabular}[t]{@{}c@{}}\textbf{+0.091} (14)\\[-1pt]{\scriptsize[+0.031, +0.163]}\end{tabular} & \begin{tabular}[t]{@{}c@{}}\textbf{+0.100} (15)\\[-1pt]{\scriptsize[+0.041, +0.165]}\end{tabular}\\
EMA & $\zeta=0.5$, $\rho=0$ & P & \begin{tabular}[t]{@{}c@{}}\textbf{+0.359} (16)\\[-1pt]{\scriptsize[+0.098, +0.708]}\end{tabular} & \begin{tabular}[t]{@{}c@{}}\textbf{+0.382} (14)\\[-1pt]{\scriptsize[+0.143, +0.718]}\end{tabular} & \begin{tabular}[t]{@{}c@{}}\textbf{+0.358} (18)\\[-1pt]{\scriptsize[+0.128, +0.666]}\end{tabular} & \begin{tabular}[t]{@{}c@{}}$-$0.041 (7)\\[-1pt]{\scriptsize[$-$0.101, +0.012]}\end{tabular} & \begin{tabular}[t]{@{}c@{}}+0.021 (10)\\[-1pt]{\scriptsize[$-$0.046, +0.087]}\end{tabular}\\
EMA & $\zeta=1$, $\rho=0$ & P & \begin{tabular}[t]{@{}c@{}}\textbf{+0.473} (18)\\[-1pt]{\scriptsize[+0.203, +0.793]}\end{tabular} & \begin{tabular}[t]{@{}c@{}}\textbf{+0.567} (19)\\[-1pt]{\scriptsize[+0.220, +1.002]}\end{tabular} & \begin{tabular}[t]{@{}c@{}}\textbf{+0.440} (17)\\[-1pt]{\scriptsize[+0.183, +0.762]}\end{tabular} & \begin{tabular}[t]{@{}c@{}}+0.008 (9)\\[-1pt]{\scriptsize[$-$0.035, +0.062]}\end{tabular} & \begin{tabular}[t]{@{}c@{}}+0.059 (13)\\[-1pt]{\scriptsize[$-$0.001, +0.127]}\end{tabular}\\
EMA & $\zeta=0$, $\rho=0.3$ & P & \begin{tabular}[t]{@{}c@{}}\textbf{+0.190} (18)\\[-1pt]{\scriptsize[+0.093, +0.307]}\end{tabular} & \begin{tabular}[t]{@{}c@{}}\textbf{+0.156} (16)\\[-1pt]{\scriptsize[+0.049, +0.300]}\end{tabular} & \begin{tabular}[t]{@{}c@{}}\textbf{+0.148} (16)\\[-1pt]{\scriptsize[+0.066, +0.246]}\end{tabular} & \begin{tabular}[t]{@{}c@{}}\textbf{+0.106} (14)\\[-1pt]{\scriptsize[+0.038, +0.185]}\end{tabular} & \begin{tabular}[t]{@{}c@{}}+0.054 (12)\\[-1pt]{\scriptsize[$-$0.006, +0.122]}\end{tabular}\\
EMA & $\zeta=0$, $\rho=0.7$ & P & \begin{tabular}[t]{@{}c@{}}$-$0.140 (11)\\[-1pt]{\scriptsize[$-$0.447, +0.066]}\end{tabular} & \begin{tabular}[t]{@{}c@{}}$-$0.193 (9)\\[-1pt]{\scriptsize[$-$0.585, +0.080]}\end{tabular} & \begin{tabular}[t]{@{}c@{}}$-$0.135 (12)\\[-1pt]{\scriptsize[$-$0.445, +0.066]}\end{tabular} & \begin{tabular}[t]{@{}c@{}}+0.105 (13)\\[-1pt]{\scriptsize[$-$0.034, +0.319]}\end{tabular} & \begin{tabular}[t]{@{}c@{}}$-$0.029 (9)\\[-1pt]{\scriptsize[$-$0.245, +0.180]}\end{tabular}\\
EMA & $\zeta=1$, $\rho=0.7$ & P & \begin{tabular}[t]{@{}c@{}}\textbf{+0.376} (14)\\[-1pt]{\scriptsize[+0.052, +0.775]}\end{tabular} & \begin{tabular}[t]{@{}c@{}}\textbf{+0.417} (11)\\[-1pt]{\scriptsize[+0.076, +0.821]}\end{tabular} & \begin{tabular}[t]{@{}c@{}}\textbf{+0.313} (14)\\[-1pt]{\scriptsize[+0.036, +0.651]}\end{tabular} & \begin{tabular}[t]{@{}c@{}}+0.117 (8)\\[-1pt]{\scriptsize[$-$0.078, +0.385]}\end{tabular} & \begin{tabular}[t]{@{}c@{}}+0.103 (12)\\[-1pt]{\scriptsize[$-$0.050, +0.287]}\end{tabular}\\
EMA & Interaction $\delta=0.25$ & P & \begin{tabular}[t]{@{}c@{}}\textbf{+0.339} (20)\\[-1pt]{\scriptsize[+0.248, +0.440]}\end{tabular} & \begin{tabular}[t]{@{}c@{}}\textbf{+0.389} (20)\\[-1pt]{\scriptsize[+0.273, +0.520]}\end{tabular} & \begin{tabular}[t]{@{}c@{}}\textbf{+0.313} (20)\\[-1pt]{\scriptsize[+0.211, +0.430]}\end{tabular} & \begin{tabular}[t]{@{}c@{}}\textbf{+0.090} (18)\\[-1pt]{\scriptsize[+0.057, +0.127]}\end{tabular} & \begin{tabular}[t]{@{}c@{}}\textbf{+0.153} (20)\\[-1pt]{\scriptsize[+0.101, +0.216]}\end{tabular}\\
EMA & Interaction $\delta=0.5$ & P & \begin{tabular}[t]{@{}c@{}}\textbf{+0.308} (20)\\[-1pt]{\scriptsize[+0.222, +0.403]}\end{tabular} & \begin{tabular}[t]{@{}c@{}}\textbf{+0.360} (20)\\[-1pt]{\scriptsize[+0.253, +0.481]}\end{tabular} & \begin{tabular}[t]{@{}c@{}}\textbf{+0.293} (20)\\[-1pt]{\scriptsize[+0.204, +0.396]}\end{tabular} & \begin{tabular}[t]{@{}c@{}}\textbf{+0.076} (19)\\[-1pt]{\scriptsize[+0.043, +0.114]}\end{tabular} & \begin{tabular}[t]{@{}c@{}}\textbf{+0.153} (19)\\[-1pt]{\scriptsize[+0.096, +0.215]}\end{tabular}\\
EMA & Interaction $\delta=1$ & P & \begin{tabular}[t]{@{}c@{}}\textbf{+0.188} (17)\\[-1pt]{\scriptsize[+0.104, +0.277]}\end{tabular} & \begin{tabular}[t]{@{}c@{}}\textbf{+0.233} (18)\\[-1pt]{\scriptsize[+0.146, +0.327]}\end{tabular} & \begin{tabular}[t]{@{}c@{}}\textbf{+0.166} (16)\\[-1pt]{\scriptsize[+0.082, +0.255]}\end{tabular} & \begin{tabular}[t]{@{}c@{}}\textbf{+0.062} (12)\\[-1pt]{\scriptsize[+0.008, +0.131]}\end{tabular} & \begin{tabular}[t]{@{}c@{}}\textbf{+0.125} (18)\\[-1pt]{\scriptsize[+0.075, +0.188]}\end{tabular}\\
EMA & Spatial $\phi=0.5$ & P & \begin{tabular}[t]{@{}c@{}}+0.007 (11)\\[-1pt]{\scriptsize[$-$0.221, +0.204]}\end{tabular} & \begin{tabular}[t]{@{}c@{}}+0.003 (11)\\[-1pt]{\scriptsize[$-$0.169, +0.147]}\end{tabular} & \begin{tabular}[t]{@{}c@{}}$-$0.020 (12)\\[-1pt]{\scriptsize[$-$0.194, +0.129]}\end{tabular} & \begin{tabular}[t]{@{}c@{}}$-$0.002 (10)\\[-1pt]{\scriptsize[$-$0.078, +0.086]}\end{tabular} & \begin{tabular}[t]{@{}c@{}}+0.009 (14)\\[-1pt]{\scriptsize[$-$0.035, +0.048]}\end{tabular}\\
EMA & Spatial $\phi=1$ & P & \begin{tabular}[t]{@{}c@{}}$-$0.098 (9)\\[-1pt]{\scriptsize[$-$0.446, +0.262]}\end{tabular} & \begin{tabular}[t]{@{}c@{}}$-$0.198 (11)\\[-1pt]{\scriptsize[$-$0.508, +0.033]}\end{tabular} & \begin{tabular}[t]{@{}c@{}}$-$0.131 (12)\\[-1pt]{\scriptsize[$-$0.492, +0.178]}\end{tabular} & \begin{tabular}[t]{@{}c@{}}+0.082 (10)\\[-1pt]{\scriptsize[$-$0.084, +0.332]}\end{tabular} & \begin{tabular}[t]{@{}c@{}}$-$0.135 (9)\\[-1pt]{\scriptsize[$-$0.341, +0.035]}\end{tabular}\\
EMA & Interaction $\delta=0.25$\textsuperscript{a} & C & \begin{tabular}[t]{@{}c@{}}\textbf{+0.315} (10)\\[-1pt]{\scriptsize[+0.185, +0.467]}\end{tabular} & \begin{tabular}[t]{@{}c@{}}\textbf{+0.340} (10)\\[-1pt]{\scriptsize[+0.190, +0.505]}\end{tabular} & \begin{tabular}[t]{@{}c@{}}\textbf{+0.274} (10)\\[-1pt]{\scriptsize[+0.150, +0.426]}\end{tabular} & – & –\\
EMA & Matched, biased prior & P & \begin{tabular}[t]{@{}c@{}}\textbf{+0.710} (20)\\[-1pt]{\scriptsize[+0.509, +0.933]}\end{tabular} & \begin{tabular}[t]{@{}c@{}}\textbf{+0.800} (20)\\[-1pt]{\scriptsize[+0.605, +1.014]}\end{tabular} & \begin{tabular}[t]{@{}c@{}}\textbf{+0.624} (20)\\[-1pt]{\scriptsize[+0.430, +0.856]}\end{tabular} & \begin{tabular}[t]{@{}c@{}}\textbf{+0.187} (19)\\[-1pt]{\scriptsize[+0.138, +0.240]}\end{tabular} & \begin{tabular}[t]{@{}c@{}}\textbf{+0.302} (20)\\[-1pt]{\scriptsize[+0.223, +0.391]}\end{tabular}\\
EMA & Two-purpose, biased prior & P & \begin{tabular}[t]{@{}c@{}}\textbf{+0.630} (20)\\[-1pt]{\scriptsize[+0.450, +0.828]}\end{tabular} & \begin{tabular}[t]{@{}c@{}}\textbf{+0.672} (20)\\[-1pt]{\scriptsize[+0.504, +0.849]}\end{tabular} & \begin{tabular}[t]{@{}c@{}}\textbf{+0.525} (20)\\[-1pt]{\scriptsize[+0.351, +0.727]}\end{tabular} & \begin{tabular}[t]{@{}c@{}}\textbf{+0.196} (19)\\[-1pt]{\scriptsize[+0.138, +0.255]}\end{tabular} & \begin{tabular}[t]{@{}c@{}}\textbf{+0.258} (20)\\[-1pt]{\scriptsize[+0.192, +0.328]}\end{tabular}\\
EMA & Matched, capacity error & P & \begin{tabular}[t]{@{}c@{}}\textbf{+0.299} (18)\\[-1pt]{\scriptsize[+0.178, +0.433]}\end{tabular} & \begin{tabular}[t]{@{}c@{}}\textbf{+0.320} (19)\\[-1pt]{\scriptsize[+0.196, +0.462]}\end{tabular} & \begin{tabular}[t]{@{}c@{}}\textbf{+0.281} (17)\\[-1pt]{\scriptsize[+0.153, +0.421]}\end{tabular} & \begin{tabular}[t]{@{}c@{}}\textbf{+0.054} (14)\\[-1pt]{\scriptsize[+0.002, +0.104]}\end{tabular} & \begin{tabular}[t]{@{}c@{}}\textbf{+0.121} (16)\\[-1pt]{\scriptsize[+0.055, +0.185]}\end{tabular}\\
EMA & Two-purpose, capacity error & P & \begin{tabular}[t]{@{}c@{}}\textbf{+0.252} (19)\\[-1pt]{\scriptsize[+0.150, +0.359]}\end{tabular} & \begin{tabular}[t]{@{}c@{}}\textbf{+0.291} (19)\\[-1pt]{\scriptsize[+0.182, +0.416]}\end{tabular} & \begin{tabular}[t]{@{}c@{}}\textbf{+0.243} (15)\\[-1pt]{\scriptsize[+0.131, +0.366]}\end{tabular} & \begin{tabular}[t]{@{}c@{}}\textbf{+0.045} (15)\\[-1pt]{\scriptsize[+0.001, +0.086]}\end{tabular} & \begin{tabular}[t]{@{}c@{}}\textbf{+0.088} (17)\\[-1pt]{\scriptsize[+0.042, +0.132]}\end{tabular}\\
Chicago & Matched & C & \begin{tabular}[t]{@{}c@{}}\textbf{+0.371} (20)\\[-1pt]{\scriptsize[+0.275, +0.480]}\end{tabular} & \begin{tabular}[t]{@{}c@{}}\textbf{+0.445} (20)\\[-1pt]{\scriptsize[+0.334, +0.569]}\end{tabular} & \begin{tabular}[t]{@{}c@{}}\textbf{+0.347} (20)\\[-1pt]{\scriptsize[+0.221, +0.499]}\end{tabular} & – & –\\
Chicago & Two-purpose $(1,3)$ & C & \begin{tabular}[t]{@{}c@{}}\textbf{+0.296} (20)\\[-1pt]{\scriptsize[+0.213, +0.391]}\end{tabular} & \begin{tabular}[t]{@{}c@{}}\textbf{+0.361} (20)\\[-1pt]{\scriptsize[+0.261, +0.478]}\end{tabular} & \begin{tabular}[t]{@{}c@{}}\textbf{+0.267} (20)\\[-1pt]{\scriptsize[+0.163, +0.395]}\end{tabular} & – & –\\
Chicago & $\zeta=1$, $\rho=0.7$ & P & \begin{tabular}[t]{@{}c@{}}\textbf{+1.160} (19)\\[-1pt]{\scriptsize[+0.814, +1.504]}\end{tabular} & \begin{tabular}[t]{@{}c@{}}\textbf{+1.300} (18)\\[-1pt]{\scriptsize[+0.918, +1.683]}\end{tabular} & \begin{tabular}[t]{@{}c@{}}\textbf{+1.142} (18)\\[-1pt]{\scriptsize[+0.780, +1.514]}\end{tabular} & \begin{tabular}[t]{@{}c@{}}+0.042 (12)\\[-1pt]{\scriptsize[$-$0.004, +0.098]}\end{tabular} & \begin{tabular}[t]{@{}c@{}}+0.039 (12)\\[-1pt]{\scriptsize[$-$0.010, +0.094]}\end{tabular}\\
Chicago & Interaction $\delta=0.25$\textsuperscript{a} & C & \begin{tabular}[t]{@{}c@{}}\textbf{+0.238} (10)\\[-1pt]{\scriptsize[+0.162, +0.317]}\end{tabular} & \begin{tabular}[t]{@{}c@{}}\textbf{+0.309} (10)\\[-1pt]{\scriptsize[+0.239, +0.379]}\end{tabular} & \begin{tabular}[t]{@{}c@{}}\textbf{+0.210} (10)\\[-1pt]{\scriptsize[+0.120, +0.306]}\end{tabular} & – & –\\
Chicago & Matched, capacity error & P & \begin{tabular}[t]{@{}c@{}}\textbf{+0.303} (20)\\[-1pt]{\scriptsize[+0.216, +0.407]}\end{tabular} & \begin{tabular}[t]{@{}c@{}}\textbf{+0.395} (20)\\[-1pt]{\scriptsize[+0.296, +0.509]}\end{tabular} & \begin{tabular}[t]{@{}c@{}}\textbf{+0.276} (19)\\[-1pt]{\scriptsize[+0.167, +0.408]}\end{tabular} & \begin{tabular}[t]{@{}c@{}}\textbf{+0.138} (19)\\[-1pt]{\scriptsize[+0.107, +0.167]}\end{tabular} & \begin{tabular}[t]{@{}c@{}}\textbf{+0.195} (20)\\[-1pt]{\scriptsize[+0.159, +0.231]}\end{tabular}\\
Chicago & Two-purpose, capacity error & P & \begin{tabular}[t]{@{}c@{}}\textbf{+0.251} (19)\\[-1pt]{\scriptsize[+0.166, +0.344]}\end{tabular} & \begin{tabular}[t]{@{}c@{}}\textbf{+0.323} (20)\\[-1pt]{\scriptsize[+0.225, +0.433]}\end{tabular} & \begin{tabular}[t]{@{}c@{}}\textbf{+0.220} (17)\\[-1pt]{\scriptsize[+0.117, +0.344]}\end{tabular} & \begin{tabular}[t]{@{}c@{}}\textbf{+0.118} (20)\\[-1pt]{\scriptsize[+0.091, +0.145]}\end{tabular} & \begin{tabular}[t]{@{}c@{}}\textbf{+0.161} (20)\\[-1pt]{\scriptsize[+0.126, +0.197]}\end{tabular}\\
\end{longtable}